\global\def\draftcontrol{0}

   \def\versionno{Lectures on Open Effective Field Theory -- draft   }

\catcode`\@=11

\expandafter\ifx\csname draftcontrol\endcsname\relax\global\def\draftcontrol{0}
\fi

{\count255=\time\divide\count255 by 60
\xdef\hourmin{\number\count255}
\multiply\count255 by-60\advance\count255 by\time
\xdef\hourmin{\hourmin:\ifnum\count255<10 0\fi\the\count255}}
\def\draftdate{\number\month/\number\day/\number\year\ \ \ \hourmin }

\newcommand\makepapertitle{\par
  \begingroup
    \renewcommand\thefootnote{\@fnsymbol\c@footnote}%
    \def\@makefnmark{\rlap{\@textsuperscript{\normalfont\@thefnmark}}}%
    \long\def\@makefntext##1{\parindent 1em\noindent
            \hb@xt@1.8em{%
                \hss\@textsuperscript{\normalfont\@thefnmark}}##1}%
     \newpage
     \global\@topnum\z@   
     \@makepapertitle
     \thispagestyle{empty}\@thanks
  \endgroup
  \setcounter{footnote}{0}%
  \global\let\thanks\relax
  \global\let\makepapertitle\relax
  \global\let\@makepapertitle\relax
  \global\let\@thanks\@empty
  \global\let\@author\@empty
  \global\let\@date\@empty
  \global\let\@title\@empty
  \global\let\title\relax
  \global\let\author\relax
  \global\let\date\relax
  \global\let\and\relax
  \def\version{\let\version\@version\@gobble}
}
\def\@makepapertitle{%
  \newpage
   \ifnum\draftcontrol=1 {}
   \version\versionno
   \vskip 3em%
   \else
   \hfill\hbox to 3cm {\parbox{4cm}{\@pubnum}\hss}%
   \vskip 3em%
   \fi
   \begin{center}%
   \let \footnote \thanks
     {\LARGE {\@title}}%
     \vskip 1.5em%
     {\normalsize
       \lineskip .5em%
       \begin{tabular}[t]{l}%
         \@author
       \end{tabular}\par}%
     \vskip 1.5em%
     {\@bstract}%
     \end{center}%
     \vskip 1.5em
     \@date%
   \par
}

\gdef\@pubnum{}
\def\pubnum#1{%
  \gdef\@pubnum{#1}}

\gdef\@bstract{}
\def\Abstract#1{%
  \gdef\@bstract{%
   \parbox{\textwidth-0pc}{%
   \centerline{\bf Abstract}\penalty1000%
\kern.2cm%
\noindent
\renewcommand\baselinestretch{1.0}%
{#1}}}
}

\def\ps@paper{\let\@mkboth\@gobbletwo%
     \ifnum\draftcontrol=1
    \def\@oddfoot{\hbox to \textwidth{\tiny \versionno \hfil\tiny\draftdate}%
    \hskip -\textwidth \hbox to \textwidth{\hfil\rm\thepage\hfil}}%
     \else\def\@oddfoot{\hbox to \textwidth{\hfil\rm\thepage\hfil}}
     \fi
     \let\@evenfoot\@oddfoot
}

\def\body{\clearpage
          \pagestyle{paper}
    }
\newenvironment{acknowledgements}{%
\vskip 3.25ex
\noindent {\bf Acknowledgements}
}

\def\@version#1{\ifnum\draftcontrol=1
\typeout{}\typeout{#1}\typeout{}
\vskip3mm\centerline{\hbox{\fbox{\normalsize{\tt DRAFT -- #1 -- }
                   {\draftdate}}}}\vskip3mm
\fi}
\let\version\@version
\long\def\eqlabel#1{\ifnum\draftcontrol=1
                    \tag@false  
                    \tag*{(\theequation) \hbox to -0.2cm{\hspace{0cm}\small{#1}\hss}}
                    \refstepcounter{equation}
                    \edef\@currentlabel{\theequation}
                    \ltx@label{#1}          
                    \else
                    \label{#1}
                    \fi
                    }
\let\st@bibitem\@bibitem
\let\st@lbibitem\@lbibitem
\ifnum\draftcontrol=1
  \def\@bibitem#1{%
    \st@bibitem{#1}\a@@label{#1}\ignorespaces}
  \def\@lbibitem[#1]#2{%
    \st@lbibitem[#1]{#2}\a@@label{#2}\ignorespaces}
  \def\a@@label#1{%
    \gdef\a@lab{\smash{\normalfont\small#1}}
    \ifvmode
      \if@inlabel
        \global\setbox\@labels\hbox{%
          \llap{\a@lab\let\a@lab\relax
                \kern\@totalleftmargin\kern\marginparsep}%
          \box\@labels}%
      \fi
    \fi}
\fi

\documentclass[10pt]{article}

\usepackage{amsmath,amssymb,array,calc,epsfig,bm}
\usepackage{graphicx}        
\usepackage{multicol}        

\usepackage{amsmath,amssymb,bm,epsfig}

\usepackage{cite}
\usepackage[
      colorlinks=true,
      linkcolor=red,  
      urlcolor=blue,    
      filecolor=red,     
      linkcolor=blue,
      citecolor = red,
      pdfstartview=FitV,
      pdftitle={},
        pdfauthor={Paolo Benincasa},
        pdfsubject={},
        pdfkeywords={},
        pdfpagemode=None,
        bookmarksopen=true    
      ]{hyperref}

\usepackage{centernot}      
\usepackage{mathpazo}
\usepackage{stmaryrd}
\usepackage{pifont}
\usepackage{colortbl}
\usepackage{multirow,bigdelim}
\usepackage{arydshln}
\usepackage{tikz, pict2e}
\usepackage{geometry}
\usepackage{lmodern}
\usepackage{wrapfig}
\usepackage{float}
\usepackage{enumitem}
\usepackage[framemethod=default]{mdframed}
\usepackage{array}
\usepackage{mathrsfs}

\usepackage{placeins}
\usepackage{circuitikz}

\usepackage[T1]{fontenc}
\usepackage[utf8]{inputenc}
\usepackage[english]{babel}

\PassOptionsToPackage{table}{xcolor}

\usepackage{xspace}
\usepackage{ifthen}
\usepackage{epstopdf}
\usepackage{amsthm}

\usepackage{relsize}
\usepackage{tabularx}
\usepackage{booktabs}
\usepackage{soul}
\usepackage{pdfpages}
\usepackage[most]{tcolorbox}
\usepackage{capt-of}
\usepackage{xfrac}
\usepackage{empheq}
\usepackage{slashed}
\usepackage{cancel}
\usepackage{upgreek}
\usepackage[skip=1em]{parskip}

\usepackage{latexsym}
\usepackage{ifoddpage}

\usepackage{subcaption}

\usepackage{cancel}
\usepackage{comment}
\usepackage{shuffle}

\usepackage{savesym}
\savesymbol{Ref}
\savesymbol{erf}
\usepackage{physics}
\restoresymbol{phys}{Ref}
\restoresymbol{phys}{erf}

\usepackage[compat=1.1.0]{tikz-feynman}
\tikzfeynmanset{compat=1.1.0}

\definecolor{myblue}{RGB}{166, 193, 222}
\definecolor{myblue2}{RGB}{68, 114, 196}
\definecolor{mygreen}{RGB}{175, 218, 162}
\definecolor{mygreen2}{RGB}{133, 210, 110}
\definecolor{myorange2}{RGB}{255, 192, 0}

\newtcolorbox{shade}[1][]{
  top=-7pt,bottom=6pt,
  sharp corners=all,
  colback=white!95!black,
  colframe=white,
}

\newcolumntype{P}[1]{>{\centering\arraybackslash}p{#1}}
\newcolumntype{L}{>{$}l<{$}}

\usetikzlibrary{calc,backgrounds}
\usetikzlibrary{arrows,shapes,positioning,shadows,trees}
\usetikzlibrary{arrows.meta, shapes.geometric}
\usetikzlibrary{mindmap}
\usetikzlibrary{decorations.pathreplacing}
\usetikzlibrary{decorations.pathmorphing}
\usetikzlibrary{decorations.markings}
\usetikzlibrary{matrix}
\usetikzlibrary{patterns.meta}

\tikzstyle arrowstyle=[scale=1]
\tikzstyle directed=[postaction={decorate,decoration={markings,
    mark=at position .5 with {\arrow[arrowstyle]{stealth}}}}]
\tikzstyle reverse directed=[postaction={decorate,decoration={markings,
    mark=at position .5 with {\arrowreversed[arrowstyle]{stealth};}}}]

\ifnum\draftcontrol=1
\fi

\renewcommand\baselinestretch{1.25}
\renewcommand\section{\@startsection {section}{1}{\z@}%
                                   {-3.5ex \@plus -1ex \@minus -.2ex}%
                                   {2.3ex \@plus.2ex}%
                                   {\normalfont\large\bfseries}}
\renewcommand\subsection{\@startsection{subsection}{2}{\z@}%
                                   {-3.25ex\@plus -1ex \@minus -.2ex}%
                                   {1.5ex \@plus .2ex}%
                                   {\normalfont\normalsize\bfseries}}
\renewcommand\subsubsection{\@startsection{subsubsection}{3}{\z@}%
                                   {-3.25ex\@plus -1ex \@minus -.2ex}%
                                   {1.5ex \@plus .2ex}%
                                   {\normalfont\normalsize\it}}
\renewcommand\paragraph{\@startsection{paragraph}{4}{\z@}%
                                   {-3.25ex\@plus -1ex \@minus -.2ex}%
                                   {1.5ex \@plus .2ex}%
                                   {\normalfont\normalsize\bf}}

\numberwithin{equation}{section}

\def\revise#1       {\raisebox{-0em}{\rule{3pt}{1em}}%
                     \marginpar{\raisebox{.5em}{\vrule width3pt\
                     \vrule width0pt height 0pt depth0.5em
                     \hbox to 0cm{\hspace{0cm}{%
                     \parbox[t]{4em}{\raggedright\footnotesize{#1}}}\hss}}}}

\def\sqr#1#2{{\vcenter{\vbox{\hrule height.#2pt
 \hbox{\vrule width.#2pt height#1pt \kern#1pt
 \vrule width.#2pt}\hrule height.#2pt}}}}

\catcode`\@=12

\newcommand{\eg}{\textsl{e.g.}\xspace}

\newcommand{\ee}{e}

\newcommand{\boldmathsymbol}[1]{{\ensuremath{\boldsymbol{#1}}}}

\newcommand{\bmk}{\boldmathsymbol{k}}
\newcommand{\bmp}{\boldmathsymbol{p}}
\newcommand{\bmq}{\boldmathsymbol{q}}
\newcommand{\bmx}{\boldmathsymbol{x}}

\newcommand{\cs}{c_{_\mathrm{S}}}

\newcommand{\beq}{\begin{equation}}
\newcommand{\eeq}{\end{equation}}
\newcommand{\bea}{\begin{equation}\begin{aligned}}
\newcommand{\eea}{\end{aligned}\end{equation}}

\newlength{\wsingfig}
\newlength{\wdblefig}
\newlength{\wquadfig}
\newlength{\wtriplefig}
\newcommand{\Eq}[1]{Eq.~(\ref{#1})}
\newcommand{\Eqs}[1]{Eqs.~(\ref{#1})}
\newcommand{\Fig}[1]{Fig.~{\ref{#1}}}
\newcommand{\Figs}[1]{Figs.~{\ref{#1}}}

\newcommand{\Sec}[1]{Sec.~\ref{#1}}

\newcommand{\bs}[1]{\boldsymbol{#1}}

\newcommand{\exx}[1]{\langle\!\langle #1 \rangle\!\rangle}
\newcommand{\ex}[1]{\langle #1 \rangle}
\newcommand{\T}{\mathcal{T}}

\newcommand{\Mpl}{M_{\mathrm Pl}}

\renewcommand{\H}{\mathcal{H}}
\renewcommand{\O}{\mathcal{O}}

\def\pia{\pi_{a}}
\def\pir{\pi_{r}}
\def\dpir{\dot{\pi}_{r}}

\def\phia{\varphi_{a}}
\def\phir{\varphi_{r}}
\def\dphir{\dot{\varphi}_{r}}
\def\dphia{\dot{\varphi}_{a}}

\newcommand{\fini}{f}

\def\bfk{\textbf{k}}

\newcommand{\bfx}{{\mathbf{x}}}

\newcommand{\stuck}{St\"uckelberg }

\newcommand   \lptl{\raise .8ex\hbox{$^\leftarrow$} \hspace{-9pt} \partial}
\newcommand   \lrptl{\raise .8ex\hbox{$^\leftrightarrow$} \hspace{-9pt} \partial}

\renewcommand \Re    {\mathop{\mathrm{Re}}}
\renewcommand \Im    {\mathop{\mathrm{Im}}}

\newcommand   \be {\mathbf{e}}

\def\be#1\ee{\begin{align}#1\end{align}}

\newcounter{ssooeexercise}[section]
\renewcommand{\thessooeexercise}{\thesection.\arabic{ssooeexercise}}

\newtcolorbox{exbox}{
  colback=gray!10,
  colframe=gray!20,
  boxrule=1pt,
  arc=5pt,
  breakable,
  left=1em,
  right=1em,
  top=0.8em,
  bottom=0.8em
}

\newcommand{\newexercise}[3][]{%
  \refstepcounter{ssooeexercise}%
  \expandafter\gdef\csname ex@content@#2\endcsname{#3}%
  \expandafter\xdef\csname ex@number@#2\endcsname{\thessooeexercise}%
  \begin{exbox}%
    \textbf{Exercise~\thessooeexercise:} #3
    \ifx&#1&\else\label{#1}\fi
  \end{exbox}%
}

\newcommand{\printexercise}[1]{%
  \begin{exbox}%
    \textbf{Exercise~\csname ex@number@#1\endcsname:}
    \csname ex@content@#1\endcsname
  \end{exbox}%
}

\begin{document}

\title{\underline{\Large\bf The Disordered Universe 2025:}\\
        \Large\bf Open Effective Field Theories}

\pubnum{%
}

\author{%
	{\scshape Lecturer:} {\bf Thomas Colas}${}^{\circledast\,\circledcirc}$,%
    \\
    {\scshape Notes written by:} {\bf Alessandro Fumagalli}${}^{\star},$ {\bf Francesco Paronetto}${}^{\diamond}$, {\bf Leonardo Stanzani}${}^{\diamond},$\\
      \hspace{3.375cm} {\bf Mattia Varrone}${}^{\ast},$ {\bf Lucas Vicente Garc{\'i}a-Consuegra}${}^{\odot}$
    \\
    {\scshape Edited by:} {\bf Paolo Benincasa}${}^{\dagger\, \ddagger}$, {\bf Guilherme Leite Pimentel}${}^{\diamond}$
    \\ [0.4cm]
    \ttfamily ${}^{\circledast}$ Department of Applied Mathematics and Theoretical Physics, 
    \\
    \ttfamily $\phantom{{}^{\dagger}}$ University of Cambridge, Wilberforce Road, Cambridge, CB3 0WA, UK
    \\
    \ttfamily ${}^{\circledcirc}$ Universit{\'e} Paris Cit{\'e}, CNRS, 
    \\
    \ttfamily $\phantom{{}^{\dagger}}$ Astroparticule et Cosmologie, F-75013 Paris, France 
    \\
    \ttfamily ${}^{\star}$ Institute for Theoretical Physics,
    \\
    \ttfamily $\phantom{{}^{\dagger}}$ University of Amsterdam, 1090 GL Amsterdam, The Netherlands
    \\
    \ttfamily ${}^{\diamond}$ Scuola Normale Superiore,
    \\
    \ttfamily $\phantom{{}^{\ast}}$ Piazza dei Cavalieri 7, 56126, Pisa, Italy
    \\
    \ttfamily ${}^{\ast}$ Laboratory for Theoretical Fundamental Physics, Institute of Physics,
    \\
    \ttfamily $\phantom{{}^{\diamond}}$ {\'E}cole Polytechnique F{\'e}d{\'e}rale de Lausanne, Switzerland
    \\
    \ttfamily ${}^{\odot}$ Department of Physics,
    \\
    \ttfamily \ $\phantom{{}^{\odot}}$King's College London, Strand, London, WC2R 2LS, UK
    \\
    \ttfamily ${}^{\dagger}$ Asociaci{\'o}n Cultural La Ricotta, Espazo La Ricotta,
    \\
    \ttfamily $\phantom{{}^{\dagger}}$ As Barreiras, Castro Caldelas, E-32769 Galicia-Spain
    \\
    \ttfamily ${}^{\ddagger}$ Instituto Galego  de F{\'i}sica de Altas Enerx{\'i}as IGFAE,
    \\
    \ttfamily $\phantom{{}^{\ddagger}}$ Universidade de Santiago de Compostela, E-15782 Galicia-Spain
    \\ [0.4cm]
    \small \ttfamily tcolas@apc.in2p3.fr, a.fumagalli@uva.nl,  francesco.paronetto@sns.it, leonardo.stanzani@sns.it, \\
    \small \ttfamily   mattia.varrone@epfl.ch, lucas.vicente\_garcia-consuegra@kcl.ac.uk, \\
    \small \ttfamily pablowellinhouse@sindominio.net, guilherme.leitepimentel@sns.it\\[0.75cm]}

\Abstract{%
Effective field theories (EFTs) offer a powerful method to unify diverse models under a small set of control parameters, allowing systematic expansions around well-established theories. While these techniques were traditionally developed for particle physics experiments with clean, isolated initial states, realistic macroscopic environments are often noisy and dissipative. Recognising the limitations of traditional EFTs has recently sparked intense progress at the interface of high-energy physics and condensed matter. These developments motivate a new approach to gravitation and cosmology, one that models the gravitational sector as evolving in the presence of an unobservable medium. Open Effective Field Theories provide a systematic, controllable framework for this endeavour. This chapter presents this versatile framework and illustrates how the Schwinger--Keldysh formalism captures the influence of unknown environments — manifesting as dissipation and noise — on observational probes. These notes are based on a series of lectures held during the \href{https://indico.sns.it/event/91/}{{\it La Ricotta} Summer School: {\it The Disordered Universe 2025}} from 21~July to 1~August~2025.}

\makepapertitle

\body

\version\versionno

\newpage

\section*{Preface}
This article is a chapter from the book {\it The Disordered Universe: Open Systems, Scattering Amplitudes, de Sitter Space, and Disordered Systems in Cosmology}~\cite{Benincasa:2027tdu}, which collects the lectures delivered at the \href{https://indico.sns.it/event/91/}{\it La Ricotta Summer School: The Disordered Universe 2025}, held from 21~July to 1~August~2025 at the \href{https://blogs.sindominio.net/laricotta/#}{\it Espazo La Ricotta} in As Barreiras (Castro Caldelas, Spain). These lecture notes are intended as a pedagogical introduction to topics that are becoming increasingly relevant in cosmology. 

\newpage

\tableofcontents


\section{Introduction}\label{sec:intro}

Theoretical cosmology lies at the intersection of general relativity, quantum field theory, and statistical mechanics. Its predictive power relies on their simultaneous interplay: the expansion of the universe, quantum processes, and non-equilibrium dynamics all contribute to phenomena such as Big Bang nucleosynthesis, cosmic microwave background anisotropies, and large-scale structure formation.

At the same time, many of the central questions in cosmology — dark matter, dark energy, and the physics of inflation — remain inaccessible to direct experiment. They are instead probed indirectly through their gravitational and electromagnetic imprints. In this sense, they define an effective medium through which cosmological signals propagate, whose properties must be inferred rather than measured in isolation.

A natural framework for organising such ignorance is effective field theory. The EFT of Inflation \cite{Cheung:2007st} and the EFT of Dark Energy \cite{Gubitosi:2012hu} provide systematic, symmetry-based descriptions of cosmological dynamics, allowing one to parameterise deviations from minimal models in a controlled expansion. This approach efficiently captures locality, symmetry breaking patterns, and scale separation.

However, cosmology is \textbf{not} a closed system: observables are constructed from a restricted set of long-wavelength degrees of freedom, while short modes, heavy fields, and horizon-scale correlations are traced over. This coarse-graining induces dissipation, stochasticity, and decoherence, and plays a central role in the emergence of classical fluctuations. While traditional effective field theories provide a powerful local and unitary description, they must be extended to capture the intrinsically open, non-equilibrium nature of cosmological dynamics. 

This chapter addresses this question by developing a framework for open effective field theories describing matter coupled to gravity.
\begin{itemize}
	\item In \textbf{Section~\ref{sec:SK}}, we introduce the \emph{Schwinger--Keldysh formalism}, an approach for handling non-equilibrium and open quantum systems.
	\item In \textbf{Section~\ref{sec:scalar}}, we construct a first example of an open effective theory for relativistic scalar fields.
	\item In \textbf{Section~\ref{sec:lec3}}, we apply this framework to the scalar sector of single-clock inflation.
	\item In \textbf{Section~\ref{sec:lec4}}, we discuss how to extend the formalism to study dissipative and stochastic effects in gauge theories and gravity.
\end{itemize}

 
\paragraph{\it Nomenclature.} There are a lot of variations in the terminology used by different communities to describe open and non-equilibrium dynamics. In these notes, we mostly follow the nomenclature from \cite{breuerTheoryOpenQuantum2002} used in the open systems community. Unitarity refers to unitary evolution operator $\mathcal{U}(t,t_0)$ such that
\begin{align}\label{eq:unitevol}
	\rho(t_0) \rightarrow \rho(t) = \mathcal{U}(t,t_0)\,  \rho(t_0)\, \mathcal{U}^\dag(t,t_0),
\end{align}
with 
\begin{align}
	\mathcal{U}^\dag(t,t_0)\, \mathcal{U}(t,t_0)  = \mathcal{U}(t,t_0)\,    \mathcal{U}^\dag(t,t_0) =  \mathrm{Id},
\end{align}
and initial condition $\mathcal{U}(t_0,t_0) = \mathrm{Id}$. We further consider density matrices which are (i) normalised $\mathrm{Tr} \rho =1$, (ii) Hermitian $\rho^\dag = \rho$ and (iii) positive definite $\rho > 0$. As we will see at length, not all physical evolution can be written under the form of \Eq{eq:unitevol}. Whenever some degrees of freedom are experimentally inaccessible, initially available information can get distributed in the unknown environment and eventually lost. Such evolution maps pure to mixed states which can be computed by performing a Schwinger-Keldysh path integral. This path integral is controlled by an effective functional that we decompose into
\begin{align}\label{eq:Seffintro}
	S_{\mathrm{eff}}\left[ \varphi_+, \varphi_-\right] = S_{\mathrm{unit}}\left[\varphi_+\right] - S_{\mathrm{unit}}\left[\varphi_-\right] + S_{\mathrm{non-unit}}\left[\varphi_+,\varphi_- \right],
\end{align}
for a generic scalar $\varphi$, where $S_{\mathrm{unit}}$ alone leads to evolution of the form of \Eq{eq:unitevol} while $S_{\mathrm{non-unit}}$ contains terms mixing the $+$ and $-$ branches of the path integral. These terms are the ones responsible for driving pure states into mixed states. Later on, we will encounter a convenient basis known as the Keldysh basis, in which the field $\varphi$ is decomposed into a retarded $\varphi_r$ and advanced component $\varphi_a$ through
\begin{align}
	\varphi_r = \frac{\varphi_+ + \varphi_-}{2}, \qquad \varphi_a = \varphi_+ - \varphi_-.
\end{align}
We will often call \textit{stochastic} the operators $\mathcal{O}(\varphi_a^{2p})$ that are even in powers of $\varphi_a$. These contributions never come from the unitary part of the functional and relate to the noise sourcing the open systems. The operators  $\mathcal{O}(\varphi_a^{2p + 1})$ that are odd in powers of $\varphi_a$ can always be decomposed into a unitary contribution of the form $S_{\mathrm{unit}}\left[\varphi_+\right] - S_{\mathrm{unit}}\left[\varphi_-\right]$ and a remaining part, intrinsically non-unitary, that we refer to as being dissipative. Note that this terminology has nothing to do with the eventual conservation laws that the system may have.\\


\section{Schwinger-Keldysh formalism}\label{sec:SK}

Primordial cosmology relies on the correspondence between late-time observables and early-time cosmological correlators. While the former involves classical statistics of cosmological tracers (temperature anisotropies of the CMB, density contrast of the LSS, $\cdots$), the latter features quantum expectation values of quantum field theoretic operators. Schematically, 
\begin{align} \label{eq:corresp}
	\langle \prod_{i = 1}^n \delta(\bmk_i) \rangle \quad \leftrightarrow \quad \langle \prod_{i = 1}^n \widehat{\phi} (\bmk_i)\rangle.
\end{align}
While many lecture notes (e.g. \cite{Baumann:2009ds}) and textbooks (e.g. \cite{Dodelson:2003ft}) develop the physical origin of the relation between the right and left hand sides of the above equation, our goal in this Section consists in reviewing the perturbative computation of the right hand side only - that is the computation of cosmological correlators through the Schwinger-Keldysh formalism. The presentation closely follows \cite{Chen:2017ryl}.


\subsection{Cosmological correlators}

Let us consider a homogeneous and isotropic universe described by a FLRW background metric
\begin{align}
	\dd s^2 = a^2(\eta) \left( - \dd \eta^2 + \dd \bmx^2 \right).
\end{align}
The scale factor $a(\eta)$ expressed in terms of the conformal time $\eta$ controls the expansion of the universe. During inflation, approximating the universe by a de Sitter geometry, we have 
\begin{align}
	a(\eta) = - \frac{1}{H \eta}, \qquad \eta \in \left] - \infty, \eta_0 \right],
\end{align}  
where the Hubble parameter $H$ is a constant (up to slow-roll corrections). The conformal time $\eta_0$ corresponds to the end of inflation and the beginning of the hot Big-Bang, which is often taken to be $\eta_0 \rightarrow 0$.

On the top of the background geometry, we consider a massless scalar denoted $\phi(\eta, \bmx)$ with a speed of sound $c_s^2$. Its linear action is given by
\begin{align}\label{eq:zetaaction}
	S^{(2)}_\phi &= - \frac{1}{2}\int \dd \eta \mathrm{d}^3 \bs{x}   a^2(\eta)  \left[c_s^{-2}\phi^{\prime 2} -\left(\partial_i \phi\right)^2\right],
\end{align}
and can be thought of as a proxy for scalar curvature perturbations produced in an inflationary scenario, as explained in \cite{Maldacena:2002vr}.

One can Fourier transform the field variable
\begin{align}\label{eq:Fourier}
		\phi_\bmk(\eta) = \int \dd^3 \bmx  \phi(\eta, \bmx)\ee^{i \bmk.\bmx}.
\end{align}	
in order to take advantage of the spatial homogeneity. 
Following the canonical quantisation prescription, field variables are promoted to quantum operators obeying the equal-time commutation relations 
\begin{align}
	[\widehat{\phi}_{\bs{k}}(\eta),\widehat{\Pi}_{\bs{q}}(\eta) ] = i \delta(\boldsymbol{k}+\bs{q}),
\end{align}
with $\widehat{\Pi}_{\bs{q}}$ the conjugate momentum.
Making use of the linear evolution, one can relate the field operators at time $\eta$ to $\widehat{a}_{\bs{k}}$ and $\widehat{a}^{\dag}_{-\bs{k}}$, the creation and annihilation operators of the Bunch-Davies vacuum $\ket{\Omega}$ in the asymptotic past through the mode-function decomposition
\begin{align}\label{eq:modefctdecomp}
	\widehat{\phi}_{\bs{k}}(\eta) &= \phi_{k}(\eta) \widehat{a}_{\bs{k}} +  \phi^{*}_{k}(\eta) \widehat{a}^{\dag}_{-\bs{k}} 
\end{align}
The mode-functions obey the classical equation of motion. Explicitly, rescaling $\phi$ by the scale factor $a$, the equation of motion for $\phi$ takes the familiar form of a parametric oscillator 
\begin{align}\label{eq:MSeom}
	(a\phi_{k})'' + \left(c_s^2 k^2 - \frac{a''}{a} \right) (a\phi_{k})=0.
\end{align}
Choosing the outgoing branch:
\begin{align}\label{eq:BDmode}
	\phi_{k}(\eta)=\frac{H}{\sqrt{2 c_s k^3}} \left(1+ i c_s k \eta \right)\ee^{-i c_s k \eta}.
\end{align}
The Hilbert space of the $\phi$ field is constructed out of the tensor product of each mode's Fock space, that is $\mathcal{H} = \otimes_\bmk \mathcal{H}_\bmk$ with
\begin{align}\label{eq:Hspace}
	\mathcal{H}_\bmk = \left\{ \ket{\Omega} , \widehat{a}^\dag_{\bs{k}} \ket{\Omega}, \widehat{a}^\dag_{\bs{k}} \widehat{a}^\dag_{\bs{k}} \ket{\Omega}, \cdots \right\},
\end{align}
the Bunch-Davies vacuum being defined from $\widehat{a}_{\bs{k}} \ket{\Omega} = 0$.

We are now in position to define the observables presented on the RHS of \Eq{eq:corresp} we aim to compute. Cosmological correlators are defined as the expectation value of the Bunch-Davies vacuum of equal-time product of local operators at the future conformal boundary of dS,
\begin{align}\label{eq:def}
	\langle \prod_{i = 1}^n \widehat{\phi}(\bmk_i)\rangle \equiv \lim_{\eta_{0} \rightarrow 0} \langle \Omega | \prod_{i = 1}^n  \widehat{\phi}_{\bmk_i}(\eta_0)|\Omega \rangle.
\end{align}
Note that the terminology \textit{observable} may be misleading. Contrarily to scattering amplitudes, cosmological correlators are not field redefinition invariant. Instead, correlators dictate the summary statistics of the field $\phi$ at time $\eta_0$ (its mean, variance, skewness, kurtosis, $\cdots$), which might be more familiar in the context of statistical field theory. The knowledge of the infinite tower of cosmological correlators is equivalent to the knowledge of the full Probability Distribution Function (PDF) for $\phi$.\footnote{Which is \textit{not} the complete characterisation of the quantum state $ \widehat{\rho}_\Omega = |\Omega \rangle \langle \Omega |$, but only the diagonal density matrix element in the field basis $P_\Omega(\phi_\bmk, \eta_0) = \langle \phi_\bmk, \eta_0 |\widehat{\rho}_\Omega | \phi_\bmk, \eta_0 \rangle$. Off-diagonal density matrix elements $\langle \widetilde{\phi}_\bmk, \eta_0 |\widehat{\rho}_\Omega  | \phi_\bmk, \eta_0 \rangle$ or, equivalently, correlators of the conjugate momentum $\widehat{\Pi}_\bmk$ are needed to fully characterise the state \cite{Colas:2024ysu}.} 


\subsection{In-in formalism}

The \textit{in-in formalism} \cite{Weinberg:2005vy, Weinberg:2006ac, Chen:2017ryl} is a well-established framework aiming at computing perturbatively cosmological correlators. Analogously to the usual flat space perturbation theory, this formalism can be developed in the operator (or Hamiltonian) representation or using the path integral. While the former has the advantage of making computations explicit, the latter is more suitable to introduce dissipative effects.


\subsubsection{Path integral representation}
For the reason explained above, the path integral representation is the one that we will mostly use, and we review here. A detailed derivation and equivalence with the operator formalism, as developed by Weinberg \cite{Weinberg:2005vy, Weinberg:2006ac}, can be found in \cite{Boyanovsky:2015xoa, Colas:2025lectures}.
Let us consider the Heisenberg's picture of our cosmological correlators given in \Eq{eq:def}. An equivalent representation of the quantum state $\ket{\Omega}$ is given in terms of the \textit{density matrix} $\widehat{\rho}_\Omega \equiv \ket{\Omega} \bra{\Omega}$ \cite{breuerTheoryOpenQuantum2002}. The advantage of working with a density matrix is that it accommodates both \textit{pure} and \textit{mixed} states.
In terms of the density matrix, \Eq{eq:def} is restated 
\begin{align}
		\langle \prod_{i = 1}^n \widehat{\phi}(\bmk_i)\rangle = \lim_{\eta_{0} \rightarrow 0} \mathrm{Tr}_{\mathcal{H}} \left[\prod_{i = 1}^n \widehat{\phi}_{\bmk_i}(\eta_0) \widehat{\rho}_\Omega  \right]
\end{align}
where the trace is over the full Hilbert space defined above \Eq{eq:Hspace}. One can explicitly compute this trace in a given basis. For instance, considering that the theory only contains one scalar degree of freedom $\phi$, one can express this expectation value in the field basis constructed out of the eigenstates of the field operators in the Heisenberg picture
\begin{align}
	\widehat{\phi}_\bmk(\eta_0) |\phi_\bmk ,\eta_0  \rangle = \phi_\bmk(\eta_0) |\phi_\bmk ,\eta_0 \rangle.
\end{align}
In this basis, the expectation value becomes
\begin{align}
	\langle \prod_{i = 1}^n \widehat{\phi}(\bmk_i)\rangle &= \lim_{\eta_{0} \rightarrow 0} \int \dd \phi_\bmk \langle \phi_\bmk, \eta_0 |  \prod_{i = 1}^n \widehat{\phi}_{\bmk_i}(\eta_0) \widehat{\rho}_\Omega|\phi_\bmk, \eta_0\rangle \label{eq:obsref}\\
	&= \lim_{\eta_{0} \rightarrow 0} \int \dd \phi_\bmk \int \dd  \widetilde{\phi}_\bmq \langle \phi_\bmk, \eta_0 |  \prod_{i = 1}^n \widehat{\phi}_{\bmk_i}(\eta_0)  |\widetilde{\phi}_\bmq , \eta_0\rangle \langle \widetilde{\phi}_\bmq , \eta_0|\widehat{\rho}_\Omega|\phi_\bmk, \eta_0 \rangle
\end{align}
where we used in the second line a resolution of the identity
\begin{align}
    \mathbb{I} =  \int \dd \widetilde{\phi}_\bmq |\widetilde{\phi}_\bmq , \eta_0\rangle \langle \widetilde{\phi}_\bmq , \eta_0|.
\end{align}

\begin{figure}[t]
	\centering
	\includegraphics{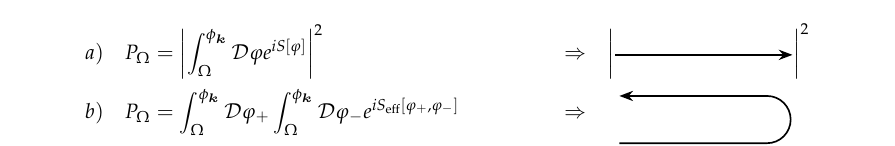}
 	\caption{Comparison between a) the single-branch path integral contour used in the computation of the wavefunction, and b) the double-branch path integral contour used in the computation of the density matrix. While the two approaches are equivalent in the case of pure states (see e.g. \cite{Donath:2024utn}), the in-in contour also accommodates mixed states, which have no single-branch analogue.}
	\label{fig:contourunit}
 \end{figure}

Since cosmological correlators are diagonal in the field basis, 
\begin{align}
	\langle \phi_\bmk ,  \eta_0|  \prod_{i = 1}^n \widehat{\phi}_{\bmk_i}(\eta_0)  |\widetilde{\phi}_\bmq, \eta_0 \rangle = \prod_{i = 1}^n \phi_{\bmk_i}(\eta_0) \delta(\phi - \widetilde{\phi} ) ,
\end{align}
all the information of interest is contained in the diagonal element of density matrix in the field basis 
\begin{align}\label{eq:PDFdef}
	P_\Omega[\phi_\bmk(\eta_0)] \equiv \langle \phi_\bmk, \eta_0 | \widehat{\rho}_\Omega | \phi_\bmk, \eta_0 \rangle.
\end{align}
This object defines a Probability Distribution Function (PDF) from which one extracts expectation values in the usual way,
\begin{align}\label{eq:exppdf}
	\langle \prod_{i = 1}^n \widehat{\phi}_{\bmk_i}\rangle &= \lim_{\eta_{0} \rightarrow 0}  \int \dd \phi_\bmk \prod_{i = 1}^n \phi_{\bmk_i}(\eta_0) P_\Omega[\phi_\bmk( \eta_0)]. 
\end{align}
When the state is pure ($\widehat{\rho}_\Omega^2 = \widehat{\rho}_\Omega$), one can think of this PDF as the modulus square of a wavefunction (or better, a wavefunctional): $P_\Omega[\phi_\bmk(\eta_0)] = |\Psi[\phi_\bmk(\eta_0)]|^2$, where the wavefunction is obtained as 
\begin{align}
	\Psi[\phi_\bmk(\eta_0)] &= \langle \phi_\bmk ,\eta_0 | \Omega \rangle = \langle \phi_\bmk | \Psi(\eta_0) \rangle = \int_{\Omega}^{\phi_\bmk} \mathcal{D} \varphi_+ \ee^{i S[\varphi_+]}.
\end{align}
Notice that this expression only features a single-branch path integral, whereas a density matrix necessitates the use of double-branch path integrals,
\begin{align}
	P_\Omega[\phi_\bmk(\eta_0)]& = \langle \phi_\bmk, \eta_0 | \Omega \rangle \langle \Omega | \phi_\bmk, \eta_0 \rangle \\
	&= \int_{\Omega}^{\phi_\bmk} \mathcal{D} \varphi_+  \int_{\Omega}^{\phi_\bmk} \mathcal{D} \varphi_- \ee^{i \left(S[\varphi_+] - S[\varphi_-]\right)}.\label{eq:unitth}
\end{align} 
Such a path integral contour is presented in \Fig{fig:contourunit}. Known as the \textit{Schwinger-Keldysh}, \textit{in-in} or \textit{closed-time} path contour, it can be intuitively understood in the following manner. Starting from some initial state $|\Omega\rangle$, the path integral evolves the dynamics forward up to $\eta_0$, then backward to initial state $\langle \Omega |$ again. Compared to the familiar \textit{in-out} formalism used in particle physics which specifies the state both in the asymptotic past and asymptotic future, the Schwinger-Keldysh formalism only relies on initial conditions, leaving the state at time $\eta_0$ unconstrained. It allows one to access transient dynamics which can be really far-from-equilibrium, which is the primary reason for introducing such a formalism.

The second important feature of the in-in contour is the apparent doubling of the degrees of freedom of the theory, one for each branch of the path integral: $\varphi_+$ for the forward branch and $\varphi_-$ for the backward branch. It is important to bear in mind that this is nothing but a trick to capture statistical and quantum fluctuations in the formalism. Physically, there is only one dynamical degree of freedom, a fact we can make manifest by computing dispersion relations and propagators. 

At last, note that the exponential in \Eq{eq:unitth} factorises between the $+$ branch contribution and the $-$ branch contribution, $S[\varphi_+] - S[\varphi_-]$. This is a peculiarity of \textit{unitary/closed} theories in which information is conserved.  When $\phi$ exchanges information with hidden sectors, or more generally, sectors that have been integrated out, this is not true, and it leads to the emergence of \textit{mixed states}. At the level of the path integral, this will correspond to investigating the appearance of terms mixing the branches of the path integral, leading to $S_{\mathrm{eff}} = S[\varphi_+] - S[\varphi_-] + F[\varphi_+, \varphi_-]$. The generic construction of these effective functionals describing pure and mixed states in a single way is the object of Lecture 2.

\subsubsection{In-in diagrammatics}\label{subsubsec:diagram}

We now aim at computing \Eq{eq:exppdf} in perturbation theory. To do so, we spell out a set of Feynman rules and associated diagrammatics. 

\begin{figure}[t]
	\centering
	\includegraphics{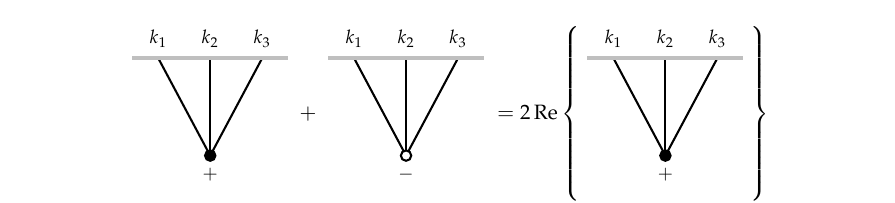}
	\caption{In-in diagrammatics of tree-level three-point correlator. The ``$+$'' and ``$-$'' diagram are related by complex conjugations for parity even interactions.
	}
	\label{fig:bispectrumpm}
\end{figure}

Invariance by spatial translation (homogeneity) implies that correlators must contain at least one Dirac delta for momentum conservation. While disconnected diagrams, proportional to two or more Dirac deltas, can always be subtracted by removing all products of lower $n$-point functions, connected correlators, proportional to a single Dirac, are the main objects of interest at a given order $n$. We introduce the notation: 
\begin{align}
		\langle \prod_{i = 1}^n \widehat{\phi}(\bmk_i)\rangle &\equiv (2\pi)^3 \delta\Big(\sum_{i = 1}^n \bmk_i\Big) B_n(\{\bmk\}). 
\end{align}
To compute this quantity in perturbation theory, we separate the free action from the interactions, 
\begin{align}
	S[\varphi] = S_{0}[\varphi] + S_{\mathrm{int}}[\varphi]
\end{align}
where $S_{\mathrm{int}}[\varphi]$ is at least cubic in $\varphi$. While the $S_{0}[\varphi]$ part can be exactly solved to obtain the propagators of the theory, we perform a systematic expansion in $S_{\mathrm{int}}$, leading to
\begin{align}
	B_n(\{\bmk\}) &= \int \dd \phi_\bmk \int_{\Omega}^{\phi_\bmk} \mathcal{D} \varphi_+  \int_{\Omega}^{\phi_\bmk} \mathcal{D} \varphi_- \bigg\{ \varphi_+^n \Big( 1 + i S_{\mathrm{int}}[\varphi_+] -  i S_{\mathrm{int}}[\varphi_-] \\
    - \frac{1}{2} S^2_{\mathrm{int}}[\varphi_+]  &- \frac{1}{2} S^2_{\mathrm{int}}[\varphi_-] \nonumber +  \frac{1}{2} S_{\mathrm{int}}[\varphi_+]  S_{\mathrm{int}}[\varphi_-] + \cdots \Big) \bigg\}  \ee^{i \left(S_0[\varphi_+] - S_0[\varphi_-]\right)} - \text{(disconnected diagrams)}
\end{align}
where we inserted $\varphi_+^n$ to compute the associated $n$-point function \cite{Chen:2017ryl}.\footnote{One could equally insert $\varphi_-^n$, given the boundary conditions considered.} In the end, it amounts to evaluate these various correlators in the free theory using Wick contractions:
\begin{align}
	B_n(\{\bmk\})& = \exx{ \varphi_+^n }^{(0)} + i \exx{ \varphi_+^n S_{\mathrm{int}}[\varphi_+] }^{(0)}  - i \exx{ \varphi_+^n S_{\mathrm{int}}[\varphi_-] }^{(0)} - \frac{1}{2} \exx{ \varphi_+^n S^2_{\mathrm{int}}[\varphi_+] }^{(0)} \nonumber \\
	&\qquad  - \frac{1}{2} \exx{ \varphi_+^n S^2_{\mathrm{int}}[\varphi_-] }^{(0)} + \exx{ \varphi_+^n S_{\mathrm{int}}[\varphi_+] S_{\mathrm{int}}[\varphi_-]}^{(0)} + \cdots - \text{(disconnected diagrams)},
\end{align}
where $\exx{ \cdots }^{(0)}$ stands for correlators evaluated in the free theory. Reproducing this computation every time we have to evaluate a correlator is tedious; this is why people develop a set of rules to perform these steps in a systematic manner:

\begin{tcolorbox}[%
	enhanced, 
	colback=gray!10,
	colframe=gray!10,
	breakable,
	skin first=enhanced,
	skin middle=enhanced,
	skin last=enhanced,
	before upper={\parindent15pt},
	]{}
	
	\vspace{0.05in}

	
\paragraph{\it Feynman rules.} 
\begin{enumerate}
	\item To compute $B_n$ in perturbation theory, draw a diagram with $V$ vertices, $I$ internal lines, each connecting two vertices, and $n$ external lines connecting a vertex to the future boundary $\eta_0 \rightarrow 0$ represented by a horizontal line at the top of the diagram, see e.g. \Fig{fig:bispectrumpm}.  Times run from bottom at $\eta \rightarrow - \infty$ to top at $\eta \rightarrow \eta_0$.
	\item Each vertex can either be a ``$+$'' vertex, coming from $S_{\mathrm{int}}[\varphi_+]$ or a ``$-$'' vertex, coming from $S_{\mathrm{int}}[\varphi_-]$. The final expression for $B_n$ is obtained by summing over the $2^V$ ways to label the $V$ vertices.
	\item To each of the $n$ external lines, associate a spatial momentum $\bmk_i$. Internal momenta are associated to $p_m$, with $m = 1, \cdots,I$.
	\item Internal lines represent four different types of bulk-to-bulk propagators:\footnote{We follow the notations in \cite{Chen:2017ryl}. Notice that the four bulk-to-bulk propagators are not linearly independent due to the identity $D_{++}+D_{--} = D_{+-} + D_{-+}$.}
	\begin{align}
		\label{eq_dSDmp}
		&~D_{-+}(k;\eta_1,\eta_2) = \phi_k(\eta_1)\phi^*_k(\eta_2) ,\quad ~D_{+-}(k;\eta_1,\eta_2) = \big[D_{-+}(k;\eta_1,\eta_2)\big]^* ,\\
		\label{eq_dSDpp}
		&~D_{\pm\pm}(k;\eta_1,\eta_2) = D_{\mp\pm}(k;\eta_1,\eta_2)\theta(\eta_1-\eta_2) + D_{\pm\mp}(k;\eta_1,\eta_2)\theta(\eta_2-\eta_1), \bigg.
	\end{align}
	depending whether  $+$ or $-$ vertices are connected by these propagators. $\phi_k(\eta_1)$ is the Bunch-Davies mode function found in \Eq{eq:BDmode}. External lines are associated to the bulk-to-boundary propagators
	\begin{align}
		\label{eq_flatK}
		K_+(k,\eta) = \phi^*_k(\eta)\phi_k(\eta_0) ,\qquad K_-(k,\eta)= \big[ K_+(k,\eta) \big]^*.
	\end{align}
	\item Vertices controlled by a coupling constant $g$ are associated to 
	\begin{align}
		\pm i g \int^{\eta_0}_{-\infty(1 \mp i \epsilon)} \frac{\dd \eta_A}{(H \eta_A)^4} , \qquad \text{for}~A = 1 , \cdots V, 
	\end{align}
	where the $\mp i \epsilon$ are there to project onto the interacting vacuum in the asymptotic past in the $+$ and $-$ contour respectively \cite{Kaya:2018jdo} and the $(H \eta_A)^{-4}$ comes from the volume factor $\sqrt{-g}$. The $\pm$ in front depend whether it is a $+$ or a $-$ vertex that is considered. 
	Moreover, each vertex comes with a delta conserving spatial momenta, a consequence of spatial homogeneity. All internal momenta $\bmp_m$ should be integrated over. Using the topological identity $I - V + 1 = L$, there should remain in the end $L$ internal momenta integrals to perform, corresponding to the $L$ loops. 
	\item Combinatorial factors may need to be added, when permutations generate the same contribution. Check \cite{Chen:2017ryl} for a detailed discussion.
\end{enumerate}
In the absence of parity odd operators, there exists a simple relation between diagrams with $V$ ``$+$'' vertices and $\bar{V}$ ``$-$'' vertices denoted $B^{V \tilde{V}}_n(\{\bmk\})$, and the same diagram with exchanged ``$+$'' and ``$-$'' vertices, $B^{\bar{V} V}_n(\{\bmk\}) = (B^{V \bar{V}}_n(\{\bmk\}))^*$, see \Fig{fig:bispectrumpm}.
\end{tcolorbox} 

Let's illustrate these diagrammatic rules through a simple example. Consider the leading cubic interactions in single-clock models of inflation \cite{Cheung:2007st}
\begin{align}
	S_{\mathrm{int}} = \int \dd^3 \bmx  \int \frac{\dd \eta}{(H \eta)^4} (- H \eta^3)\left[ -\frac{\lambda_1}{3!} \varphi^{\prime 3} - \lambda_2 \varphi' (\partial_i \varphi)^2 \right].
\end{align}
We explicitly treat the $\lambda_1$ interaction. The associated contact bispectrum, represented in \Fig{fig:bispectrumpm}, is given by 
\begin{align}
	B_3^{\lambda_1} = 2 \Re \left[ - i \frac{\lambda_1}{3!} \times 3! \times \int_{-\infty(1 - i \epsilon)}^{\eta_0} \frac{\dd \eta}{(H \eta)^4} (- H\eta)^3 K'_+(k_1,\eta) K'_+(k_2,\eta) K'_+(k_3,\eta) \right].
\end{align}
Let us discuss term by term the various contributions. First, the real part comes from the fact that the ``$-$'' diagram is nothing but the complex conjugate of the ``$+$'', as shown in \Fig{fig:bispectrumpm}. The $- \lambda_1/3!$ is the vertex contribution, while the $3!$ is the combinatorial factor, coming from the indistinguishability between $k_1$, $k_2$ and $k_3$. The integrand also contains the $(-H\eta)^3$ factor coming from the vertex and the bulk-to-boundary propagators for the three field insertions. The latter feature a conformal time derivative since the interaction term had time derivatives of fields $\varphi^{\prime 3}$ and not the field itself $\varphi^3$. Injecting the expression of $\phi_k(\eta)$ found in \eqref{eq:BDmode} into the bulk-to-boundary propagator expression \eqref{eq_flatK}, we obtain 
\begin{align}
	B_3^{\lambda_1} &=  2 \Re \left[ i \lambda_1  \int_{-\infty(1 - i \epsilon)}^{\eta_0} \frac{\dd \eta}{(H \eta)} \prod_{i=1}^3 \frac{H^2}{2 c_s k_i^3} c_s^2 k_i^2 \eta \ee^{i c_s k_i \eta} \right] \\
	&=  2 \Re \left[ i\frac{ \lambda_1 H^5 c_s^3}{8 k_1 k_2 k_3} \int_{-\infty(1 - i \epsilon)}^{0} \dd \eta \eta^2 \ee^{i E_T \eta} \right] \\
	&= - \frac{\lambda_1 H^5 c_s^3}{2 k_1 k_2 k_3 E_T^3},
\end{align}
where in the last line we defined the total ``energy'' $E_T \equiv c_s (k_1 + k_2 + k_3)$ and took the $\eta_0 \rightarrow 0$ limit. The contribution from the $\lambda_2$ interaction can be computed through similar methods, though leading to much more involved results \cite{Pajer:FTC}. In the \textit{Problem Set} part \ref{subsec:prob1} of this Lecture, the reader can practice in-in diagrammatics through simpler examples.


\subsection{\textit{Problem set}}\label{subsec:prob1}


\begin{exbox}
\paragraph{\it Flat space in-in diagrams}

Consider a massless scalar field in flat spacetime where the metric is given by
\begin{equation}
	\mathrm ds^2 = -\mathrm d t^2+ \mathrm d \bm x^2, \qquad \quad t \in ] - \infty, 0],\quad  \bm x \in \mathbb{R}^3.
\end{equation}
The action is given by 
\begin{align}
	S = \int \dd^4 x  \left\{\frac{1}{2}\left[\dot{\varphi}^2 - (\partial_i \varphi)^2 \right] + \mathcal{L}_{\mathrm{int}} \right\}.
\end{align}
\begin{enumerate}
	\item Write down the mode function of the field.
	\item Write down the propagators.
	\item Compute $B_3(k_1, k_2,k_3)$ at order $\lambda$ for $\mathcal{L}_{\mathrm{int}} = -(\lambda/3!) \varphi^3$.
	\item Compute $B_3(k_1, k_2,k_3)$ at order $\lambda_t$ for $\mathcal{L}_{\mathrm{int}} = (\lambda_t/2) \dot{\varphi}^2 \varphi$.
	\item Compute $B_3(k_1, k_2,k_3)$ at order $\lambda_s$ for $\mathcal{L}_{\mathrm{int}} = - (\lambda_s/2) (\partial_i \varphi)^2 \varphi$.
	\item Comment the common structure of these correlators. It might be useful to consider $B_n(k_1,\cdots,k_n)$ at order $\lambda_n$ for $\mathcal{L}_{\mathrm{int}} = -(\lambda_n/n!) \varphi^n$.
\end{enumerate}
\end{exbox}


\paragraph{\textit{Solutions}}

\begin{enumerate}
\item Upon Fourier transforming the field, the equations of motion are $(\partial_0^2+k^2)\varphi=0$ with $k^2=\sum_i k_i^2$, hence the (forward propagating) mode function is
\begin{align}
    \phi_k(k,t)=A e^{-i k t}\,.
\end{align}
where $A$ is a complex constant and $k=\sqrt{k^2}$, this has to be chosen to normalise the Wronskian:
\begin{align}
    2i \text{Im}\left[\phi_k(t)\dot \phi^*_k(t)\right]=i \implies \phi_k(t)=\frac{e^{-i kt}}{\sqrt{2k}}\,.
\end{align}
\item The bulk to boundary propagator is (we choose $t_0=0$) \begin{align}
    K_+(k,t)=\phi_k^*(t)\phi_k^*(t_0=0)= \frac{e^{i kt}}{\sqrt{2k}}\,,\quad K_-(k,t)=\frac{e^{-i kt}}{\sqrt{2k}}\,.
\end{align}
The bulk to bulk propagator instead is
\begin{align}
   & D_{-+}(k;t_1,t_2)=\phi_k(t_1)\phi_k^*(t_2)=\frac{e^{ik(t_1-t_2)}}{2k}\,,\quad D_{+-}=[D_{-+}(k;t_1,t_2)]^*\\
   & D_{\pm \pm}(k;t_1,t_2)=D_{\mp \pm}(k;t_1,t_2) \theta(t_1-t_2)+D_{\pm \mp}(k;t_1,t_2) \theta(t_2-t_1)\,.
\end{align}
\item Let us start with the simplest interaction $\mathcal{L}= -\frac{\lambda}{3!}\varphi^3$, the trispectrum is given by the following diagrams 
\begin{equation}
    B_3(k_1,k_2,k_3)=  \raisebox{-33pt}{
\begin{tikzpicture}[line width=1. pt, scale=2.2]
\draw[black, line width=1.pt] (0,0) -- (-0.4,0.75);
\draw[black, line width=1.pt] (0,0) -- (0.4,0.75);
\draw[black, line width=1.pt] (0,0) -- (0,0.75);
\draw[lightgray, line width=2.pt] (-0.6,0.75) -- (.6,0.75);
\draw[fill=black] (0,0) circle (.04cm);
\node[scale=.9] at (0,-0.15) {$+$};
\node[scale=.9] at (-0.4,0.9) {$k_1$};
\node[scale=.9] at (0,0.9) {$k_2$};
\node[scale=.9] at (0.4,0.9) {$k_3$};
\end{tikzpicture}
} +
        \raisebox{-33pt}{
\begin{tikzpicture}[line width=1. pt, scale=2.2]
\draw[black, line width=1.pt] (0,0) -- (-0.4,0.75);
\draw[black, line width=1.pt] (0,0) -- (0.4,0.75);
\draw[black, line width=1.pt] (0,0) -- (0,0.75);
\draw[lightgray, line width=2.pt] (-0.6,0.75) -- (.6,0.75);
\draw[fill=white] (0,0) circle (.04cm);
\node[scale=.9] at (0,-0.15) {$-$};
\node[scale=.9] at (-0.4,0.9) {$k_1$};
\node[scale=.9] at (0,0.9) {$k_2$};
\node[scale=.9] at (0.4,0.9) {$k_3$};
\end{tikzpicture}
} +\text{permutations}\,.
\end{equation}
The sum over permutations gives exactly $3!$ terms of the same type, which cancels with the normalisation of the coupling $\lambda$. Let us compute the first diagram, this is
\begin{align}
  &\raisebox{-33pt}{
\begin{tikzpicture}[line width=1. pt, scale=2.2]
\draw[black, line width=1.pt] (0,0) -- (-0.4,0.75);
\draw[black, line width=1.pt] (0,0) -- (0.4,0.75);
\draw[black, line width=1.pt] (0,0) -- (0,0.75);
\draw[lightgray, line width=2.pt] (-0.6,0.75) -- (.6,0.75);
\draw[fill=black] (0,0) circle (.04cm);
\node[scale=.9] at (0,-0.15) {$+$};
\node[scale=.9] at (-0.4,0.9) {$k_1$};
\node[scale=.9] at (0,0.9) {$k_2$};
\node[scale=.9] at (0.4,0.9) {$k_3$};
\end{tikzpicture}
}= -i \lambda \int_{- \infty (1-i \epsilon)}^0dt\,K_+(k_1,t)K_+(k_2,t)K_+(k_3,t)=\\
=&-i \frac{\lambda}{8 k_1 k_2 k_3} \int_{- \infty (1-i \epsilon)}dte^{-(k_1+k_2+k_3)t}=\frac{\lambda}{8 k_1 k_2 k_3}\left[\frac{1}{k_T}e^{-ik_T t}\right]_{-\infty(1-i \epsilon)}^0=\frac{\lambda}{8 k_1 k_2 k_3}\frac{1}{k_T}
\end{align}
Here, we defined $k_T= k_1+k_2+k_3$. Notice that in the last step the contribution from early time vanishes thanks to the $i\epsilon$ prescription.
As discussed in the main text, it can be checked that the other term gives the complex conjugated answer, so that the final result reads
\begin{align}
     B_3(k_1,k_2,k_3)=\frac{\lambda}{4 k_1 k_2 k_3}\frac{1}{k_T}\,.
\end{align}
\item For the second interaction the time derivatives imply that instead of having $K_\pm(k_i,t)$ one has the time derivative of the propagator $\dot K_\pm(k_i,t)$ on two of the legs at a time. Exchanges of the two legs give the same answer and cancel with the factor that multiplies the coupling constant, we are left with 
\begin{align}
    B_3(k_1,k_2,k_3)&= 2 \text{Re}\left[\frac{i}{\lambda_t} \int_{-\infty(1-i\epsilon)}^0 dt \dot K_+(k_1,t) \dot K_+(k_2,t)K_+(k_3,t)+(1\leftrightarrow 3)+(2\leftrightarrow 3))\right]=\nonumber\\
    &=\frac{\lambda_t}{4 k_1 k_2 k_3}\frac{1}{k_T}(k_1 k_2+k_2 k_3+k_1 k_3)
\end{align}
\item For the last interaction one must remember that in Fourier space $\varphi(t,\bm x)=\int \frac{d^3 \bm k}{(2k)^3} e^{-i \bm k \cdot \bm x}\varphi(k,t)$ so that taking a derivative w.r.t. $x^i$ corresponds in multiplying the momentum space field and therefore the mode function times $i k^i$ (here the Greek index indicates the spatial component), this in turn implies that two legs at a time come with $(-\bm k_1 \cdot \bm k_2)K_{\pm}(k_1,t)K_{\pm}(k_2,t)$ instead of just $K_{\pm}(k_1,t)K_{\pm}(k_2,t)$. Combinatorial factors cancel out with the normalisation of the coupling as in the previous example. Taking this into account we get,
\begin{align}
    &B_3(k_1,k_2,k_3)=\\
    &=2 \text{Re}\left[-i \lambda_s(-\bm k_1 \cdot \bm k_2)\int^0_ {\infty (1-i \epsilon)}dt\,K_+(k_1,t)K_+(k_2,t)K_+(k_3,t) +(1\leftrightarrow 3)+(2\leftrightarrow 3)) \right]=\nonumber\\
    &=-\frac{\lambda_t}{4 k_1 k_2 k_3}\frac{1}{k_T}[(\bm k_1 \cdot \bm k_2)+(\bm k_2 \cdot \bm k_3)+(\bm k_1 \cdot \bm k_3)]\,.
\end{align}
\item Considering the interaction as suggested, it is easy to check that the result is
\begin{align}
    B_n(k_1, \dots,k_n)= \frac{\lambda_n}{2^{n-1}}\frac{1}{\prod_i k_i}\frac{1}{k_T}\,,
\end{align}
with $k_T=\sum_i k_i$. This $1/k_T$ factor is what is common in all these correlators, and is present also in the de Sitter example of the main text. It is related to the fact that there exist particle production from the vacuum.
\end{enumerate}


\section{Open scalar theory}\label{sec:scalar}

\begin{figure}[t]
	\centering
	\includegraphics{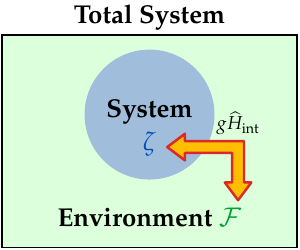}
	\caption{Schematic setup of open systems. The system, made of the degree of freedom, $\zeta$, is embedded in the environment, $\mathcal{F}$. Their interaction is specified by $g \widehat{H}_{\mathrm{int}}$. From the point of view of the system, this interaction renormalises its energy levels, generates energy loss through dissipation and information exchanges through noise. Figure adapted from \cite{manzanoShortIntroductionLindblad2020}\protect\footnotemark.}
	\label{fig:setup}
\end{figure}

Open systems describe physical set-ups in which energy and information are exchanged with an environment. We use the word \emph{system} to refer to experimentally accessible degrees of freedom. By contrast, in this construction, the \emph{environment} remains inaccessible and only partially specified; it typically encodes unobservable degrees of freedom that nonetheless affect dynamics. As a particular example, cosmological colliders aim to probe the impact of heavy fields and higher-spin particles (\emph{environment}) on curvature perturbations in the early universe (\emph{system}) \cite{Colas:2025ind}. Figure \ref{fig:setup} summarises the setup considered.

\footnotetext{\cite{manzanoShortIntroductionLindblad2020}; $\copyright$ 2020D.~Manzano. Licensed under a Creative Commons Attribution (CC BY) license. \href{https://creativecommons.org/licenses/by/4.0/deed.en}{https://creativecommons.org/licenses/by/4.0/deed.en}}

Open theories provide a set of effective tools to describe the dynamical evolution of systems losing energy and information.
\begin{figure}[t]
	\centering
	\includegraphics[width=1\textwidth]{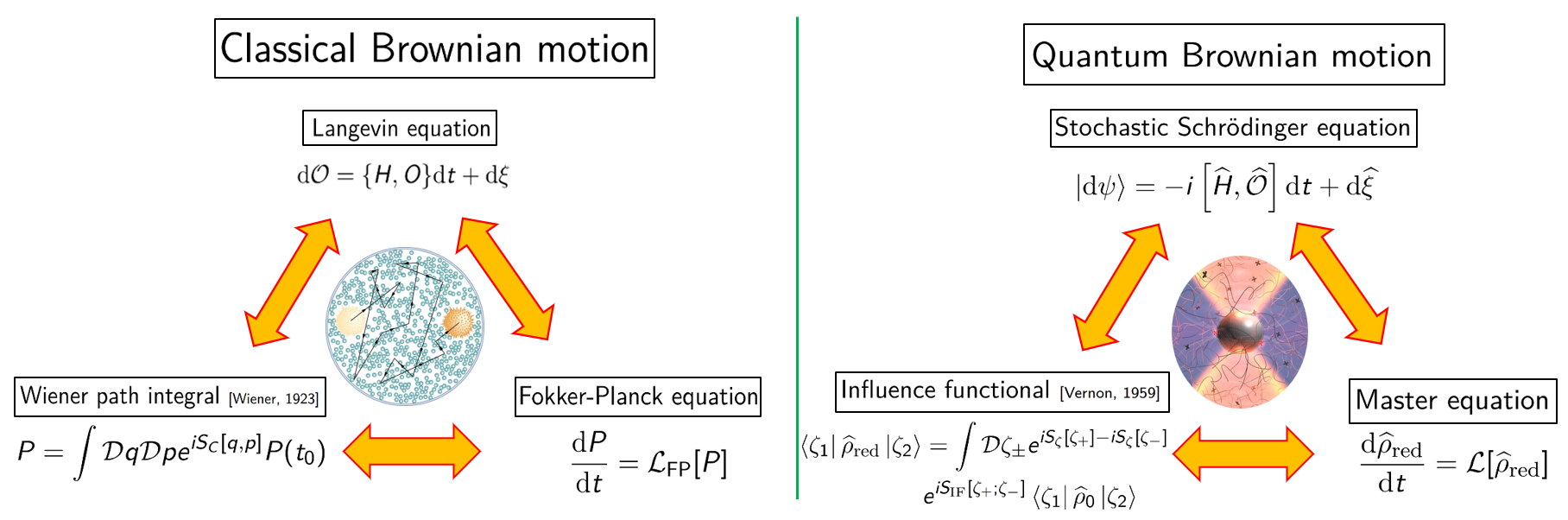}
	\caption{Classical and quantum open system correspondence. Classically, stochastic effects are captured by the Langevin equation, the Fokker--Planck equation, and the Martin--Siggia--Rose path integral. Their quantum counterparts are the stochastic Schrödinger equation, the master equation, and the influence functional.}
	\label{fig:trinity}
\end{figure} 	
To illustrate these ideas, consider a Brownian particle immersed in a thermal bath. We model the system by a scalar variable $\varphi(x)$ and describe its dynamics in the presence of an environment. The effect of the surrounding medium is encoded by stochastic variables, called noises, which promote deterministic equations of motion to stochastic differential equations such as the Langevin equation,
\begin{equation}\label{eq:langevinintro}
	\ddot{\varphi} + \gamma \dot{\varphi} + c_s^2 k^2 \phi = \xi.
\end{equation}
The new variable $\xi(x)$ has to be understood in a statistical sense, for instance obeying the Gaussian statistics
\begin{equation}\label{eq:noisecontactintro}
\ex{\xi(x)}=0, \qquad	\langle\xi(x) \xi(y)\rangle=2\beta\, \delta(x-y).
\end{equation}
In flat space, the late-time dynamics generated by \Eq{eq:langevinintro} is governed by an equilibrium between the drift term $\gamma \dot{\varphi}$, which damps the motion, and the stochastic source $\xi$, which drives fluctuations. This balance gives rise to standard Brownian motion.

There exist several equivalent ways to describe this dynamics, which can be mapped onto one another \cite{breuerTheoryOpenQuantum2002}. By averaging over many stochastic realisations of the Langevin equation, one finds a dynamical equation for the probability distribution of being at a given position and time, known as the Fokker-Planck (FP) equation. This probability distribution admits a path integral representation, known as the Martin-Siggia-Rose (MSR) path integral \cite{PhysRevA.8.423}. These formulations rely on the same underlying physics and represent different aspects of the same problem. The Langevin equation focuses on the equations of motion and is well-suited for numerical simulations. The MSR path integral on the other hand provides a natural framework for relativistic systems in a manifestly covariant way, while the Fokker-Planck equation has been widely studied for its ability to implement resummations. As shown in the right panel of \Fig{fig:trinity}, there exists a correspondence in the quantum framework where the Fokker-Planck equation is supplemented by master equations, the MSR path integral by the influence functional and the Langevin equation by a stochastic lifting of Schrödinger's equation \cite{breuerTheoryOpenQuantum2002}. 

Since the focus of these notes is relativistic quantum field theory, the path-integral formulation of open-system dynamics will prove most suitable. In particular, we aim to identify what general rules obeyed by the influence functional are independent of the microscopic description of the environment. One may attempt to model the environment explicitly and derive its impact on the system; we refer to this as a \emph{top-down} approach. A canonical example is the Caldeira--Leggett model \cite{Caldeira:1981rx, Caldeira:1982iu, Caldeira:1982uj}, which couples a harmonic oscillator $\varphi$ to a set of $N$ other harmonic oscillators $\{\sigma_n|n = 1, \dots, N\}$,
\begin{align}
	S[\varphi,\sigma_n] = \int \dd t \int \frac{\dd\bmk^3}{(2\pi)^3} \bigg\{\frac{1}{2}\left[\dot{\varphi}^2 - k^2 \varphi^2 \right] + \sum_{n = 1}^{N} \frac{1}{2} \left[ \dot{\sigma}^2_n - \left(k^2 + m_n^2\right)\sigma_n^2\right] + \sum_{n = 1}^{N} g_n \varphi \sigma_n \bigg\}.
\end{align}
Here, $m_n$ denote the masses of the environmental oscillators and $g_n$ their couplings to the system oscillator $\varphi$. By an appropriate choice of $\{m_n^2, g_n\}$ such that the $N$ oscillators form an effective thermal bath, one recovers the Langevin equation \eqref{eq:langevinintro} for the dynamics of $\varphi$ \cite{Lau:2024mqm}. While such constructions are often insightful, they can obscure the fact that many properties of open-system dynamics are independent of the environmental model. Instead, these properties are dictated by general physical principles such as symmetries, locality, and unitarity. This is the central message of the Lindblad theorem \cite{Lindblad:1975ef}, which states that \emph{any Markovian (memoryless) open quantum system evolves according to a rigid structure}, the Gorini--Kossakowski--Sudarshan--\textbf{Lindblad} (GKS\textbf{L}) equation. By developing a \emph{bottom-up} approach that is agnostic to the microphysical details of the environment, we aim to uncover analogous constraints in relativistic quantum field theory.

\subsection{Effective functional}

The Schwinger--Keldysh formalism \cite{Schwinger:1960qe, Keldysh:1964ud} provides a framework to compute expectation values in far-from-equilibrium and/or open quantum systems. In this section, we adopt an EFT viewpoint and seek a general description of the system that is maximally agnostic about the environment. We assume a separation of scales between system and environment, allowing environmental effects to be encoded by a finite set of local, possibly dissipative, interactions of the system. We therefore avoid an explicit modelling of the environment and instead work with a self-consistent effective description of the system dynamics. This strategy mirrors the Lindblad approach to open quantum systems \cite{Lindblad:1975ef, 10.1063/1.522979}, which focuses on the structure of the reduced dynamics rather than the physics of the environment. For definiteness, we assume that the environment is homogeneous, isotropic, and time-translation invariant, and we neglect any backreaction of the system. The construction of a local EFT then proceeds in three steps:  i) identify the relevant degrees of freedom; ii) impose the symmetries of the problem; iii) retain a finite set of operators consistent with radiative stability.

We consider the case in which the system of interest interacts with an unspecified environment, assuming that the combined $\{\textit{system}+\textit{environment}\}$ evolves unitarily. Starting from any pure state, tracing out the environment will eventually lead to a mixed state for the system, described by a density matrix $\widehat{\rho}(t)$. In an open quantum system, equal-time expectation values of operators are schematically given by
\begin{equation}\label{eq:expect}
	\langle \widehat{\mathcal O}(t) \rangle
	= \Tr\!\left[\widehat{\rho}(t)\widehat{\mathcal O}(t)\right]
	= \int \dd\phi \, \dd\phi' \, \rho_{\phi\phi'}(t)\,
	\bra{\phi'}\widehat{\mathcal O}(t)\ket{\phi}\,,
\end{equation}
where
\begin{equation}\label{rhomatrix}
	\rho_{\phi\phi'}(t)
	\equiv \bra{\phi}\widehat{\rho}(t)\ket{\phi'}
	= \int^{\phi}\!\mathcal{D}\varphi_{+}
	\int^{\phi'}\!\mathcal{D}\varphi_{-}\,
	e^{iS_{\mathrm{eff}}[\varphi_{+},\varphi_{-}]}\,,
\end{equation}
where the functional $S_{\mathrm{eff}}$ is known as the \emph{open effective action}. This quantity contains the unitary dynamics of the system together with an influence functional encoding environmental effects \cite{FEYNMAN1963118}. Here $\dd\phi$ denotes an integral over boundary field configurations at time $t$, while $\mathcal{D}\varphi_{\pm}$ denote path integrals over field histories on the forward and backward branches of the contour \cite{Boyanovsky:2015xoa}. Initial conditions are specified by an initial density matrix, which we take to be a pure state in the infinite past. We will then focus on operators diagonal in the field basis, $\bra{\phi}\widehat{\mathcal O}\ket{\phi'} \propto \delta(\phi-\phi')$, such as products of fields evaluated at different spacetime points. Operators involving conjugate momenta fall outside this class. Accordingly, we will be interested in the diagonal elements $\rho_{\phi\phi}(t)$ appearing in \Eq{eq:expect}.

\begin{figure}[t]
	\centering
	\includegraphics{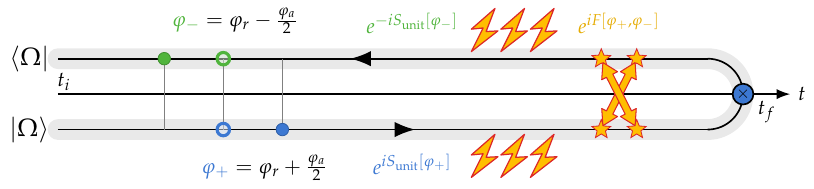}
	\caption{Illustration of the Schwinger-Keldysh path integral, where time is running from left to right in both contours and the arrows represent path ordering (time ordering in $\ket{\Omega}$ and anti-time-ordering in $\bra{\Omega}$). We consider an initially pure state $\ket{\Omega}\bra{\Omega}$ at initial time $t_i$. Unitary time evolution, which preserves the purity of the state, is related by complex conjugation on each branch, captured by $S_{\mathrm{unit}}$. Dissipative and stochastic effects are then captured by $F$ which has no unitary counterpart and cannot be captured through a single-branch contour.}
	\label{fig:contour}
\end{figure}

The functional $S_{\mathrm{eff}}$ integrates a local density over space (e.g.\ $\mathbb{R}^3$) and over a time interval. It is often convenient to decompose it as
\begin{align}\label{eq:splitting}
	S_{\mathrm{eff}}[\varphi_{+},\varphi_{-}]&= S_{\mathrm{unit}}[\varphi_{+}]-S_{\mathrm{unit}}[\varphi_{-}] + F[\varphi_{+},\varphi_{-}]\,,
\end{align}
as summarised in Fig.~\ref{fig:contour}. The functional $F$ is the Feynman--Vernon influence functional and captures all effects induced by the environment \cite{FEYNMAN1963118}. 

\subsubsection{Non-equilibrium constraints}\label{subsec:NEQconst}

By definition, $\widehat\rho$ obeys the following constraints \cite{breuerTheoryOpenQuantum2002}
\begin{align}\label{eq:NEQ}
	\Tr \widehat\rho&=1\,, & \widehat{\rho}^{\dagger}&=\widehat\rho\,, & \widehat\rho&\geq0\,,
\end{align}
where the last is a shorthand notation for $ \bra{\psi}\widehat \rho \ket{\psi}\geq 0$ for all $ \ket{\psi}\in \H$. These conditions are crucial to ensure that the quantum mechanical formalism makes meaningful statistical predictions. These three defining properties of a density matrix impose additional constraints on the functional $ S_{\mathrm{eff}}$ \cite{Calzetta:2008iqa, Liu:2018kfw} that are respectively
\begin{align}
	S_{\mathrm{eff}} \left[\varphi_+,\varphi_+\right] &= 0\,,  \label{eq:norm1}\\
	S_{\mathrm{eff}} \left[\varphi_+,\varphi_-\right] &= - 	S^*_{\mathrm{eff}} \left[\varphi_-,\varphi_+\right] \,, \label{eq:herm1}\\
	\Im S_{\mathrm{eff}} \left[\varphi_+,\varphi_-\right] &\geq 0 \label{eq:im1}\,.
\end{align}
We often work in the Keldysh basis 
\begin{align}
	\varphi_r = \frac{\varphi_+ + \varphi_-}{2},\qquad \varphi_a = \varphi_+ - \varphi_- \qquad \Leftrightarrow \qquad \varphi_+ = \varphi_r + \frac{\varphi_a}{2}, \qquad \varphi_- = \varphi_r - \frac{\varphi_a}{2} .\label{eq:RAbasis}
\end{align}
The retarded and advanced components $\varphi_r$ and $\varphi_a$, which respectively correspond to the mean and the difference of the field inserted along each branch of the path integral, turn out to conveniently organise the perturbative expansion \cite{Kamenev:2009jj}. In this basis, the above non-equilibrium (NEQ) constraints become 
\begin{align}
	S_{\mathrm{eff}} \left[\varphi_r,\varphi_a = 0\right] &= 0 \,,\label{eq:norm} \\
	S_{\mathrm{eff}} \left[\varphi_r,\varphi_a\right] &= - 	S^*_{\mathrm{eff}} \left[\varphi_r,-\varphi_a\right] \,,\label{eq:herm} \\
	\Im S_{\mathrm{eff}} \left[\varphi_r,\varphi_a\right] &\geq 0. \label{eq:pos}
\end{align} 

\noindent Despite their simple appearance, the constraints \eqref{eq:norm1}, \eqref{eq:herm1} and \eqref{eq:im1} impose important model-independent and non-perturbative restrictions on $S_{\mathrm{eff}}$: 
\begin{enumerate}
	\item Following \eqref{eq:norm}, $S_{\mathrm{eff}}$ starts linear in $\varphi_a$;
	\item Following \eqref{eq:herm}, odd powers of $\varphi_a$ are purely real and even powers of $\varphi_a$ purely imaginary;
	\item Following \eqref{eq:pos}, some EFT coefficients have to obey a positivity bound.
\end{enumerate}
If the theory is unitary, that is, if $ F[\varphi_{+}, \varphi_{-}] = 0 $ in \eqref{eq:splitting}, it is straightforward to show that $ S_{\mathrm{eff}} $ contains only odd powers of $ \varphi_a $. However, the converse is not true, as there exist operators that are odd in $ \varphi_a $ but do not originate from a unitary theory. At last, even powers of $\varphi_a$ only appear in stochastic field theory.

A natural way to organise open EFTs is in powers of the advanced field
\begin{align}
	S_{\mathrm{eff}} = \sum_{n = 1}^{\infty} S_n \quad \text{with} \quad S_n = \mathcal{O}(\varphi^n_a)
\end{align}
The reason is that if the scalar field can be decomposed along a background profile, 
\begin{align}
	\varphi_{\pm} = \bar{\varphi}(t) + \delta \varphi_\pm
\end{align} 
the retarded component carries the background value of the field while the advanced component starts linear in perturbations, 
\begin{align}
	\varphi_r \equiv \frac{\varphi_{+} + \varphi_{-}}{2} = \bar{\varphi}(t) + \delta \varphi_r, \qquad \varphi_a \equiv \varphi_{+} - \varphi_{-} =  \delta \varphi_a
\end{align}
As a result, any operator in $S_{n \geq 3}$ will be at least cubic in the perturbations. While in a vacuum QFT, $ \delta \varphi_r$ and $ \delta \varphi_a$ scale the same, the power counting of NEQ QFTs strongly depends on the occupation number of the state, $N_\varphi = \langle \widehat{a}^\dag_\varphi \widehat{a}_\varphi\rangle$ \cite{2016RPPh...79i6001S, Berges:2012ty, Rosenhaus:2025bgy, Radovskaya:2020lns}. At large occupation $N_\varphi \gg 1$, we find $\varphi_a \sim \mathcal{O}(\hbar)$, whereas $\varphi_r \sim \mathcal{O}(1) $. Therefore, operators contained in $S_{n \geq 3}$ become irrelevant and the influence functional reduces to its semiclassical limit, $S_{\mathrm{eff}} = S_1 + S_2$ known as the Martin-Siggia-Rose (MSR) formalism.

\subsubsection{Free functional} \label{subsec:BCs}

To explore the basic features of this formalism, let us first consider a single scalar degree of freedom $\varphi$ evolving in flat space according to 
\begin{align}\label{eq:BM}
	S_{\mathrm{eff}}[\phir,\phia] = \int \dd^4x \left(\dphir \dphia - c_s^2 \partial_i \phir \partial^i \phia + \gamma \dphir \phia + i \beta \phia^2\right),
\end{align}
where a dot represents a time derivative and $\beta,\gamma\in\mathbb{R}$. First, note that the first two terms can be written in a factorised form in the original $+/-$ basis,
\begin{align}
	S_{\mathrm{unit}}[\varphi_+] - S_{\mathrm{unit}}[\varphi_-] \qquad \mathrm{with} \qquad S_{\mathrm{unit}}[\varphi] = \frac{1}{2} \int \dd^4x \left[\dot{\varphi}^2 - c_s^2 (\partial_i \varphi)^2 \right],
\end{align}
and hence encode the unitary evolution. 
Conversely, the last two terms have no analogue in the unitary case and encode the dissipative and diffusive effects of $F[\varphi_+,\varphi_-]$.  

Let us first discuss the first non-unitary contribution $\gamma \dphir \phia$ appearing in \Eq{eq:BM}. This dissipative operator is crucial in describing the loss of energy of the system $\varphi$ into its surrounding environment. In the original $+/-$ basis, it follows from the boundary term
\begin{align}
	S_{\mathrm{unit}}[\varphi_+] - S_{\mathrm{unit}}[\varphi_-] \qquad \mathrm{with} \qquad S_{\mathrm{unit}}[\varphi] = -\frac{1}{4} \int \dd^4 x \left(\frac{\dd}{\dd t} \left[\varphi^2\right] \right),
\end{align}
and the mixing between the two branches of the path integral 
\begin{align}
	F[\varphi_+,\varphi_-] = - \frac{1}{2} \int \dd^4 x \left(\dot{\varphi}_+ \varphi_- - \varphi_+ \dot{\varphi}_- \right).
\end{align}
Upon specifying the boundary conditions of the diagonal density matrix element appearing in \Eq{eq:expect}
\begin{align}
	\varphi_+(\bmx, t_0) = \varphi_-(\bmx, t_0) = \phi(\bmx),
\end{align}
it is clear that the boundary term does not contribute to the diagonal elements of $\widehat \rho$. The mixing term $F[\varphi_+,\varphi_-] $, however, has no reason to vanish. This effect has no unitary counterpart and is precisely the one responsible for the damping of $\varphi$ fluctuations. 

To obtain the equations of motion for $\varphi_r$, we vary $S_{\mathrm{eff}}$ with respect to $\varphi_a$ and set $\varphi_a = 0$. One could naively think that this implies only the functional $ S_1 $, linear in $\varphi_a$, contributes to the equations of motion of $\varphi_r$. However, it turns out that this is not the case, because of the boundary conditions of the path integral. The boundary conditions for $  \varphi_{a} $ are those responsible for $  \varphi_{a} $ not propagating, despite the fact that $  \varphi_{a} $ appears in the quadratic action in a way that is very similar to $  \varphi_{r} $. If it was not for the boundary conditions, the theory would inevitably suffer from an instability.\footnote{To see this, imagine to incorrectly assume that $  \varphi_{a} $ has standard initial conditions (as opposed to satisfying a boundary condition problem) and integrate it out. Its equation of motion is given by $  \varphi_{a}=\Box \varphi_{r} $, which, once substituted back into the action would give terms as $  \varphi_{r}\Box^{2}\varphi_{r} $. This is the quintessential example of a ghostly degree of freedom.}
Because of the boundary conditions, $  \varphi_{a} $ does not propagate and the theory remains free from pathologies. However, something peculiar happens as a result: the $  \varphi_{a} $ path integral does not admit a stationary phase! This is because, in the presence of fluctuations $  i \beta \varphi_{a}^{2} $, the stationary phase would be $  \varphi_{a}\sim \Box \varphi_{r} $, which is incompatible.

Things change if $  \varphi_{a} $ appears linearly. Then, its stationary phase is some equation of motion for the \textit{other} modes, not involving $  \varphi_{a} $, and that would always be compatible with the vanishing of $  \varphi_{a} $ on the boundary. This is indeed the magic that ensues from the Hubbard-Stratonovich (HS) transformation \cite{Stratonovich1957,Hubbard1959}. Upon introducing an auxiliary noise field $ \xi   $, the open functional becomes \textit{linear} in $  \varphi_{a} $. It is only at this point that the path integral in $  \varphi_{a} $ is well approximated by a stationary phase and it is only now that $  \delta S/ \delta \varphi_{a} $ can be interpreted as an equation of motion, namely the Langevin equation. Explicitly, the key trick is the mathematical identity
\begin{equation}\label{eq:HStrick}
	\exp\left(-\int d^{4}x\, \beta\phia^{2}\right)=\mathcal{N}_{0}\int\mathcal{D}\xi\;\exp\left[\int d^{4}x \left(-\frac{\xi^{2}}{4\beta}+i\xi\phia\right)\right],
\end{equation}
with $\mathcal{N}_{0}$ being the normalisation constant. Note that because of the positivity constraint \Eq{eq:pos}, convergence is always ensured.
After this transformation, the path integral in $  \varphi_{a} $ is well approximated by a stationary phase which generates the Langevin equation
\begin{align}\label{eq:langevin}
	\ddot{\varphi}_r + \gamma \dphir + c_s^2 k^2 \phir = \xi.
\end{align}
The new variable $\xi$, satisfying $\ex{\xi(x)}=0$, behaves as a Gaussian field with a prescribed two-point function
\begin{equation}\label{eq:noisecontact}
	\langle\xi(x) \xi(y)\rangle=2\beta \delta(x-y).
\end{equation}

We conclude that \Eq{eq:BM} provides a path integral representation of the Langevin equation introduced in \eqref{eq:langevinintro}. We stress that we are not assuming that the system or the environment are in thermodynamical equilibrium. In particular $\beta$ is not in general related to the inverse temperature.\footnote{In other words, Schwinger-Keldysh path integral considered prepares a density matrix that is not a thermal state.} If desired, one could impose additional constraints, known as Kubo–Martin–Schwinger (KMS) conditions (see \cite{Kubo:1966fyg, PhysRev.115.1342} and \cite{Hongo:2018ant} for a review), to specify our construction to the thermal case. We do not follow this approach because our goal is to develop a formalism for problems in cosmology and gravity where thermalisation does not necessarily take place.

\subsubsection{Interactions}

To get familiar with interactions in the Keldysh basis, let us first consider the simple toy model of a massive scalar with $  \lambda \varphi^{3} $ interaction
\begin{align}
	S_{\mathrm{eff}}[\phir,\phia] = S_{\mathrm{unit}}[\varphi_+] - S_{\mathrm{unit}}[\varphi_-] \quad \mathrm{with} \quad S_{\mathrm{unit}}[\varphi] = \int \dd^4x \left[ -\frac{1}{2}\left(\partial_{\mu}\varphi\right)^{2}-\frac{1}{2}m^{2}\varphi^{2} -\lambda \varphi^{3}\right], \label{eq:unitarySphi}
\end{align}
that is 
\begin{align}
	S_{\mathrm{eff}}[\phir,\phia] &=\int \dd^4x \left(-\partial_{\mu}\varphi_{r}\partial^{\mu}\varphi_{a}-m^{2}\varphi_{r}\varphi_{a}-3\lambda \varphi_{r}^{2} \varphi_{a}-\frac{\lambda}{4} \varphi_{a}^{3} \right)\,.\label{eq:unitaryIphi}
\end{align}
Note that the theory fulfils the NEQ constraints presented above. The unitary interaction $\lambda(\varphi_{+}^3 - \varphi_-^3)$ generates two types of vertices, $\varphi_r^2 \varphi_a$ and $\varphi_a^3$.

If we vary $S_{\mathrm{eff}}$ with respect to $\varphi_a$, terms linear in $  \varphi_{a} $ generate the classical equations of motion, with all the right coefficients. This type of interactions can be understood as classical non-linearities. One may wonder what the role of the $\varphi_a^3$ term is. It is not easy to see how it contributes to the Langevin equation because we have to perform a more complex version of the Hubbard-Stratonovich transformation, as done in \cite{Salcedo:2024smn} for $\varphi_a^3$ (see also \cite{Chakrabarty:2019aeu}). If one manages to do so, it appears that $\varphi_a^3$ can be understood as a white non-Gaussian noise, 
\begin{equation}\label{eq:noisecontactv2}
	\langle\xi(x) \xi(y) \xi(z)\rangle \propto \delta(x-y) \delta(y-z).
\end{equation}
We will come back to these terms below when we compute some flat-space correlators, for which these contributions are crucial to recover the flat space vacuum results.

So far, we considered \textit{unitary interactions}. Note that if we detune the fine balance between $\varphi_r^2 \varphi_a$ and $\varphi_a^3$, for instance considering
\begin{align}
		S_{\mathrm{eff}}[\phir,\phia] \supset  \int \dd^4x\left[-3\lambda \varphi_{r}^{2} \varphi_{a}-\frac{\lambda}{4} (1 + \delta) \varphi_{a}^{3}\right],
\end{align}
the theory stops being unitary, $S_{\mathrm{eff}} \neq S_{\mathrm{unit}}[\varphi_+] - S_{\mathrm{unit}}[\varphi_-]$, as long as $\delta \neq 0$. Importantly, the theory still fulfils the constraints \eqref{eq:norm}, \eqref{eq:herm} and \eqref{eq:pos}. The theory can still make perfect sense if understood as an open theory. At last, we could also have considered some interactions of the form $\mathcal{O}(\varphi_r) \varphi_a^{2p}$ with $p$ an integer. These interactions couple the operator $\mathcal{O}(\varphi_r)$ to the noise variables. For instance, let us consider 
\begin{align}\label{eq:thfull}
	S_{\mathrm{eff}}[\phir,\phia] =\int \dd^4x \bigg[& \dphir \dphia - c_s^2 \partial_i \phir \partial^i \phia + \gamma \dphir \phia + i \beta \phia^2  \\
	-&3\lambda \varphi_{r}^{2} \varphi_{a}  + i \widetilde{\beta} \varphi_r\varphi_a^2 -\frac{\lambda}{4}(1 + \delta) \varphi_{a}^{3}  \bigg],
\end{align}
where the first line contains the linear operators discussed above and the second line some interactions.
The associated equations of motion are given by the non-linear Langevin equation \cite{Salcedo:2024smn}
\begin{align}
	\ddot{\varphi}_r + \gamma \dphir + c_s^2 k^2 \phir =   3 \lambda \varphi_r^2 + \left(1 +  \frac{\widetilde{\beta}}{\beta} \phir \right)\xi , 
\end{align}
with noise statistics
\begin{align}
	\langle\xi(x) \rangle &= 0,  \\
	\langle\xi(x) \xi(y) \rangle &= 2   \beta \delta(x-y) \\
	\langle\xi(x) \xi(y) \xi(z) \rangle &= 6 \lambda (1 + \delta) \delta(x-y)\delta(y-z).
\end{align}

It is interesting to understand if we can \textit{complete} an operator in order to obtain a unitary combination. A proposal to unitarise an operator is the following. Let us consider an operator $\mathcal{O}\left(\varphi_r,\varphi_a\right)$ made of field insertions of $\varphi_r$, $\varphi_a$ and their (at most single) derivatives. The combination of operators
\begin{align}\label{eq:unitarize}
	\mathcal{U}\left[\mathcal{O}\left(\varphi_r,\varphi_a\right)\right] = \mathcal{O}\left(\frac{\varphi_+}{2},\varphi_+\right) + \mathcal{O}(\frac{\varphi_-}{2},-\varphi_-) - \mathcal{O}\left(\frac{\varphi_-}{2},\varphi_-\right) - \mathcal{O}(\frac{\varphi_+}{2},-\varphi_+)
\end{align}
is unitary by construction. For operators such as $\varphi_a^2$ (the diffusion operators) that are intrinsically non-unitary, the above combination vanishes. Also note that some operators such as $\dot{\varphi}_r \varphi_a$ (the dissipation operators) are unitarised into total derivatives and so do not add any net contribution to the open effective functional. Lastly, some of the contributions obtained out of \Eq{eq:unitarize} eventually violate symmetries of the problem (\eg the fact that $\varphi_r$ non-linearly realises time-translations and boosts) and must then be discarded.

\subsection{Free theory}\label{subsec:freeth}

\subsubsection{Path ordering}\label{subsec:path}

Ultimately, the path integral aims at computing expectation values of observables. To achieve this task, we need to acknowledge what the path integral is actually computing. In this Section, we discuss the link between path integral and quantum operators. 

To make the following discussion more streamlined, it is useful to introduce the following shorthand notation for the path integral average over some function 
\begin{align}
	\exx{ \O[\varphi_+,\varphi_-]}\equiv \int \dd \phi \int^{\phi }_{\text{I.C.}} \mathcal{D}\varphi_+\int^{\phi}_{\text{I.C.}} \mathcal{D}\varphi_-\, \O[\varphi_+,\varphi_-] e^{iS_{\mathrm{eff}}[\varphi_+,\varphi_-]}\,.
\end{align}
Let us now focus on the closed-time contour of the Schwinger-Keldysh formalism, which has only two timefolds, one going forwards and one backwards in time. Along the ``$+$'' branch of the path integral, expectation values are time ordered (denoted $\T$), whereas along the ``$-$'' branch, expectation values are anti-time ordered (denoted $\bar \T$), \cite{Kamenev:2009jj}
\begin{align}
	\exx{\varphi_{+}(t)\varphi_{+}(t')}&=\ex{\T \widehat \phi(t)\widehat \phi(t')}\,,\\
	\exx{\varphi_{-}(t)\varphi_{-}(t')}&=\ex{\bar\T \widehat \phi(t)\widehat \phi(t')}\,.
\end{align}
When there is one insertion on each branch of the path integral, the expectation value is neither time nor anti-time ordered and we obtain a Wightman function, 
\begin{align}
	\exx{\varphi_{+}(t)\varphi_{-}(t')}&=\ex{ \widehat \phi(t')\widehat \phi(t)}\,,\\
	\exx{\varphi_{-}(t)\varphi_{+}(t')}&=\ex{\widehat \phi(t)\widehat \phi(t')}\,.
\end{align}
Using these results repeatedly it is easy to check that
\begin{align}
	\exx{\varphi_{r}(t)\varphi_{a}(t')}&=\frac{1}{2} \exx{\left[ \varphi_{+}(t)+\varphi_{-}(t) \right]\left[ \varphi_{+}(t')-\varphi_{-}(t') \right] }\\
	&=\frac{1}{2} \left[\ex{\T \widehat \phi(t)\widehat \phi(t')}-\ex{\bar\T \widehat \phi(t)\widehat \phi(t')} + \ex{  \widehat \phi(t)\widehat \phi(t')}  - \ex{  \widehat \phi(t')\widehat \phi(t)} \right]\\
	&=\theta(t-t') \ex{[\widehat \phi(t),\widehat \phi(t')]}\,,
\end{align}
which is the definition of the retarded propagator. Similarly one finds the reversed time argument to correspond to the advanced propagator: 
\begin{align}
	\exx{\varphi_{a}(t)\varphi_{r}(t')}&=\theta(t'-t) \ex{[\widehat \phi(t'),\widehat \phi(t)]}.
\end{align}
The two-point function of the retarded propagator corresponds to the anticommutator of the field operator:
\begin{align}
	\exx{\varphi_{r}(t)\varphi_{r}(t')}&=\frac{1}{4}\exx{\left[\varphi_{+}(t)+\varphi_{-}(t)\right]\left[\varphi_{-}(t)+\varphi_{+}(t)\right]} \\
	&=\frac{1}{4}\left[\langle\mathcal{T}\widehat{\phi}(t)\widehat{\phi}(t')\rangle+\langle\widehat{\phi}(t)\widehat{\phi}(t')\rangle+\langle\widehat{\phi}(t')\widehat{\phi}(t)\rangle+\langle\bar{\mathcal{T}}\widehat{\phi}(t)\widehat{\phi}(t')\rangle\right]\\
	&=\frac{1}{2} \ex{\{\widehat \phi(t),\widehat \phi(t')\}},
\end{align}
where we have expanded the anti-time and time orderings in terms of theta functions and used $\theta(x)+\theta(-x)=1$. At last, a similar calculation gives
\begin{align}
	\exx{\varphi_{a}(t)\varphi_{a}(t')}&= \exx{\left[ \varphi_{+}(t)-\varphi_{-}(t) \right] \left[ \varphi_{+}(t')-\varphi_{-}(t') \right]}\\
	&=\ex{\T \widehat \phi(t)\widehat \phi(t')}+\ex{\bar\T \widehat \phi(t)\widehat \phi(t')}-\ex{  \widehat \phi(t)\widehat \phi(t')}-\ex{  \widehat \phi(t')\widehat \phi(t')}=0\,.
\end{align}
This illustrates the fact that $\varphi_{a}(t)$ cannot be considered as a propagating degree of freedom. These four two-point functions constitute the propagators of our theory, that we will now derive in an explicit model.

\subsubsection{Gaussian generating functional}

Let us consider the quadratic part of \Eq{eq:thfull}.

The form of \Eq{eq:thfull} allows us to write the action as a bilinear on the fields
\begin{equation}\label{eq:presc}
	S_{\text{eff}}^{(2)} = -\frac{1}{2}\int d^{4}x \left(\phir,\phia\right)\begin{pmatrix}
		0 & \widehat{D}_{A} \\ \widehat{D}_{R} & -2i\widehat{D}_{K}
	\end{pmatrix}\begin{pmatrix}
		\phir \\ \phia
	\end{pmatrix},
\end{equation}
the matrix being a second order differential operator acting to the right, which is made of
\begin{align}
	\widehat{D}_{A}& \equiv \partial_t^2 - \gamma \partial_t - c_s^2 \partial_i^2 ,\\
	\widehat{D}_{R}&\equiv \partial_t^2 + \gamma \partial_t - c_s^2 \partial_i^2 ,\\
	\widehat{D}_{K}&\equiv \beta.
\end{align}
The following path integral of the free theory computes the diagonal of the density matrix of the system
\begin{align}
	\rho_{\phi \phi}(t)= &\int_{\Omega}^\phi \mathcal{D}\phir  \int_{\Omega}^{0}\mathcal{D}\phia \exp\left\{i  S_{\text{eff}}^{(2)}[\phir,\phia]\right\}.
\end{align}
Upon the introduction of sources, we obtain the generating function
\begin{align}
	\mathcal{Z}[J_{r},J_{a}]=\int_{\Omega}^\phi \mathcal{D}\phir  \int_{\Omega}^{0}\mathcal{D}\phia \exp\Big\{-\frac{i}{2}  &\int d^{4}x \left(\phir,\phia\right)\begin{pmatrix}
		0 & \widehat{D}_{A} \\ \widehat{D}_{R} & -2i\widehat{D}_{K}
	\end{pmatrix}\begin{pmatrix}
		\phir \\ \phia
	\end{pmatrix} \nonumber\\
	&+\int d^{4}x \left(J_{a}\phir+J_{r}\phia\right)\Big\}
\end{align}
Completing the square for $\phia$ and $\phir$ allows us to factorise the dependence on the sources $J_{a}$ and $J_{r}$. This is done by introducing a shift in the path integral
\begin{equation}
	\begin{pmatrix}
		\phir \\ \phia
	\end{pmatrix}=\begin{pmatrix}
		\Pi_{r} \\ \Pi_{a}
	\end{pmatrix}+\int d^{4}y \begin{pmatrix}
		A_{11}(x,y) & A_{12}(x,y) \\  A_{21}(x,y) & 0
	\end{pmatrix}\begin{pmatrix}
		\phir \\ \phia
	\end{pmatrix}
\end{equation}   
Demanding that the terms linear in $\Pi_{r/a}$ vanish, we find\footnote{There exists a similar equation where we integrate by parts on $\widehat{D}_{A/R/K}$ which yields the same information.}
\begin{equation}
	- \frac{i}{2} \begin{pmatrix}
		0 & \widehat{D}_{A} \\ \widehat{D}_{R} & -2i\widehat{D}_{K}
	\end{pmatrix}\begin{pmatrix}
		A_{11}(x,y) & A_{12}(x,y) \\  A_{21}(x,y) & 0
	\end{pmatrix}+\frac{1}{2}\begin{pmatrix}
		\delta(x-y) & 0 \\  0 & \delta(x-y)
	\end{pmatrix}=\begin{pmatrix}
		0 & 0 \\  0 & 0
	\end{pmatrix}\,,
\end{equation}
where the Dirac delta $\delta(x-y)$ is a tensor density. These equations can be rewritten in a covariant form such that it becomes straightforward to recognise the advanced and retarded propagators
\begin{align}
	\widehat{D}_{R}(x)A_{12}(x,y)&=-i\delta(x-y)\;,\quad \;A_{12}(x,y)=-iG^{R}(x,y)\;,\quad \; G^{R}(x^{0}<y^{0})=0,\\
	\widehat{D}_{A}(x)A_{21}(x,y)&=-i\delta(x-y)\;,\quad \;A_{21}(x,y)=-iG^{A}(x,y)\;,\quad \; G^{A}(x^{0}>y^{0})=0.
\end{align}
A fundamental property of the retarded and advanced propagator is that they are mapped to each other under the exchange of the arguments:
\begin{equation}\label{eq:Rxy_Ayx}
	G^{R}(x,y)=G^{A}(y,x)\,.
\end{equation}
The Keldysh propagator is obtained from the matrix element $A_{11}(x,y)=-iG^{K}(x,y)$, which obeys the differential equation
\begin{equation}
	\widehat{D}_{R}(x)A_{11}(x,y)=2\widehat{D}_{K}(x)G^{A}(x,y).
\end{equation}
Notice that $G^{K}$ is thus not a Green's function of some equation of motion. We choose to symmetrise in $x\leftrightarrow y$ such that 
\begin{align}\label{eq:Keldyshreff}
	G^{K}(x,y)=&i\int d^{4}z \sqrt{-g(z)} G^{R}(x,z)\widehat{D}_{K}(z)G^{A}(z,y) \nonumber \\
	&+i\int d^{4}z \sqrt{-g(z)} G^{R}(y,z)\widehat{D}_{K}(z)G^{A}(z,x)
\end{align}
where we have used the property \eqref{eq:Rxy_Ayx} to write $A_{11}(x,y)$ in the most symmetric way\footnote{The generalisation to non-local noises is straightforward from here by upgrading the local $\widehat{D}_{K}(z)$ to a non-local $\widehat{D}_{K}(z_{1}-z_{2})$, where now we have the integrand to be $G^{R}(x,z_{1})\widehat{D}_{K}(z_{1}-z_{2})G^{A}(z_{2},y)$ with the appropriate factors of the square root of the metric.}. Note that causality is implemented in a natural way as the retarded propagator requires $x^{0}>z^{0}$ and the advanced propagator requires $z^{0}<y^{0}$.

This leaves the partition function to be 
\begin{equation}
	\mathcal{Z}[J_{r},J_{a}]= \exp\left\{ - \frac{i}{2}\int d^{4}x \int d^{4}y \left(J_{a}(x),J_{r}(x)\right)\begin{pmatrix}
		G^{K}(x,y) & G^{R}(x,y) \\ G^{A}(x,y) & 0
	\end{pmatrix}\begin{pmatrix}
		J_{a}(y) \\ J_{r}(y)
	\end{pmatrix}\right\}
\end{equation}
where we used the fact that $\mathcal{Z}[0,0] = 1$ in the closed time contour from trace normalisation \cite{kamenev_2011}. The two-point function reduces to
\begin{align}\label{eq:extract}
	&\exx{\phir(x_{1})\phir(x_{2})}=\left.\frac{\delta^{2}}{\delta J_{a}(x_{1})\delta J_{a}(x_{2})}\mathcal{Z}[J_{a},J_{r}]\right|_{J_{r,a}=0}=-i G^K(x_{1},x_{2}).
\end{align}
Evaluating the two-point function at coincident times in Fourier space provides an expression for the power spectrum of the system
\begin{align}
	&\exx{\phir(\bfk, \eta_{0})\phir(\bfk', \eta_{0})}=P_{k}(\eta_{0})(2\pi)^{3}\delta(\bfk+\bfk') \quad \mathrm{with} \quad P_{k}(\eta_{0}) = -i G^K(k; \eta_0,\eta_0).\bigg.
\end{align}      

\subsubsection{Propagators}\label{subsec:M4Pk}

Let us now derive the propagators in the explicit theory prescribed by \Eq{eq:presc}. The equations of motion for the propagators read  
\begin{align}
	\left(\partial_t^2 \pm \gamma \partial_t + c_s^2 k^2 \right) G^{R/A} (k; t_1, t_2) =  \delta(t_1-t_2)\,,			 
\end{align}
and, from \Eq{eq:Keldyshreff},
\begin{align}
	G^K (k;t_1,t_2) = 2 i \beta \int \dd t' G^{R} (k;t_1,t')G^{A} (k;t',t_2).
\end{align}
One must first solve for the retarded propagator to deduce the advanced and Keldysh ones.

The retarded and advanced propagators are easily found in frequency space, where 
\begin{align}
	G^{R/A} (k; \omega) &= - \frac{1}{\omega^2 \pm i \gamma \omega - c_s^2k^2}.
\end{align}
The dispersion relation 
\begin{align}\label{eq:disprelref}
	\omega^2 + i \gamma \omega - c_s^2 k^2 = 0
\end{align}
characterises the propagation in the system. Explicitly, it is the fact that \Eq{eq:disprelref} admits non-trivial solutions for the frequencies $\omega(k,\gamma)$ that ensures the existence of a dissipative degree of freedom. In this case, the solutions are 
\begin{align}\label{eq:soldispeg}
	\omega_\pm \equiv - i \frac{\gamma}{2} \pm E_k^{\gamma} \qquad \qquad \mathrm{and} \qquad \qquad 	E_k^{\gamma} \equiv \sqrt{c_s^2 k^2 - \frac{\gamma^2}{4}}.
\end{align}
which have both a real and imaginary part whenever $2c_s k > \gamma$. Note that in the limit $k \rightarrow 0$ with $\gamma$ finite, the dispersion relation is purely imaginary. We notice a gapless mode, corresponding to $\omega_+ \rightarrow 0$ for $k \rightarrow 0$ and a gapped mode associated to $\omega_- \rightarrow i \gamma$ for $k \rightarrow 0$. Importantly, for $\gamma > 0$ both modes have a negative imaginary part
\begin{align}
	\gamma > 0 \qquad  \Rightarrow \qquad \Im \omega_\pm < 0,
\end{align}
so that the retarded Green's function 
\begin{align}
	G^{R} (k; \omega) &= - \frac{1}{\omega^2 + i \gamma \omega - c_s^2k^2}= -\frac{1}{(\omega_- - \omega_+)} \left[ \frac{1}{\omega- \omega_-} - \frac{1}{\omega- \omega_+}   \right],
\end{align}
is manifestly convergent, that is
\begin{align}\label{eq:retardedM4dissip}
	G^{R} (k; t - t') &= \int \frac{\dd \omega}{2\pi} \ee^{i \omega (t-t')} G^{R} (k; \omega) = - \frac{\sin \left[E_k^{\gamma} (t-t')\right]}{E_k^{\gamma}} \ee^{- \frac{\gamma}{2}(t-t')}\theta\left(t-t'\right),
\end{align}
where we defined $E_k^{\gamma} = \sqrt{c_s^2 k^{2}-(\gamma/2)^{2}}$. Note that the causality property encoded in the above Heaviside theta function indeed implies analyticity of $G^R$ in the upper-half complex frequency plane.
Therefore, $\gamma > 0$ ensures causality and stability, via late-time convergence. 

Note that the noise term $ i \beta \varphi_a^2$ appearing in \Eq{eq:presc} does not affect the propagation of the dissipative degree of freedom \cite{kamenev_2011}. Indeed, the noise sources the dynamics of the system but does not directly transform the propagation of information within it (said differently, the former relates to the Keldysh function whereas the latter to the Green's functions). The Keldysh component is also easily obtained in frequency space where
\begin{align}
	G^{K} (k; \omega) &= 2 G^{R} (k; \omega) \widehat{D}_K (k; \omega) G^{A} (k; \omega) = \frac{2 i  \beta}{\left(\omega^2-c_s^2k^2\right)^2+\gamma^2 \omega^2}.
\end{align}
We assume $E_k^{\gamma}$ to be real (\textit{damped regime}), keeping in mind that the \textit{overdamped regime} for which $E_k^{\gamma} \in i\mathbb{R}$ can be obtained by analytic continuation. One can analyse the pole structure and obtain the real-space propagator from 
\begin{align}
	G^{K} (k; \tau) &= 2 i \beta \int \frac{\dd \omega}{2\pi} \frac{\ee^{-i\omega \tau}}{\prod_{i=1}^{4}(\omega-\omega_i)}
\end{align}
with the poles satisfying $\omega_2 = - \omega_1$, $\omega_3 = - \omega^*_1$, $\omega_4 = \omega^*_1$ and 
\begin{align}
	\omega_1 = E_k^{\gamma} + i \frac{\gamma}{2}.
\end{align}
A long but straightforward derivation leads to 
\begin{align}\label{eq:distribM4dissip}
	G^{K} (k; \tau) &= i \frac{\ee^{-\frac{\gamma}{2}\tau}}{c_s^2 k^2} \bigg[\frac{2 \beta}{\gamma} \cos\left(E_k^{\gamma} \tau\right) + \beta \frac{\sin\left(E_k^{\gamma} \tau\right)}{E_k^{\gamma}}\bigg]
\end{align}
In the coincident time limit, we obtain the dissipative power spectrum
\begin{align}
	P_k = \frac{2 \beta}{\gamma c_s^2k^2}.
\end{align} 
From the fact that the equal-time power spectrum must be non-negative we conclude that the $\beta$'s must be positive. Note that the vacuum contribution to the power spectrum, the usual $P_k=1/2E_k^{\gamma = 0}$ term, is absent because of the exponential decay in time caused by dissipation. Nevertheless, with hindsight, one could still recover the standard vacuum Minkowski power spectrum by taking both $\beta_1$ and $\gamma$ to zero while keeping their ratio fixed. This will be the object of the next Section.

This computation highlights several features of the formalism. While $G^R(k, \tau)$ encodes the dynamics but is oblivious to the state of the system, $G^K(k, \tau)$ captures the state of the environment by probing the statistics of fluctuations \cite{kamenev_2011}. The final outcome is an interplay between the dissipation of the system into its surroundings and the fluctuations of the environment getting imprinted onto the observable sector. Crucially, these two effects cannot be easily disentangled from one another.

\subsection{Interactions and diagrammatics}\label{subsec:flatinter}

Beyond the Gaussian statistics, higher-point functions are computed following the usual perturbative approach. Once the propagators are known, we can derive a new set of Feynman rules from which we construct correlators order by order in perturbation theory. In this section, after reviewing the standard in-in treatment of interactions, we study the structure of the bispectrum (three-point function in Fourier space) in flat space where analytic results are easily obtained.

\subsubsection{Feynman rules}

\begin{figure}[t]
	\centering
	\includegraphics{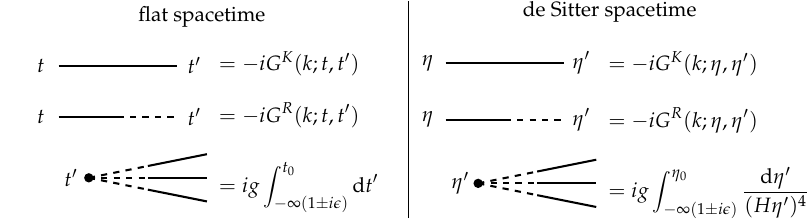}
	\caption{Feynman rules in the Keldysh basis.}
	\label{fig:rules}
\end{figure}

Interactions are treated as in the familiar in-in approach presented in \Sec{subsubsec:diagram}. This provides a comforting unified treatment for the cases of an open and closed system. The only small difference from some references is that we find it convenient to work in the Keldysh basis, $\varphi_{r,a}$ instead of the $\varphi_\pm$ basis. Expectation values of $\widehat{Q}(\eta) \equiv \widehat{\varphi}(\eta, \bmx_1) \cdots \widehat{\varphi}(\eta, \bmx_n)$ are defined through 
\begin{align}\label{eq:expvalQ}
	\langle \widehat{Q} \rangle =  \int \dd \phi \int_{\mathrm{I.C.}}^{\phi} \mathcal{D}\phir \int_{\mathrm{I.C.}}^{0} \mathcal{D}\phia\,  \left[\phir(\eta, \bmx_1) \cdots \phir(\eta, \bmx_n) \right] \ee^{i S_{\mathrm{eff}}\left[ \phir, \phia\right]},
\end{align}
where initial conditions lie on the boundaries of the path integral. Once the generating functional is known,  $\langle \widehat{Q} \rangle$ is extracted out of functional derivatives\footnote{Note that one needs to vary with respect to the \textit{advanced} source to obtained a \textit{retarded} field insertion. This comes from the NEQ constraint which imposes $\varphi_r J_a + \varphi_a J_r$ \cite{Liu:2018kfw}.}
\begin{align}
	\langle \widehat{Q} \rangle =  \left.\frac{\delta}{\delta J_{a}(\eta, \bmx_{1})} \cdots \frac{\delta}{\delta J_{a}(\eta, \bmx_{n})}\mathcal{Z}[J_{r},J_{a}]\right|_{J_{r,a}=0}
\end{align}
just as we did in \Eq{eq:extract} for the two-point function. The Feynman rules are derived as in \cite{Chen:2017ryl} and lead to \Fig{fig:rules}. Compared to the previous rules, the main differences are the following:
\begin{itemize}
	\item There are two propagators that are $-iG^K(k; \eta, \eta')$ (continuous line) which connects $\phir(k;\eta)$ to $\phir(k;\eta')$ and $-iG^R(k; \eta, \eta')$ (continuous-to-dashed line), which connects $\phir(k;\eta)$ to $\phia(k;\eta')$. Notice that the latter is directional, with the continuous line being attached to the $\phir(k;\eta)$ insertion and the dashed part to the $\phia(k;\eta')$. This leads to certain properties of the Feynman diagrams known as causality flows \cite{Radovskaya:2020lns, Ema:2024hkj}. There is no propagator connecting $\phia(k;\eta)$ to $\phia(k;\eta')$ which is a consequence of the causality structure of the closed time contour \cite{kamenev_2011}.
	\item Then, diagrams evaluation follows the exact same rules as in \cite{Chen:2017ryl}. As seen from \Eq{eq:expvalQ}, external legs connecting to the conformal boundary $\eta_0 \rightarrow 0^-$ are continuous, corresponding to $\phir$ insertions. 
\end{itemize}
An example is given in \Fig{fig:Btree} for a contact bispectrum. One can easily be convinced of these Feynman rules by recovering some known results as we do in \textit{Problem Set} \ref{subsec:pb2}.

\subsubsection{Flat-space dissipative bispectrum}\label{subsec:flatB}

Before turning our attention to primordial cosmology in Lecture \ref{sec:lec3}, it is instructive to discuss the generic structure of the contact three-point functions in Minkowski in the presence of dissipation. We derive these results following the Feynman rules enumerated in \Fig{fig:rules}, the propagators being given in \Eqs{eq:retardedM4dissip} and \eqref{eq:distribM4dissip}. For the sake of clarity, we temporarily set $c_s = 1$. 

\begin{figure}[t]
	\centering
	\includegraphics{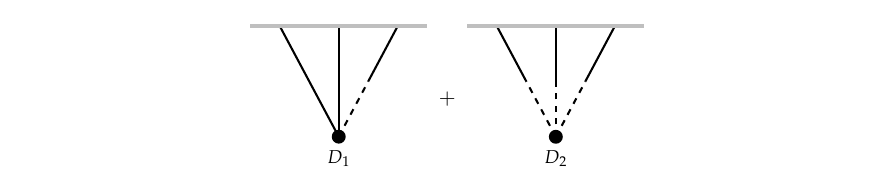}
	\caption{The two diagrams to compute the interaction given in \Eq{eq:cubicdot}. \textit{Left}: diagram corresponding to the vertex $\dot{\varphi}_r^2 \dot{\varphi}_a$, resulting in \Eqs{eq:D1dot}. \textit{Right}: diagram corresponding to the vertex $\dot{\varphi}_a^3$, resulting in \eqref{eq:D2dissip}.}
	\label{fig:Btree}
\end{figure}

\paragraph{\it $\dot{\varphi}^3$ interactions} Let us first consider 
\begin{align}\label{eq:cubicdot}
	\mathcal{L}_{\mathrm{int}} = -\frac{\alpha}{3!}\left(\dot{\varphi}^3_+ - \dot{\varphi}^3_-\right) = -\frac{\alpha}{2} \left(\dot{\varphi}_r^2 \dot{\varphi}_a + \frac{1}{12} \dot{\varphi}_a^3\right).
\end{align}
where the minus sign in front comes from the Lorentzian signature, assuming $\alpha>0$. We aim at computing the bispectrum 
\begin{align}\label{eq:bispecdef}
	\langle \varphi_{\bmk_1}  \varphi_{\bmk_2}  \varphi_{\bmk_3} \rangle \equiv (2\varphi)^3 \delta(\bmk_1 + \bmk_2 + \bmk_3) B_\varphi(k_1,k_2,k_3),
\end{align}
for which we have two contact diagrams to consider presented in \Fig{fig:Btree}, $B_\varphi(k_1,k_2,k_3) = D^{\dot{\varphi}^3}_1 + D^{\dot{\varphi}^3}_2$. 

The first one leads to
\begin{align}\label{eq:D1dot}
	D^{\dot{\varphi}^3}_1 &=  \frac{\alpha}{2} \int_{-\infty(1\pm i \epsilon)}^{t_0} \dd t \left[\partial_t G^K(k_1; t_0,t)\right] \left[\partial_t G^K(k_2; t_0,t)\right] \left[\partial_tG^R(k_3; t_0,t)\right] + 5~\mathrm{perms}. 
\end{align}
Upon injecting \Eqs{eq:retardedM4dissip} and \eqref{eq:distribM4dissip} in \Eq{eq:D1dot} and  summing over the six possible permutations, we observe that $D^{\dot{\varphi}^3}_1 = 0$. The second diagram is made of the $\phia$ components only, and reads 
\begin{align}\label{eq:D2dissip}
	D^{\dot{\varphi}^3}_2 = \frac{\alpha}{24} \int_{-\infty(1\pm i \epsilon)}^{t_0} \dd t \left[\partial_t G^R(k_1; t_0,t) \right] \left[\partial_t G^R(k_2; t_0,t)\right] \left[ \partial_t G^R(k_3; t_0,t)\right] + 5~\mathrm{perms}.
\end{align}
Under the same procedure, this diagram leads to a non-zero contribution to the bispectrum such that
\begin{align}\label{eq:D2resultdissip}
	B_\varphi(k_1,k_2,k_3) &=  \frac{\alpha \gamma}{2} \frac{  \mathrm{Poly}_6\left(e_1^\gamma, e_2^\gamma,e_3^\gamma \right)}{ \mathrm{Sing}_\gamma} ,
\end{align}
where we defined the singularity structure
\begin{align}\label{eq:singdissip}
	\mathrm{Sing}_\gamma =& \left| E_1^{\gamma} + E_2^{\gamma} + E_3^{\gamma} + \frac{3}{2}i \gamma\right|^2 \left| -E_1^{\gamma} + E_2^{\gamma} + E_3^{\gamma} + \frac{3}{2}i \gamma\right|^2 \nonumber\\
	&\times \left| E_1^{\gamma} - E_2^{\gamma} + E_3^{\gamma} + \frac{3}{2}i \gamma\right|^2 \left| E_1^{\gamma} + E_2^{\gamma} - E_3^{\gamma} + \frac{3}{2}i \gamma\right|^2, 
\end{align}
remembering that $E_k^{\gamma} = \sqrt{c_s^2 k^2 - \gamma^2/4}$. This singularity structure captures most of the specificities of the non-unitary dynamics. It emerges from time integrals of the form
\begin{align}\label{eq:M4int}
	\int_{-\infty(1\pm i \epsilon)}^{t_0} \dd t \ee^{ \pm i E_1^{\gamma} (t_0 - t) } \ee^{\pm i E_2^{\gamma} (t_0 - t) } \ee^{\pm i E_3^{\gamma} (t_0 - t) } \ee^{- \frac{3}{2}\gamma(t_0-t)}\,,
\end{align}
which follow from the structure of the propagators given in \Eqs{eq:retardedM4dissip} and \eqref{eq:distribM4dissip}.
Physically, it represents $3 \leftrightarrow 0$ and $2 \leftrightarrow 1$ interactions mediated by the environment. Fluctuations generate folded singularities while dissipation displaces the pole and regularises the divergence. Consequently, the singularity is not located in the physical plane and the bispectrum remains under perturbative control over the whole kinematic space. As we will see below, this singularity structure is generic and does not depend on the details of the interactions. The details of the particular interaction are imprinted into $\mathrm{Poly}_n$, which is a $n^{\mathrm{th}}$-order polynomial of the elementary symmetric polynomials
\begin{align}\label{eq:var}
	e_1^\gamma = E_1^{\gamma} + E_2^{\gamma} + E_3^{\gamma}, \quad e_2^\gamma = E_1^{\gamma} E_2^{\gamma} + E_2^{\gamma}  E_3^{\gamma} + E_1^{\gamma} E_3^{\gamma} \quad  e_3^\gamma = E_1^{\gamma} E_2^{\gamma} E_3^{\gamma}.
\end{align}
For this specific case, 
\begin{align}
	\mathrm{Poly}_6\left(e_1^\gamma, e_2^\gamma,e_3^\gamma \right) &= 243\gamma^6 - 792 \gamma^4 e_2^\gamma + 396 \gamma^4 \left(e_1^\gamma\right)^2 + 576 \gamma^2 \left(e_2^\gamma\right)^2 -1088 \gamma^2 e_2^\gamma\left(e_1^\gamma\right)^2 \nonumber \\
	&+272 \gamma^2 \left(e_1^\gamma\right)^4 + 1024 \gamma^2e_1^\gamma e_3^\gamma + 512 \left(e_1^\gamma\right)^2 \left(e_2^\gamma\right)^2 - 384 \left(e_1^\gamma\right)^4 e_2^\gamma \\
	&-1024 e_1^\gamma e_2^\gamma e_3^\gamma + 64 \left(e_1^\gamma\right)^6 + 512 e_3^\gamma \left(e_1^\gamma\right)^3 + 768 \left(e_3^\gamma\right)^2. \nonumber
\end{align}
Note that in a unitary shift symmetric theory, the operators $\dot{\varphi}^3$ and $(\partial_i \varphi)^2 \dot{\varphi}$ do not generate any contact bispectrum in Minkowski due to the time reversal symmetry $t\rightarrow-t$ and $\varphi \rightarrow -\varphi$ (one can check explicitly that there is zero contribution to the bispectrum, each diagram vanishing independently). Dissipation spontaneously breaks this symmetry and we observe that the diagrams now lead to a non-zero contribution to the bispectrum.

In \cite{Salcedo:2024smn}, it was found that other cubic interactions follow the same structure. The generic structure is:
\begin{tcolorbox}[
	enhanced, 
	colback=gray!10,
	colframe=gray!10,
	breakable,
	skin first=enhanced,
	skin middle=enhanced,
	skin last=enhanced,
	before upper={\parindent15pt},
	]{}
	
	\paragraph{\it Dissipative bispectrum in Minkowski}
	\begin{align}\label{eq:gen}
		B_\varphi(k_1,k_2,k_3) =f(\mathrm{EFT}) \frac{  \mathrm{Poly}_n\left(e_1^\gamma, e_2^\gamma,e_3^\gamma \right)}{ \mathrm{Sing}_\gamma} 
	\end{align}
	where $f(\mathrm{EFT})$ a rational function of the EFT coefficients (and possibly the kinematics for spatial derivative interactions), $\mathrm{Poly}_n$ are polynomials of the variables given in \Eq{eq:var} and $\mathrm{Sing}_\gamma$ is the singularity structure expressed in \Eq{eq:singdissip}.
	
\end{tcolorbox}
The simplicity of the structure, which originates from integrals of the form of \Eq{eq:M4int}, might suggest the future development of bootstrap techniques for this kind of local dissipative dynamics. It also suggests the physics is well captured from the interpretation of $\mathrm{Sing}_\gamma$ controlling the amplitude of $3 \leftrightarrow 0$ and $2 \leftrightarrow 1$ interactions mediated by the environment.

\begin{figure}[t]
	\centering 
	\makebox[\textwidth][c]{\includegraphics[width=.5\textwidth]{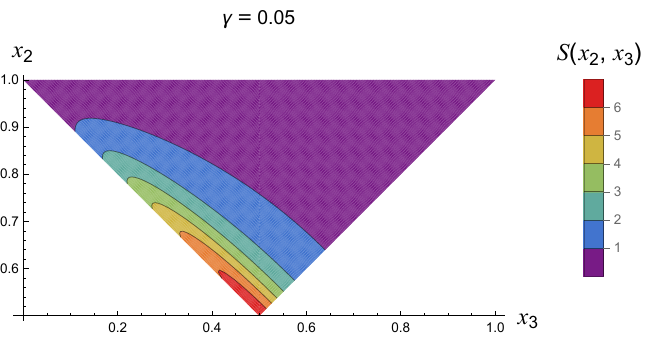}
		\includegraphics[width=.5\textwidth]{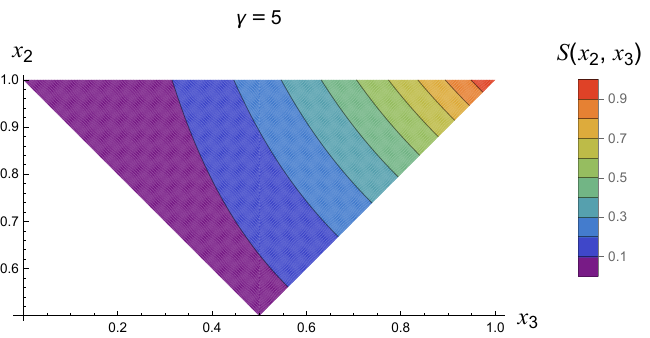}}
	\caption{\label{fig:shapeMink} Shape function of the contact bispectrum generated by $\dot{\varphi}_a^3$ in Minkowski. \textit{Left:} In the small dissipation regime, $\mathrm{Sing}_\gamma$ becomes small in the folded region $x_2 + x_3 \simeq 1$, which enhances the bispectrum. \textit{Right:} In the large dissipation regime, the imaginary contributions from the dissipation dominates $\mathrm{Sing}_\gamma$ such that no bispectrum enhancement is observed in the $x_2 + x_3 \simeq 1$ folded region.}
\end{figure}

Momentum conservation in \Eq{eq:bispecdef} forces the momenta to form a close triangle. As different inflationary models predict maximal signals in different triangular configurations (see e.g. \cite{Babich:2004gb, Chen:2006nt}), the shape function
\begin{align}\label{eq:shaperef}
	S(x_2,x_3) \equiv (x_2 x_3)^2 \frac{B(k_1, x_2 k_1, x_3 k_1)}{B(k_1,k_1,k_1)}
\end{align}
is an informative probe of the mechanism generating primordial non-Gaussianities. The variables $x_2 \equiv k_2/k_1$ and $x_3 \equiv k_3/k_1$ control the shape of the triangles and are restricted by $\delta(\bmk_1 + \bmk_2 + \bmk_3)$ to the region $\max(x_3, 1- x_3) \leq x_2 \leq 1$. The singularity structure $\mathrm{Sing}_\gamma$ presented in \Eq{eq:singdissip} exhibits two different behaviours depending on the magnitude of the dissipation coefficient $\gamma$ as can be seen in \Fig{fig:shapeMink}

\subsection{\textit{Problem set}}\label{subsec:pb2}

\begin{exbox}
\paragraph{\it A condensed matter approach to the $i \epsilon$ prescription}

\noindent Consider a massless scalar field $\varphi$ with free functional in the Keldysh basis given by 
\begin{equation}\label{eq:freeieps}
	S_{\mathrm{eff}} =\int \dd^4x \left(\dot{\varphi}_r \dot{\varphi}_a  -\partial^{i}\varphi_{r}\partial_{i}\varphi_{a} -2\epsilon\dot{\varphi}_{r} \varphi_{a} -\epsilon^{2}\varphi_{r}\varphi_{a}+i\epsilon \fini\varphi_{a}^{2} \right)\,,
\end{equation}
with $\epsilon$ a small parameter and $f$ a function to be determined.
\begin{enumerate}
	\item Interpret the various terms appearing in $S_{\mathrm{eff}}$.
	\item We first find the retarded Green's function.
	\begin{enumerate}
		\item What is the equation of motion obeyed by the retarded Green's function $G^{R} (k; t_1, t_2)$?
		\item Solve it in frequency space. Draw the poles of $G^{R} (k; \omega)$ in the complex plane. What is the role of $\epsilon$?
		\item Give the expression of $G^{R} (k; t_1, t_2)$ in real space. Take the limit $\epsilon \rightarrow 0$ and comment. 
	\end{enumerate}
	\item We now find the Keldysh propagator.
	\begin{enumerate}
		\item Express the Keldysh propagator in terms of the retarded Green's function. 
		\item Give the expression of $G^{K} (k; t_1, t_2)$ in real space. What is the role of $\epsilon$? Take the limit $\epsilon \rightarrow 0$ and comment.
		\item Which value $f$ should take to recover the vacuum power spectrum of a massless scalar, $1/2k$? Can you guess which physical parameter is controlled by $f$?
	\end{enumerate} 
\end{enumerate}
You have recovered the vacuum propagators! The $i \epsilon$ prescription is a way to implement the system's initial conditions in a vacuum or a thermal state.
\end{exbox}
\paragraph{\textit{Solutions}}

\begin{enumerate}
    \item We start from the free functional
\begin{equation}
	S_{\mathrm{eff}}[\phir,\phia] =\int \dd^4x \left(\dot{\varphi}_r \dot{\varphi}_a  -\partial^{i}\varphi_{r}\partial_{i}\varphi_{a} - 2\epsilon\dot{\varphi}_{r} \varphi_{a} - \epsilon^{2}\varphi_{r}\varphi_{a} + i\epsilon \fini\varphi_{a}^{2} \right)\,.
\end{equation}
The term $\dot{\varphi}_r \dot{\varphi}_a  -\partial^{i}\varphi_{r}\partial_{i}\varphi_{a}$ corresponds to the canonical kinetic term for a scalar field in the Keldysh basis and reproduces the classical equations of motion upon variation with respect to $\varphi_a$. The term $2\epsilon\dot{\varphi}_{r} \varphi_{a}$ introduces dissipation and implements the retarded $i\epsilon$ prescription, ensuring convergence of time integrals and causal propagation. The $\epsilon^2$-term acts as a regulator that accompanies the linear damping term and guarantees the correct analytic structure of the propagators by shifting the poles away from the real frequency axis. Finally, the last term represents stochastic noise and is required for the consistency of the SK formulation. It renders the functional integral well-defined and complements dissipation so that causal, unitary evolution is preserved.
\item We now look at the retarded Green's function:
\begin{enumerate}
    \item To find the retarded equation of motion, we vary $S_{\text{eff}}$ with respect to the advanced field and find its extremum. In spatial Fourier space, this gives
    \begin{equation}    \left[(\partial_0+\epsilon)^2+k^2\right]\varphi_r(k,t)=0,
    \end{equation}
    The retarded Green's function is defined as the solution of
    \begin{equation}
    \left[(\partial_{t_1}+\epsilon)^2+k^2\right]G^R(k;t_1,t_2)=\delta(t_1-t_2),
    \end{equation}
    supplemented with the retarded boundary condition $ G^R(k;t_1,t_2)=0$ for $t_1<t_2$.
    \item Introducing $\tau\equiv t_1-t_2$ and Fourier transforming in time,
    \begin{equation}
    G^R(k;\tau)=\int\frac{d\omega}{2\pi}\,e^{-i\omega\tau}\,G^R(k;\omega),
    \end{equation}
    the differential equation becomes
    \begin{equation}
    \left[(-i\omega+\epsilon)^2+k^2\right]G^R(k;\omega)=1 .
    \end{equation}
    Solving for $G^R$ gives
    \begin{equation}
    G^R(k;\omega)=-\frac{1}{(\omega+i\epsilon)^2-E_k^2}=\frac{1}{2E_k}\left(\frac{1}{\omega-E_k+i\epsilon}-\frac{1}{\omega+E_k+i\epsilon}\right),
    \end{equation}
    where $E_k\equiv k$ for a massless scalar.
    
    The retarded Green's function has simple poles at
    \begin{equation}
    \omega=\pm E_k-i\epsilon ,
    \end{equation}
    which lie entirely in the lower half of the complex $\omega$-plane. The parameter $\epsilon>0$ enforces the retarded boundary condition, ensures convergence of time integrals, and specifies how the poles are approached in the limit $\epsilon\to0^+$.
\item The retarded Green's function in real time is obtained by inverse Fourier
transforming $G^R(k;\omega)$,
\begin{equation}
G^R(k;\tau)=\int\frac{d\omega}{2\pi}\,e^{i\omega\tau}\,G^R(k;\omega),\qquad \tau\equiv t_1-t_2 .
\end{equation}
For $\tau>0$ the integration contour is closed in the lower half of the complex $\omega$-plane, while for $\tau<0$ it is closed in the upper half-plane. Since both poles of $G^R(k;\omega)$ lie below the real axis, only the case
$\tau>0$ contributes, yielding
\begin{equation}
G^R(k;\tau)=\theta(\tau)\,e^{\epsilon\tau}\,\frac{\sin(E_k\tau)}{E_k}.
\end{equation}

Taking the limit $\epsilon\to0^+$, one finds
\begin{equation}
\lim_{\epsilon\to0^+} G^R(k;\tau)=\theta(\tau)\,\frac{\sin(E_k\tau)}{E_k}.
\end{equation}
The $\epsilon$ prescription is therefore essential to enforce causality ($G^R=0$ for $\tau<0$), while the physical propagator is recovered smoothly in the limit $\epsilon\to0^+$.

\end{enumerate}
\item We now focus on the Keldysh propagator.
\begin{enumerate}
    \item In the Schwinger--Keldysh formalism, the Keldysh propagator is related to the
retarded and advanced Green's functions through
\begin{equation}
G^K(k;t_1,t_2)= -\,i f\,\epsilon \int dt'\,G^R(k;t_1,t')\,G^A(k;t',t_2),
\end{equation}
where $f$ parametrises the occupation of the initial state.

The advanced Green's function is obtained by reversing the $i\epsilon$ prescription relative to the retarded one. In frequency space,
\begin{equation}
G^{A}(k;\omega)=-\frac{1}{(\omega-i\epsilon)^2-E_k^2}=\frac{1}{2E_k}\left(\frac{1}{\omega-E_k-i\epsilon}-\frac{1}{\omega+E_k-i\epsilon}\right),
\end{equation}
and satisfies $G^A(k;\omega)=[G^R(k;\omega)]^\ast$.

\item
In real time, the retarded and advanced Green's functions read
\begin{equation}
G^{R/A}(k;\tau)=\pm\,\theta(\pm\tau)\,e^{\mp\epsilon\tau}\,\frac{\sin(E_k\tau)}{E_k},\qquad \tau\equiv t_1-t_2 .
\end{equation}
Substituting these expressions into the convolution formula, the integral over $t'$ has support only for $t_1>t'>t_2$. Performing the integral yields
\begin{equation}
G^K(k;\tau)= i\,\frac{f}{4E_k^2}\,\cos(E_k\tau).
\end{equation}

The parameter $\epsilon$ ensures convergence of the time integral and enforces the retarded/advanced boundary conditions. In the limit $\epsilon\to0^+$, the Keldysh propagator remains finite, while causality is preserved.

\item
For a free massless scalar field in the vacuum, the equal-time power spectrum is
\begin{equation}
\langle \varphi(k)\varphi(-k)\rangle = \frac{1}{2E_k}.
\end{equation}
Comparing with the Keldysh propagator, this result is recovered by choosing
\begin{equation}
f(E_k)=2E_k .
\end{equation}
More generally, one may write $f(E_k)=2E_k(1+2n_k)$, where $n_k$ is the occupation number. The Keldysh propagator thus encodes the statistical properties of the initial state, while causality and time evolution are entirely controlled by the retarded and advanced sectors.
\end{enumerate}
\end{enumerate}


\section{Open inflation}\label{sec:lec3}

Despite the variety of microphysical models compatible with the current cosmological observations, the EFT of Inflation (EFToI) \cite{Cheung:2007st} and its modern avatar through the cosmological bootstrap \cite{Baumann:2022jpr} have extracted a vast number of model-independent results. Solely based on symmetries (scale invariance, homogeneity and isotropy), principles (unitarity, locality) and scale hierarchies (slow roll expansion), cosmological correlators can still be constrained to a large extent. In this Section, after reviewing the original construction of the EFToI in \Sec{subsec:EFToI}, we develop its extension to capture local dissipation and noise in \Sec{subsec:open}. As we will see in \Sec{subsec:pheno}, model-independent phenomenological implications can still be derived in this case, some of which constitute a set of smoking-gun signatures for dissipative and stochastic dynamics.


\subsection{The Effective Field Theory of Inflation}\label{subsec:EFToI}

We first briefly review the original construction presented in \cite{Cheung:2007st, Gubitosi:2012hu, Piazza:2013coa}. We consider a single scalar field $\phi(t,\bfx)$ evolving unitarily in a perturbed Friedmann–Lema\^itre–Robertson–Walker (FLRW) geometry. Both the scalar field and the metric are expanded around their background value 
\begin{align}
	\phi(t,\bfx) = \bar \phi(t) + \delta \phi(t, \bfx), \qquad\qquad g_{\mu\nu}(t,\bfx) = \bar{g}_{\mu\nu}(t) + \delta g_{\mu \nu}(t,\bfx),
\end{align}
and we aim at understanding the dynamics of the fluctuations. The theory is conveniently constructed in the unitary gauge, which is defined by the condition\footnote{In this subsection, we use $\doteq$ to denote an equality that holds in unitary gauge.} $\delta \phi \doteq 0$. In this case, all the perturbations are absorbed into the metric and the homogeneous scalar field $\phi(t, \bfx) \doteq \bar{\phi}(t)$ can be used as a clock. Indeed, one can parametrise the time slicing in terms of the homogeneous value of the scalar field $t = t(\phi(t,\bfx))$ and construct geometrical objects based on this time foliation of spacetime. For instance, the unit vector perpendicular to the foliation is
\begin{align}\label{eq:n_mu_def}
	n_\mu \equiv - \frac{\partial_\mu \phi}{\sqrt{-g^{\mu\nu} \partial_\mu \phi \partial_\nu \phi}} \doteq - \frac{\delta^0_{~\mu}}{\sqrt{-g^{00}}},
\end{align}
where the second equality holds in the unitary gauge\footnote{A related useful expression is $n^\mu=-g^{\mu 0}/\sqrt{-g^{00}}$ and $\sqrt{-g^{00}}\doteq n^0 \doteq -1/n_0$. Note that $n^\mu$ is a future-pointing time-like vector.}.


\paragraph{\it Effective action.} This formulation allows us to write down the most general (unitary) EFT compatible with the symmetries of the problem \cite{Cheung:2007st}. The presence of the inflaton background $\bar \phi(t)$ spontaneously breaks time-translation symmetry, such that the resulting action is made of terms that are invariant under spatial diffeomorphisms only. Of course allowed terms include $4d$ covariant operators such as $R$, which are necessarily also $3d$ covariant. But one should also allow time-dependent functions ($\Lambda(t),~\cdots$), $g^{00}$ (which is a scalar w.r.t. spatial transformations), or geometrical objects constructed out of the foliation such as contractions with $n_\mu$ or the extrinsic curvature 
\begin{align}\label{eq:extrinsiccurv}
	K_{\mu\nu} \equiv \left(\delta_{\mu}^{\sigma} + n_\mu n^\sigma \right) \nabla_\sigma n_\nu \doteq \delta_\mu^i \delta_\nu^j \Gamma_{ij}^0 (-g^{00})^{-1/2}\,.
\end{align}
The most generic action takes the form \cite{Cheung:2007st}
\begin{align}
	S = \int \dd^4 x \sqrt{-g} F(R_{\mu\nu\rho\sigma}, g^{00}, K_{\mu\nu}, \nabla_\mu; t),
\end{align}
where $F$ is an arbitrary function. Expanding the metric in powers of the perturbations we obtain \cite{Gubitosi:2012hu}
\begin{align}\label{eq:universal}
	S = \int \dd^4 x \sqrt{-g} \left[ \frac{M^2_{\mathrm{Pl}}}{2} R - \Lambda(t) - c(t) g^{00}\right] + S^{(2)},
\end{align}
where $R$ is the Ricci scalar and $\Lambda$ and $c$ are functions of time, and $S^{(2)}$ starts at second order in perturbations, and consequently does not affect the background Friedmann equations 
\begin{align}\label{eq:backgroundeq}
	3\Mpl^2 H^2 = \Lambda(t) + c(t), \qquad 2\Mpl^2 \dot{H} = -2c(t),
\end{align}
together with the continuity equation 
\begin{align}\label{eq:contiEFToI}
	\dot{\Lambda}(t) + \dot{c}(t) + 6Hc(t) = 0.
\end{align}
The first three terms in \Eq{eq:universal} constitute the \textit{universal part} of the EFT of Inflation. 
Details about $S^{(2)}$ can be found in \cite{Creminelli:2017sry}. At lowest order in derivatives acting on the metric, it takes the form
\begin{align}
	S^{(2)} =  \int \dd^4 x \sqrt{-g} \left[ \sum_{n=2}^\infty M^4_n(t) (1 + g^{00})^n\right].
\end{align}
\paragraph{\it St\"uckelberg trick and the Goldstone boson.}

At its core, the EFT of Inflation is a theory of the Goldstone mode of spontaneously broken time translations. Although in unitary gauge this scalar degree of freedom is entirely encoded within the metric perturbations, we can make the Goldstone mode $\pi(t, \bfx)$ manifest again by reintroducing time diffeomorphisms. 
This is achieved via the St\"uckelberg trick \cite{Cheung:2007st}, which restores full 4d diffeomorphism invariance to the action by performing the local time shift:
\begin{align}
	t \rightarrow t + \pi(t,\bfx), \qquad \bfx \rightarrow \bfx.
\end{align}
Under this transformation, objects that are not $4d$ diffeo-invariant do transform, such as time-dependent functions
\begin{align}
	c(t) \rightarrow c(t + \pi) =  c(t) + \dot{c}(t) \pi + \mathcal{O}(\pi^2)
\end{align}
or the time component of the metric 
\begin{align}
	g^{00} &\rightarrow  \frac{\partial ( t + \pi) }{\partial x^\alpha} \frac{\partial ( t + \pi) }{\partial x^\beta} g^{\alpha \beta} =  g^{00} + 2 g^{0\mu} \partial_\mu \pi + g^{\mu \nu} \partial_\mu \pi \partial_\nu \pi, \\
	g^{0i} &\rightarrow  \frac{\partial ( t + \pi) }{\partial x^\alpha} g^{\alpha i} = g^{0i} +  g^{i\nu} \partial_\nu \pi.
\end{align}
Under this transformation, the effective action 
\begin{align}\label{eq:fullaction}
	 S = \int \dd^4 x \sqrt{-g} \bigg[ &\frac{M^2_{\mathrm{Pl}}}{2} R - \Lambda(t+\pi) - c(t+\pi) \left(g^{00} + 2 g^{0\mu} \partial_\mu \pi + g^{\mu \nu} \partial_\mu \pi \partial_\nu \pi \right) \nonumber \\
	  & \quad + \sum_{n=2}^\infty M^4_n(t + \pi) (1 + g^{00} + 2 g^{0\mu} \partial_\mu \pi + g^{\mu \nu} \partial_\mu \pi \partial_\nu \pi)^n\bigg],
\end{align}
becomes manifestly $4d$ diffeo-invariant as long as $\pi$ realises time-translations non-linearly, that is when $t\rightarrow t - \epsilon$, $\pi$ transforms as
\begin{align}
	\pi(t) \rightarrow \pi'(t) = \pi(t+\epsilon)+\epsilon.
\end{align}


\paragraph{\it Decoupling.} 
\textit{Foreword.} This paragraph provides partial solutions to Problem 1 of the problem set in Section~\ref{subsec:prob3}. Readers who wish to practice on problems related to the EFT of Inflation are encouraged to attempt the problem set first, before consulting this paragraph.
\\

The primary advantage of restoring the dynamical scalar $\pi$ is the existence of a decoupling limit, where metric fluctuations can be safely neglected. This decoupling is most transparent in the spatially flat gauge, parametrised by:
\begin{equation}
	\delta g_{ij}^{\rm scalar}=0\quad,\quad\delta g_{00}=-2\phi=-\delta g^{00}\quad,\quad \delta g_{0i}=a(t)\partial_{i}F=\delta g^{0i}.
\end{equation}
Considering the universal action along with the minimal extension controlled by $M_2^4(t)$, the $00$ and $0i$ Einstein equations act as non-dynamical constraints. Solving these constraint equations for the metric fluctuations yields:
\begin{align}\label{eq:constraintsol}
	\phi = 2 \epsilon H \pi, \qquad \nabla^2 F = - \frac{a \epsilon H}{c_s^2} \left[\dot{\pi} + \left(3 c_s^2- 2\epsilon \right) H \pi \right], 
\end{align}
where we introduced the first slow-roll parameter $\epsilon \equiv -\dot{H}/H^2$ and the speed of sound $c_s^{-2} \equiv [c(t) + 2 M_2^4(t)]/c(t)$.

One can then perturb \Eq{eq:fullaction} at second order in perturbations, substitute $\phi$ and $\nabla^2F$ found in \Eq{eq:constraintsol} and compare the contributions from $\phi$ and $F$ to the self dynamics of $\pi$ itself. For instance, the kinetic term of the $\pi$ roughly scales as 
\begin{align}
 (\mathrm{I}): \qquad	c(t) (\partial_\mu \pi)^2 \quad \sim \quad\epsilon \Mpl^2 H^2 \dot{\pi}^2 \quad \sim \quad  \epsilon \Mpl^2 H^4 \pi^2
\end{align}
where we used the background expression of $c(t)$ found in \Eq{eq:backgroundeq} and the heuristic estimate of the time derivatives during inflation $\dot{\pi}\sim H\pi$ \cite{Salcedo:2024smn}. Comparing this expression with the leading mixing with gravity \cite{Cheung:2007st}
\begin{align}
(\mathrm{II}): \qquad	c(t) \delta g^{00} \dot{\pi} \quad \sim \quad  \epsilon \Mpl^2 H^2 \phi \dot{\pi} \quad \sim \quad \epsilon^2 \Mpl^2 H^4 \pi^2,
\end{align}
where in the last line we used \Eq{eq:constraintsol}, we find that $(\mathrm{II}) \ll (\mathrm{I})$ as long as $\epsilon \ll 1$. 
This explicit computation shows that the contribution of the fluctuations of the metric on the scalar dynamics of $\pi$ is suppressed by the slow-roll parameter $\epsilon \ll 1$. A more rigorous derivation of the decoupling limit may be found in \cite{Salcedo:2025ezu}, where we perform an analogue computation in the Keldysh basis, extending the results of \cite{Cheung:2007st} to include dissipation and noise. 

In this limit, the action dramatically simplifies to 
\begin{align}
	\begin{split}
	    S_\pi = \int \dd^4 x \sqrt{-g} &\left\{ \frac{1}{2} \Mpl^2 R + \epsilon \Mpl^2 H^2 \left[ \dot{\pi}^2 - \frac{(\partial_i \pi)^2}{a^2} \right] +\right.\\
    &\left.+2 M_2^4 \left[ \dot{\pi}^2 + \dot{\pi}^3 - \dot{\pi}\frac{(\partial_i \pi)^2}{a^2} \right] - \frac{4}{3} M_3^4 \dot{\pi}^3 + \cdots \right\}, 
	\end{split}
\end{align}
where the ellipse represents terms at least quartic in $\pi$. From a theory with two tensor modes and one scalar degree of freedom, we end up with a theory with a single shift symmetric scalar \cite{Finelli:2018upr}. This framework is the starting point of the study of \cite{Salcedo:2024smn} which aims at extending this framework to include dissipation and noise.


\subsection{Opening the theory}\label{subsec:open}
The starting point to study an open system is a choice of the degrees of freedom that we want to describe. Given our focus on single-clock inflation, our system will consist of a scalar field $\pi$, the St\"uckelberg field introduced above. We aim to construct an EFT for $\pi$ in the presence of dissipative and stochastic effects induced by interactions with the unknown environment. The statistics of the system is characterised by its density matrix in the field basis
\begin{align}
	\rho_{\mathrm{red}}\left[\pi, \pi';\eta_0\right] = \int_{\Omega}^\pi \mathcal{D}\pi_+ \int_{\Omega}^{\pi'} \mathcal{D}\pi_- \ee^{i S_{\mathrm{eff}}\left[ \pi_+, \pi_-\right]}.\label{eq:densitymatrix}
\end{align}
Here $\Omega$ represents the choice of initial state, which we assume to be the Bunch-Davies state. The \textit{open effective functional} $S_{\mathrm{eff}}\left[ \pi_+, \pi_-\right]$ admits a Hermitian and a non-Hermitian part
\begin{align}\label{eq:Seffref}
	S_{\mathrm{eff}}\left[ \pi_+, \pi_-\right] = S_\pi\left[\pi_+\right] - S_\pi\left[\pi_-\right] + S_{\mathrm{IF}}\left[\pi_+,\pi_- \right].
\end{align}
Following \cite{Salcedo:2024smn}, our goal here is to construct and study $S_{\mathrm{eff}}\left[ \pi_+, \pi_-\right]$ for single-clock inflation. 


\subsubsection{Symmetries and locality}


\paragraph{\it Time-translation symmetry breaking}
Symmetries further restrict the number and structure of EFT operators. Let's consider a schematic microscopic theory $S_{\mathrm{UV}}[\pi,\chi]$ with $\chi$ a collective variable capturing the presence of an environment. Following the EFToI construction presented above \cite{Cheung:2007st, Finelli:2018upr}, $S_{\mathrm{UV}}[\pi,\chi]$ is invariant under shift symmetry of the $\pi$ field. Then, the Schwinger-Keldysh action $S_{\mathrm{UV}}[\pi_+,\chi_+] - S_{\mathrm{UV}}[\pi_-,\chi_-] $ for the closed system is invariant under $\mathrm{shift}_+ \times \mathrm{shift}_+ \equiv \mathrm{shift}_r \times \mathrm{shift}_a $. Integrating out the $\chi$ field generates terms mixing the branches of the path integral. In general, these terms do not transform nicely under transformation along each branch of the path integral. This leads to a symmetry breaking pattern under which $S_{\mathrm{eff}} [\pi_+,\pi_-]$ becomes only invariant under the smaller diagonal subgroup,
\begin{align}\label{eq:SSBEFToI}
\mathrm{shift}_+ \times \mathrm{shift}_-  \quad \equiv  \quad \mathrm{shift}_r \times \mathrm{shift}_a \quad	\rightarrow  \quad \mathrm{shift}_r.
\end{align}

\begin{figure}[t]
	\centering
	\includegraphics{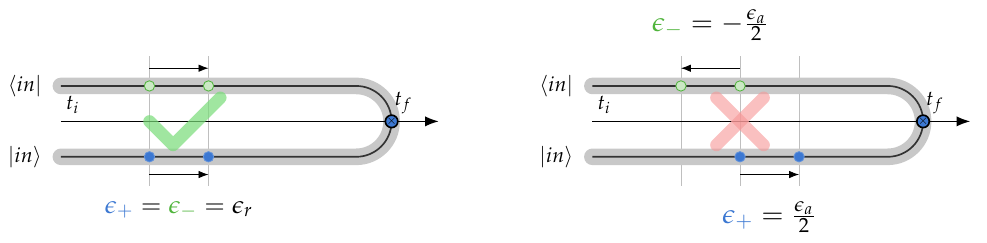}
	\caption{The $\epsilon_{r}$ and $\epsilon_{a}$ transformations on the closed time path, where time is running from left to right in both contours and the arrows represent path ordering (time ordering in $\ket{in}$ and anti-time-ordering in $\bra{in}$). The $\epsilon_{r}$ transformation translates the $``+"$ and $``-"$ variables in the same direction while $\epsilon_{a}$ transformation does it in the opposite directions. While the $\epsilon_{r}$ transformation is preserved, the $\epsilon_{a}$ one is explicitly broken due to dissipative effects \cite{Hongo:2018ant}.}
	\label{fig:sym}
\end{figure}

Explicitly, let us consider a UV-action that is invariant under independent time-translations, one transforming the $+$ branch of the path integral by $t - \epsilon_+$ and the other transforming the $-$ branch of the path integral by $t - \epsilon_-$. Under these transformations, $\pi_\pm$ \textit{non-linearly realise} the symmetry, 
\begin{align}\label{eq:SSB}
	\pi_+(t) \rightarrow \pi'_+(t) = \pi_+(t+\epsilon_+)+\epsilon_+,\quad\quad\pi_-(t) \rightarrow \pi'_-(t) = \pi_-(t+\epsilon_-)+\epsilon_-,  
\end{align}
that is they shift, while additional environment fields act linearly on the transformations. Upon tracing over the environment, the open effective functional $S_{\mathrm{eff}}\left[\pi_+,\pi_-\right]$ is not in general invariant under general $\epsilon_\pm$ translations. More precisely, the effective functional remains invariant under the translation (see the left-hand panel of \Fig{fig:sym}) \cite{Hongo:2018ant}
\begin{align}
	\epsilon_+ = \epsilon_- =\epsilon_{r}\,,
\end{align}
while the translations
\begin{align}
	\epsilon_+ =  - \epsilon_- =\frac{\epsilon_{a}}{2}
\end{align}
are explicitly broken (see the right-hand panel of \Fig{fig:sym}). In this way, out of two time-translational symmetries of the microscopic action, we are left with a single diagonal subgroup $\epsilon_+ = \epsilon_-$.

Let us consider the transformation of $\pi_r$ and $\pi_a$ under the diagonal subgroup of time translations and boosts. In the Keldysh basis, the $\epsilon_{r}$-transformations read \cite{Hongo:2018ant} (see left-hand panel of \Fig{fig:sym})\footnote{One can check that $S_{\mathrm{eff}}[\pi_r,\pi_a]$ is left invariant by the transformation given in \Eq{eq:translRsym}. To see it explicitly, one can consider the expansion
	\begin{align}\label{eq:translRsymexp}
		\pi_{r}(t,\bmx) &\rightarrow \pi_{r}(t,\bmx) + \epsilon_r\left[1+\dot{\pi}_{r}(t,\bmx)\right] + \mathcal{O}\left[(\epsilon_r)^2\right],\\
		\pi_{a}(t,\bmx) &\rightarrow  \pi_{a}(t,\bmx) + \epsilon_r\dot{\pi}_{a}(t,\bmx) + \mathcal{O}\left[(\epsilon_r)^2\right].
\end{align} } 
\begin{align}\label{eq:translRsym} 
	\pi_{r}(t,\bmx) \rightarrow \pi'_{r}(t,\bmx) &= \pi_{r}(t+\epsilon_{r},\bmx) + \epsilon_{r},\\
	\pi_{a}(t,\bmx) \rightarrow \pi'_{a}(t,\bmx) &= \pi_{a}(t+\epsilon_{r},\bmx),
\end{align} 
whereas the $\Lambda_{r}$-transformations follow 
\begin{align}\label{eq:LorentzRsym}
	\pi_{r}(t,\bmx) \rightarrow \pi'_{r}(t,\bmx) &= \pi_{r}\left(\Lambda^0_{r~\mu}x^\mu,\Lambda^i_{r~\mu}x^\mu \right)+\Lambda^0_{r~\mu}x^\mu - t, \\
	\pi_{a}(t,\bmx) \rightarrow \pi'_{a}(t,\bmx) &= \pi_{a}\left(\Lambda^0_{r~\mu}x^\mu,\Lambda^i_{r~\mu}x^\mu \right),
\end{align}
where we introduced $\Lambda^\mu_{r~\nu} \in \mathrm{SO}(1,3)$. The important point is that $\pi_r$ \textit{non-linearly realises time-translations and boosts} whereas $\pi_{a}$ \textit{transforms linearly}, just as ordinary matter \cite{Hongo:2018ant}. 


\paragraph{\it Locality.} 

In general, tracing over an environment yields a highly non-local effective functional that is intractable without knowledge of the exact UV completion. However, just as in standard EFTs, a dramatic simplification occurs in the presence of a \textit{separation of scales}. We assume that the typical length and timescales characterising the environment are much shorter than the Hubble radius and Hubble time, the scales at which we compute cosmological correlators. This hierarchy ensures that our open EFT is local in space and time; it can be expressed via operators that are products of fields at the same spacetime point and a finite number of derivatives thereof. 

Not all UV models exhibit this hierarchy. For example, well-studied models of inflation featuring tachyonic production of gauge modes via a $\phi F\tilde F$ coupling \cite{Anber:2009ua} would yield a non-local open EFT for $\pi$, because the gauge fields are predominantly produced around Hubble crossing. The fact that our open EFT cannot describe these models is not a flaw specific to open systems: standard EFTs similarly fail when attempting to integrate out light or massless fields. In contrast, our construction successfully describes models of gauge and warm inflation where particle production is strictly confined to the sub-Hubble regime, such as the framework developed in \cite{Creminelli:2023aly}, as we will show later.


\subsubsection{Open effective functional}

Now that we motivated a local open EFT, we proceed to outline the structure of the most general effective functional, by combining all possible local operators compatible with the non-equilibrium constraints \eqref{eq:norm}, \eqref{eq:herm} and \eqref{eq:pos}, and retarded shift symmetry \eqref{eq:SSB}. We can do this by using invariant combinations as fundamental building blocks \cite{Akyuz:2023lsm}. The open effective functional can be constructed out of
\begin{align}\label{eq:Pmu}
	\text{building blocks:} \qquad \pi_{a}, \quad t+ \pi_{r},\quad \partial_\mu \pi_a, \quad   \partial_\mu (t + \pi_{r}).
\end{align} 
It will turn convenient to define $	P_\mu \equiv \partial_\mu (t + \pi_{r}) = \delta^0_{~\mu}+ \partial_\mu \pi_{r}$. Working to leading order in the derivative expansion, we will restrict ourselves to operators with at most one derivative per field. 
In addition to the derivative expansion, a useful way to organise the open effective functional is in powers of $\pi_a$ so that $ S_{\mathrm{eff}} = \int \dd^4 x \sqrt{-g }  \mathcal{L}_{\mathrm{eff}}$ with 
\begin{align}\label{eq:expref}
	\mathcal{L}_{\mathrm{eff}} = \sum_{n=1}^\infty \mathcal{L}_n\quad\mathrm{with}\quad \mathcal{L}_n = \mathcal{O}(\pi_{a}^n)
\end{align}
where we used the unitarity condition \eqref{eq:norm} to notice that $\mathcal{L}_{\mathrm{eff}}$ starts from the first order term in $\pi_{a}$. 
We will illustrate this construction procedure by including the leading interactions, which are cubic in the fields. We can therefore restrict our attention to the functionals $\mathcal{L}_1$, $\mathcal{L}_2$, and $\mathcal{L}_3$.


\paragraph{\it $\mathcal{L}_1$ functional.}

Let us illustrate the procedure by first considering $\mathcal{L}_1$. We aim at using the building blocks \eqref{eq:Pmu} to construct invariant combinations that are linear in $\pi_a$. The only option consists in multiplying $\pi_a$ and $P^\mu \partial_\mu \pi_{a} = \left(-\dot{\pi}_{a} + \partial^\mu \pi_{r} \partial_\mu \pi_{a}\right)$ by powers of
\begin{align}
	\left( P_\mu P^\mu  +1\right) = -2 \dot{\pi}_{r} + \left( \partial_\mu \pi_{r}\right)^2,
\end{align}
leading to \cite{Hongo:2018ant}\footnote{The sign in front of the $\gamma_n$ term is chosen for later convenience.}
\begin{align}
	\mathcal{L}_1 = & \sum_{n=0}^\infty \left( P_\mu P^\mu  +1\right)^{n} \left[ \gamma_n \pi_{a} -\alpha_n P^\mu \partial_\mu \pi_{a}  \right].
\end{align}
The EFT coefficients $\gamma_n$ and $\alpha_n$ are in general functions of $t + \pi_{r}$ which have to be real because of the conjugate condition \eqref{eq:herm}.  In the slow-roll regime, one can assume time-independence for the EFT coefficients at leading order in slow-roll \cite{Hongo:2018ant}. 

\begin{tcolorbox}[
	enhanced, 
	colback=gray!10,
	colframe=gray!20,
	breakable,
	skin first=enhanced,
	skin middle=enhanced,
	skin last=enhanced,
	before upper={\parindent15pt},
	]{}
	\paragraph{\it Background evolution.}
	
	Before describing the dynamics of the fluctuations, let us connect with the standard background evolution of the EFToI by discussing tadpole cancellation \cite{Cheung:2007st, Collins:2012nq}. As mentioned below \Eq{eq:expref}, the unitarity condition \eqref{eq:norm} imposes that $\mathcal{L}_{\mathrm{eff}}$ starts linear in $\pi_{a}$. Therefore, the only available tadpoles are \cite{LopezNacir:2011kk,Hongo:2018ant}
	\begin{align}\label{eq:tad1}
		S_{\mathrm{eff}}  \supset  \int \dd^4 x  \sqrt{-g} \left[ \gamma_0(t) \pi_a  + \alpha_0(t) \dot{\pi}_a \right],
	\end{align}
	which leads to the continuity equation 
	\begin{align}
		- \gamma_0 + \dot{\alpha}_0 + 3H \alpha_0 = 0.
	\end{align}
	We can compare this expression with \Eq{eq:contiEFToI} to identify 
	\begin{align}\label{eq:tadpole}
		\alpha_0(t) \equiv 2 c(t) \qquad \mathrm{and} \qquad \gamma_0 (t)\equiv  \dot{c}(t) - \dot{\Lambda}(t),
	\end{align}
	such that \eqref{eq:contiEFToI} and \Eqs{eq:tad1} are equivalent. The main physical outcome is that, as noticed in \cite{LopezNacir:2011kk}, there is no new tadpole for this class of local dissipative models of inflation. The background evolution is fixed by the slicing and probes the global energy density, which does not distinguish the contributions of the inflaton from those of the unknown environment.\footnote{It would be interesting to further investigate if there exists a slicing where system and environment are distinguishable from the background dynamics, for instance through different charges.} It is only at the level of the fluctuations that the distinction between system and environment becomes relevant, as we can disentangle observable degrees of freedom associated to the hydrodynamical direction $\pi$ and unobservable degrees of freedom that have been integrated out. 
\end{tcolorbox}

Now we have fixed the background dynamics, $\mathcal{L}_1$ takes the explicit form
\begin{align}
	\mathcal{L}_1 = - \alpha_0 \partial^\mu \pi_{r} \partial_\mu \pi_{a} &-\sum_{n=1}^\infty \alpha_n \left[-2 \dot{\pi}_{r} + \left( \partial_\mu \pi_{r}\right)^2\right]^{n}\left(-\dot{\pi}_{a} + \partial^\mu \pi_{r} \partial_\mu \pi_{a}\right) \nonumber \\
	+& \sum_{n=1}^\infty \gamma_n \left[-2 \dot{\pi}_{r} + \left( \partial_\mu \pi_{r}\right)^2\right]^n\pi_{a}.
\end{align}
Notice that only $\alpha_0$, $\alpha_1$ and $\gamma_1$ provide quadratic terms in $\pi$ relevant for the dispersion relation of the Goldstone mode \cite{Hongo:2018ant}. The $\alpha_0$ term is the usual kinetic term written in the Keldysh basis. The $\alpha_1$ term generates a non-trivial speed of sound accompanied by higher order operators controlling the appearance of equilateral non-Gaussianities \cite{Cheung:2007st}. 
In contrast to the $\alpha_0$ and $\alpha_1$ terms, the $\gamma_1$ term has no unitary counterpart and leads to a dissipative term in the $\pi_{r}$ equation of motion. Interestingly, the dissipation term $\dot{\pi}_{r}\pi_{a}$ is accompanied by a cubic interaction $\left( \partial_\mu \pi_{r}\right)^2 \pi_{a}$, as first noted in \cite{LopezNacir:2011kk}, such that the combination is invariant under Lorentz boosts. 

Let us explicitly consider the quadratic and cubic contributions. Indeed, in addition to the expansion in powers of $\pi_a$, one can classify
\begin{align}
	\mathcal{L}_{\mathrm{eff}} = \sum_{n,m=1}^\infty  \mathcal{L}^{(m)}_n \quad\mathrm{with}\quad \mathcal{L}^{(m)}_n = \mathcal{O}(\pi^m,~\pi_{a}^n,~ \pi_r^{m-n})
\end{align}
where $m$ labels the number of field operators. The quadratic contributions are
\begin{align}
	\mathcal{L}_1^{(2)} &= \left( \alpha_0 -2 \alpha_1\right) \dot{\pi}_{r} \dot{\pi}_{a} - \alpha_0 \partial_i \pi_{r}  \partial^i \pi_{a} - 2 \gamma_1  \dot{\pi}_{r} \pi_a
\end{align}
where $\alpha_0$ and $\alpha_1$ control the unitary kinetic term with a non-trivial speed of sound and $\gamma_1$ encodes the linear dissipation of the system onto its environment. At cubic order, $\mathcal{L}_1$ reads
\begin{align}\label{eq:cubicinter}
	\mathcal{L}_1^{(3)} =  &\left(4 \alpha_2 -3\alpha_1 \right) \dot{\pi}^2_{r} \dot{\pi}_{a} + \alpha_1 \left(\partial_i \pi_{r} \right)^2 \dot{\pi}_a + 2 \alpha_1  \dot{\pi}_{r}	\partial_i \pi_{r}  \partial^i \pi_{a} \Big.\nonumber \\
	&\qquad + \left(4 \gamma_2 -\gamma_1 \right)  \dot{\pi}^2_{r} \pi_{a}	+ \gamma_1 	\left(\partial_i \pi_{r} \right)^2 \pi_a 
\end{align} 
where the first line corresponds to parts of the unitary operators $\dot{\pi}^3$ and $(\partial_i \pi)^2 \dot{\pi}$ and the second line to the non-linear dissipation induced by the non-linearly realised symmetries, as discussed in \cite{LopezNacir:2011kk}.


\paragraph{\it Higher-order functionals} 
One can carry on the same construction for the higher-order functionals $\mathcal{L}_2$, $\mathcal{L}_3$, \dots as well.
We summarise the results obtained at quadratic and cubic order in the following box.

\begin{tcolorbox}[
	enhanced, 
	colback=gray!10,
	colframe=gray!10,
	breakable,
	skin first=enhanced,
	skin middle=enhanced,
	skin last=enhanced,
	before upper={\parindent15pt},
	]{}
\paragraph{\it Summary}
	
	Under the symmetry breaking pattern specified in \Eq{eq:SSBEFToI}, the most generic local second order open effective functional is
	\begin{align}\label{eq:L2ref}
		&\qquad \mathcal{L}^{(2)} =\left( \alpha_0 -2 \alpha_1\right) \dot{\pi}_{r} \dot{\pi}_{a} - \alpha_0 \partial_i \pi_{r}  \partial^i \pi_{a} \nonumber   \\
		&- 2 \gamma_1  \dot{\pi}_{r} \pi_a+i \left[\beta_1 \pi_a^2 - \left(\beta_2 - \beta_4\right)\dot{\pi}_a^2 + \beta_2 \left(\partial_i \pi_{a} \right)^2\right],
	\end{align}
	where the EFT coefficients are chosen to match the notations of \cite{Hongo:2018ant}. The first line corresponds to the usual unitary dynamics with the kinetic term and an effective speed of sound. The first term of the second line controlled by $\gamma_1$ corresponds to the dissipation due to the surrounding environment. At last, the $\beta_i$ coefficients control the diffusion (noise-induced) process.
	
	\indent At cubic order, the most generic open effective functional (up to total derivatives) reads
	\begin{align}
		\mathcal{L}^{(3)} &= \left(4 \alpha_2 -3\alpha_1 \right) \dot{\pi}^2_{r} \dot{\pi}_{a} + \alpha_1 \left(\partial_i \pi_{r} \right)^2 \dot{\pi}_a + 2 \alpha_1  \dot{\pi}_{r}	\partial_i \pi_{r}  \partial^i \pi_{a} \Big. \label{eq:L13} \\
		&\qquad \quad + \left(4 \gamma_2 -\gamma_1 \right)  \dot{\pi}^2_{r} \pi_{a}	+ \gamma_1 	\left(\partial_i \pi_{r} \right)^2 \pi_a \nonumber \\ 
		+&i \Big[-(\beta_3 -2 \beta_7)\dot{\pi}_{r} \dot{\pi}_{a} \pi_a + \beta_3 \partial_i \pi_{r}  \partial^i \pi_{a} \pi_a +2 (\beta_4+\beta_6 - \beta_8) \dot{\pi}_{r} \dot{\pi}_a^2 \label{eq:L23} \\
		& \qquad \quad - 2 \beta_4   \partial_i \pi_{r}  \partial^i \pi_{a} \dot{\pi}_{a} - 2\beta_5 \dot{\pi}_{r} \pi_a^2  - 2 \beta_6 \dot{\pi}_{r}(\partial_i\pi_a)^2\Big] \nonumber \\
		+&\delta_1 \pi_a^3 + (\delta_5- \delta_2)  \dot{\pi}_{a}^2 \pi_a  + \delta_2  (\partial_i \pi_{a})^2 \pi_a - \delta_4  (\partial_i \pi_{a})^2 \dot{\pi}_a + (\delta_4-\delta_6)\dot{\pi}_{a}^3, \label{eq:L33}
	\end{align}
	where \ref{eq:L13} originates from $\mathcal{L}_1^{(3)}$, \ref{eq:L23} originates from $\mathcal{L}_2^{(3)}$ and \ref{eq:L33} from $\mathcal{L}_3^{(3)}$.
	
\end{tcolorbox}

In de Sitter, in terms of the conformal time and the scale factor $a = -1/(H\eta)$, the open effective functional up to quadratic order reads
\begin{align}
	&\quad S_{\mathrm{eff}}^{(2)} = \int \dd^4 x \Big\{\left( \alpha_0 -2 \alpha_1\right) a^2 \pi'_{r} \pi'_{a} - \alpha_0 a^2 \partial_i \pi_{r}  \partial^i \pi_{a}   \\
	-& 2 a^3 \gamma_1 \pi'_{r} \pi_a + i \left[\beta_1 a^4 \pi_a^2 - \left(\beta_2 - \beta_4\right) a^2\pi_a^{\prime2} + \beta_2 a^2 \left(\partial_i \pi_{a} \right)^2\right] \Big\}, \nonumber
\end{align}
and one can similarly compute the open effective functional at cubic order as well.


\subsubsection{Classification of the EFT operators}\label{subsubsec:unit}

The above construction exhibits a wide zoology of terms compared to its unitary counterpart: $5$ free parameters in $\mathcal{L}^{(2)}$ compared to only $1$ in the standard EFToI \cite{Cheung:2007st}; $13$ free parameters in $\mathcal{L}^{(3)}$ compared to only $1$ in \cite{Cheung:2007st}. While some operators describe faithful non-unitary effects generated by the presence of additional degrees of freedom, others are simply a consequence of writing unitary interactions in the Keldysh basis.


\paragraph{\it Recovering the EFToI} We expect the open effective functional $S_{\mathrm{eff}}$ to be able to reproduce in a certain limit the EFToI \cite{Cheung:2007st}.
This can be achieved by imposing specific equalities between the open EFT coefficients and setting some of them to zero, in order to recover unitary operators only. This limit defines the unitary direction of the parameter space of the theory.


\paragraph{\it Unitary and orthogonal directions}
What about the other operators? The open effective functional has more Wilsonian coefficients than the EFToI. Does it imply that all the remaining operators are intrinsically related to non-unitary effects?  
The symmetry structure of the theory allows one to answer this question in a systematic manner.
This restriction to the unitary direction is obtained by restoring the $\epsilon_a$ symmetry explicitly broken by the non-unitary effects (see \textit{Right} panel of \Fig{fig:sym}) \cite{Hongo:2018ant}. Indeed, in the unitary limit, the two branches of the path integral must transform equally under $\epsilon_\pm$.
By imposing this symmetry one can systematically recover the conditions on coefficients that make the theory unitary.


\subsection{Phenomenology of open inflation}\label{subsec:pheno}

Now that the theory is established, we derive the phenomenology associated to this class of models.
We first make some heuristic estimates and then work out the power spectrum, before characterising the primordial non-Gaussianities expected from dissipation and noise.
At last, we illustrate the scope of this construction by sketching the matching with an explicit model of gauge-inflation \cite{Creminelli:2023aly}.


\subsubsection{Scales and estimates}\label{subsec:energy}

The EFT coefficients are dimensionful quantities, such that $[\alpha_0] =[\alpha_1] = E^{4}$, $[\gamma_1] = E^{5}$, $[\beta_1] = E^{6}$ and $[\beta_2] =[\beta_4] = E^{4}$. One can then ask what are the relevant scales controlling the physics and the regime of validity of our EFT description. 

\paragraph{\it Energy scales and canonical normalisation}
To treat the problems of the scales, we first canonically normalise the fields such that they have dimension of energy $E$. This canonical normalisation relates the original Wilsonian coefficients to quantities of physical interest such as the speed of sound $c_s$, the dissipation scale $\gamma$ or the fluctuations of the environment $\beta_1$. We first define the energy scale
\begin{align}
	f_\pi^4 \equiv \alpha_0 -2 \alpha_1\,.
\end{align}
From this we construct the canonically normalised field
\begin{equation}\label{eq:can}
	\pi \rightarrow  \pi/f_\pi^2.
\end{equation}
It follows that at leading order, the curvature perturbations are given by the relation
\begin{align}\label{eq:zetarel}
	\zeta =  - \frac{H}{f_\pi^2} \pi.
\end{align}

In terms of the canonically normalised variables, the quadratic action takes the form
\begin{align}\label{eq:canonorm} 
	&\quad S_{\mathrm{eff}}^{(2)} = \int \dd^4 x \Big\{ a^2 \pi'_{r} \pi'_{a} - c_{s}^{2} a^2 \partial_i \pir  \partial^i \pia  \\
	-&  a^3 \gamma \pi'_{r} \pia + i \left[\beta_{1} a^4 \pia^2 - \left(\beta_2 - \beta_4\right) a^2\pia^{\prime 2} + \beta_2 a^2 \left(\partial_i \pia \right)^2\right] \Big\}, \nonumber
\end{align}
where we absorbed a factor $f_\pi^{-4}$ into the coefficients.
We can apply a similar redefinition to the cubic action as well.

\paragraph{\it Heuristic estimates}

We are now in the position to carry out a rough estimate of the non-Gaussianities sourced by the cubic operators.
For simplicity, we set $\beta_2 = \beta_4 = 0$ and focus on the leading noise term of \Eq{eq:canonorm} controlled by $\beta_1$.
The estimate relies on the following rules:
\begin{itemize}
	\item We can approximate $\pir$ from the amplitude of the primordial power spectrum $\Delta^2_{\zeta}$, accounting for the canonical normalisation and the leading-order relation with $\zeta$ such that
	\begin{equation}\label{eq:rule1}
		\pir\sim\frac{f_{\pi}^{2}}{H}\Delta_{\zeta}.
	\end{equation}
	\item  We can estimate spatial derivatives by the spatial momenta. The value of spatial derivatives in different directions is, on average, the same by isotropy:
	\begin{equation}\label{eq:rule2}
		\partial_{i}\pi_{r,a}\sim k\pi_{r,a}.
	\end{equation}
	Adiabatic perturbations of momentum $k$ freeze at a scale factor $a_*$ that depends on the dissipation parameter $\gamma$ \cite{LopezNacir:2011kk}. In \cite{Salcedo:2024smn}, we derived this freezing time by comparing the early and the late time limit of the power spectrum. This leads to the relation
	\begin{equation}\label{eq:rule3}
		c_s k\sim a_* H \sqrt{\frac{H+\gamma}{H}},
	\end{equation}
	where we use the expression $(H+\gamma)$ as a shorthand reminder of a quantity that scales to leading order in $\gamma\to\infty$ as $\gamma$ and to leading order in $\gamma \to 0$ as $H$. While freezing still occurs at wavelengths around (sound) horizon crossing at low dissipation, it is displaced to sub (sound) horizon wavelength at large dissipation.
	\item The characteristic frequencies of the retarded $\pir$ and advanced $\pia$ components are estimated to be \cite{Salcedo:2024smn}
	\begin{equation}\label{eq:rule4}
		\pi_{r,a}'\sim a H \pi_{r,a}.
	\end{equation}
	\item The retarded component $\pir$ evolves according to a dynamics sourced by the advanced component $\pia$ and controlled by the environment noise $\beta_1$ \cite{kamenev_2011}. This sourced dynamics implies that $\pir$ and $\pia$ are not of the same amplitude, their ratio being controlled by
	\begin{equation}\label{eq:rule5}
		\frac{\pir}{\pia} \sim \frac{\beta_1}{H(H+\gamma)}.
	\end{equation}
	This relation can be obtained from evaluating the equation of motion for $\pir$ at $a=a_{*}$ \cite{Salcedo:2024smn}.
\end{itemize}

These prescriptions imply some hierarchies among the quadratic operators. Comparing the kinetic terms $a^2 \pir'\pia'$ and $c_s^2 a^2 \partial_i \pir \partial^i \pia$ and the linear dissipation $a^3\gamma\pir' \pia$ with the noise $a^{4} \beta_1 \pia^2$, we observe that
\begin{align}\label{eq:quad1}
	\frac{a^2 \pir'\pia'}{a^{4} \beta_1 \pia^2} \sim \frac{H}{H+\gamma}, \qquad  \qquad  \frac{c_s^2 a^2 \partial_i \pir \partial^i \pia}{a^{4} \beta_1 \pia^2} \sim 1, \qquad \qquad \frac{a^3\gamma\pir' \pia}{a^{4} \beta_1 \pia^2} \sim \frac{\gamma}{H+\gamma}.
\end{align}
This illustrates the dynamical regimes of a driven-dissipative harmonic oscillator. At low dissipation, $a^3\gamma\pir' \pia$ is negligible and the system is mostly controlled by the other three operators. At large dissipation, we enter the overdamped regime in which $a^2 \pir'\pia'$ becomes sub-dominant compared to the other contributions. 

To estimate the size of non-Gaussianities, we can first approximate the ratio between the cubic operators of interest and the dominant quadratic terms in \Eq{eq:canonorm}. Based on the above estimate, the quadratic operator $a^{4} \beta_1 \pia^2$ (or equivalently $c_s^2 a^2 \partial_i \pir \partial^i \pia$) is a valid choice for a `reference quadratic operator' both at large and small dissipation, leading to
\begin{equation}
	f_{\text{NL}}\Delta_{\zeta}\sim\frac{\mathcal{L}_{3}}{a^{4} \beta_1 \pia^2}\,.
\end{equation}
Like in the EFToI, one of the amplitudes of the unitary cubic vertices is controlled by the speed of sound $c_s$. The associated non-Gaussianities can be estimated through the above prescriptions, leading for instance to
\begin{equation}
	\mathcal{L}_{3} \supset \frac{(c^2_s-1)}{2f_{\pi}^{2}}  a \left(\partial_i \pir \right)^2 \pi'_a \quad \rightarrow \quad f_{\text{NL}}\sim \frac{(c^2_s-1)}{c_{s}^{2}} .
\end{equation}
This matches the usual expectation from \cite{Cheung:2007st}. We can then consider dissipative vertices as well, e.g. $(\partial_{i}\pir)^{2}\pia$.
They are controlled by $\gamma$ and related to the quadratic dissipation $\pir^{\prime}\pia$ through non-linearly realised boosts. Using the above prescriptions, we find for the first operator the linear-in-$\gamma$ scaling
\begin{equation}
	\mathcal{L}_{3} \supset \frac{\gamma}{f_{\pi}^{2}}a^2(\partial_{i}\pir)^{2}\pia \quad \rightarrow \quad f_{\text{NL}}\sim \frac{1}{c_s^2} \frac{\gamma}{H} ,
\end{equation}
in agreement with the results of \cite{LopezNacir:2011kk, Creminelli:2023aly}.

Similar estimates can be obtained for all cubic operators. This heuristic derivation correctly accounts for all contributions found in \cite{Salcedo:2024smn}, hence providing a valuable insight to the rich physics of the open EFT of inflation.

\subsubsection{Power spectrum}

To derive the power spectrum, we follow the procedure introduced in \Sec{subsec:freeth}. Embedded in the inflating background, the retarded and Keldysh differential operators become
\begin{align}\label{eq:DRexp}
	\widehat{D}_{R} &= \frac{1}{H^2\eta^2} \left[\partial_\eta^2 - \frac{2 + \frac{\gamma}{H}}{\eta}\partial_\eta - c_s^2 \partial_i^2 \right]
\end{align}
and 
\begin{align}\label{eq:GKexp}
	\widehat{D}_K &=\frac{1}{H^2\eta^2}\left[\frac{\beta_1}{H^2\eta^2} - \left(\beta_2 - \beta_4\right) \left(\partial_\eta^2 -\frac{2}{\eta} \partial_\eta\right) + \beta_2 \partial_i^2  \right].
\end{align}
From now on, we set $c_s = 1$ for simplicity. The inclusion of $c_s$ is discussed in \cite{Salcedo:2024smn}.

\paragraph{\it Retarded Green function} \Eq{eq:DRexp} determines the equations of motion obeyed by the Green function  
\begin{align}\label{eq:Greenfct}
	\left(\partial_{\eta_1}^2 - \frac{2 + \frac{\gamma}{H}}{\eta_1}\partial_{\eta_1} + k^2  \right) G^{R} (k; \eta_1, \eta_2) = H^2 \eta_1^2 \delta(\eta_1-\eta_2).
\end{align}
We start by finding the two homogeneous solutions of $\widehat{D}_{R}$ in Fourier space (mode functions):
\begin{align}
	\pi_k(\eta) \propto \eta^{\nu_\gamma}  H^{(1)}_{\nu_\gamma}(- k \eta) \qquad \mathrm{and} \qquad \propto \eta^{\nu_\gamma} H^{(2)}_{\nu_\gamma}(- k \eta) \,,
\end{align}
where 
\begin{align}
	\nu_\gamma \equiv \frac{3}{2}+ \frac{\gamma}{2H}\,,
\end{align}
and $H^{(1)}$ and $H^{(2)}$ are Hankel functions\footnote{
    Selecting positive frequency mode functions of the form $\ee^{- i k \eta}$ in the asymptotic past one obtains the following scaling for the dissipative mode functions at late times:
\begin{align}
	\pi_k(z = - k \eta) &=  \mathcal{O}_+  z^{\Delta_+}  \left[1 +\mathcal{O}(z^2) \right] + \mathcal{O}_- z^{\Delta_-}  \left[1 +\mathcal{O}(z^2) \right]\,,
\end{align}
where $\mathcal{O}_+$ and $\mathcal{O}_-$ depend on $\gamma$ and $H$, and the scaling dimensions are
\begin{align}
	\Delta_+ = 0 \qquad \mathrm{and} \qquad \Delta_- = 3 + \frac{\gamma}{H}.
\end{align}
Comparing this result to the well-known relation for closed systems $\Delta_+ + \Delta_-= d$, involving representations related by a shadow transform, we note that \emph{dissipation acts in the same way as a continuation to a non-integer number of dimensions}. 
}.

The normalisation of the retarded Green function is fixed by its continuity at coincident times while its first derivative discontinuity is controlled by the time-dependent prefactor of the $\partial_{\eta_1}^2$ term in \Eq{eq:Greenfct}, that is 
\begin{align}
	G^R(k; \eta_2,\eta_2) &= 0, \qquad \mathrm{and} \qquad
	\left. \partial_{\eta_1}G^R(k; \eta_1,\eta_2) \right|_{\eta_2 = \eta_1} = H^2 \eta^2_2.
\end{align}
This imposes
\begin{align}\label{eq:GRfinsimp}
	G^R(k; \eta_1,\eta_2) =  \frac{\pi}{2} H^2 (\eta_1 \eta_2)^{\frac{3}{2}}\left(\frac{\eta_1}{\eta_2} \right)^{\frac{\gamma}{2H}}  \Im\left[H^{(1)}_{\frac{3}{2}+ \frac{\gamma}{2H}}(-k \eta_1) H^{(2)}_{\frac{3}{2}+ \frac{\gamma}{2H}}(-k\eta_2)   \right] \theta(\eta_1-\eta_2). 
\end{align}
which can also be conveniently expressed in terms of Bessel functions:
\begin{align}\label{eq:GRfinsimpv2}
	G^R(k;\eta_1,\eta_2) = \frac{\pi}{2} \frac{H^2}{k^3} \left(\frac{z_1}{z_2}\right)^{\nu_\gamma} z_2^{3}\left[Y_{\nu_\gamma}(z_1)J_{\nu_\gamma}(z_2) -J_{\nu_\gamma}(z_1)Y_{\nu_\gamma}(z_2) \right] \theta(z_2-z_1)
\end{align}
where $z_i \equiv - k \eta_i$. 

\paragraph{\it Keldysh propagator} We now turn our attention to the computation of the Keldysh propagator of the theory from which the observables such as the power spectrum can be derived.
The Keldysh function is obtained from 
\begin{equation}\label{eq:GK1new}
	G^K(k; \eta_1,\eta_2) = i \int \dd \eta' G^R(k; \eta_1,\eta') \widehat{D}_K(\eta' ) G^R(k; \eta_2,\eta') + (\eta_1 \leftrightarrow \eta_2).
\end{equation}
Notice that this equation has precisely the structure one would expect for the power spectrum of a field obeying the Langevin equation with source fluctuations possessing a power spectrum $\widehat{D}_K(\eta' )$.
Injecting \Eq{eq:GRfinsimpv2} into the above expressions, we obtain three contributions which correspond to the three noise terms appearing in \Eq{eq:GKexp}.
We will show the computation of the first contribution, proportional to $\beta_1$ (the other two are analogous):

Injecting \Eq{eq:GRfinsimpv2} in \Eq{eq:GK1new}, we obtain the Keldysh propagator 
\begin{align}\label{eq:GKeq}
	G^K_1(k;\eta_1,\eta_2) = i\frac{\pi^2 \beta_1}{4 k^3}  (z_1 z_2)^{\nu_\gamma} \bigg\{& Y_{\nu_\gamma}(z_1) Y_{\nu_\gamma}(z_2) A^{(1)}_{\nu_\gamma}(z_2) + J_{\nu_\gamma}(z_1) J_{\nu_\gamma}(z_2) C^{(1)}_{\nu_\gamma}(z_2) \\
	-& \left[J_{\nu_\gamma}(z_1)Y_{\nu_\gamma}(z_2) + J_{\nu_\gamma}(z_2)Y_{\nu_\gamma}(z_1)\right]B^{(1)}_{\nu_\gamma}(z_2) \bigg\} + (1 \leftrightarrow 2)\nonumber
\end{align}
where 
\begin{align}
	A^{(1)}_{\nu_\gamma}(z) &\equiv \int_{z}^{\infty} \dd z' z^{\prime2 - 2{\nu_\gamma}} J^2_{\nu_\gamma}(z')  \label{eq:Fbetaref}\\
	B^{(1)}_{\nu_\gamma}(z) &\equiv \int_{z}^{\infty} \dd z' z^{\prime2 - 2{\nu_\gamma}} J_{\nu_\gamma}(z') Y_{\nu_\gamma}(z') \label{eq:Gbetaref} \\
	C^{(1)}_{\nu_\gamma}(z)&\equiv \int_{z}^{\infty} \dd z' z^{\prime2 - 2{\nu_\gamma}} Y^2_{\nu_\gamma}(z') \label{eq:Hbetaref}
\end{align} 
are complicated functions which can be evaluated analytically but result in lengthy expressions given explicitly in \cite{Colas:2025lectures}.

\paragraph{\it Power spectrum} Considering that $\zeta = - H\pi/ f_\pi^2$ where $\pi$ is the canonically normalised field, the reduced power spectrum 
\begin{align}
	\Delta^2_\zeta(k) \equiv \frac{k^3}{2\pi^2}P_\zeta(k) \qquad \mathrm{with} \qquad \langle \zeta_\bmk \zeta_{-\bmk}\rangle = (2\pi)^3 \delta(\bmk + \bmk') P_\zeta(k)
\end{align}
is obtained in the coincident time limit $\eta_1 = \eta_2 = \eta_0$ of the Keldysh propagator given in \Eq{eq:GKeq}, such that
\begin{align}\label{eq:PKref}
	P_\zeta(k) = \frac{\pi^2\beta_1}{8 k^3}  \frac{H^2}{f_\pi^4} z^{2{\nu_\gamma}} \Big[& Y^2_{\nu_\gamma}(z)  A^{(1)}_{\nu_\gamma}(z) + J^2_{\nu_\gamma}(z) C^{(1)}_{\nu_\gamma}(z) -2J_{\nu_\gamma}(z)Y_{\nu_\gamma}(z) B^{(1)}_{\nu_\gamma}(z) \Big]\,.
\end{align}

In the super-Hubble regime $z = -k\eta \ll 1$, the power spectrum freezes.
E.g. the contribution of the $\beta_1$ term is
\begin{align}
	\Delta^2_\zeta(k) &= \frac{1}{4} \frac{\beta_1}{H^2}   \frac{H^4}{f_\pi^4}  2^{2\nu_\gamma} \frac{\Gamma\left(\nu_\gamma-1\right)\Gamma\left(\nu_\gamma\right)^2}{\Gamma\left(\nu_\gamma - \frac{1}{2}\right)\Gamma\left(2\nu_\gamma - \frac{1}{2}\right)}\label{eq:PK1SH}
\end{align}
which is indeed dimensionless. Keeping in mind that $\nu_\gamma \equiv \frac{3}{2}+ \frac{\gamma}{2H}$, one can expand this result in the small and large dissipation regime leading to 
\begin{tcolorbox}[
	enhanced, 
	colback=gray!10,
	colframe=gray!10,
	breakable,
	skin first=enhanced,
	skin middle=enhanced,
	skin last=enhanced,
	before upper={\parindent15pt},
	]{}
	
\paragraph{\it Dissipative power-spectrum}
	\begin{align}\label{eq:dissipPk}
		\Delta^2_\zeta(k)  &\propto  \begin{dcases}
			\frac{\beta_1}{H^2} \frac{H^4}{f_\pi^4}  + \mathcal{O}\left(\frac{\gamma}{H}\right), & \gamma \ll H,\\
			\frac{\beta_1}{H^2} \frac{H^4}{f_\pi^4}   \sqrt{\frac{H}{\gamma}} \left[1 + \mathcal{O}\left(\frac{H}{\gamma}\right) \right], & \gamma \gg H.
		\end{dcases}
	\end{align}
	The observational constraint $\Delta^2_\zeta = 10^{-9}$ is easily obtained by imposing hierarchies between the various scales of the problem. Note that if one further imposes thermal equilibrium of the environment such that the fluctuation-dissipation relation holds, the dynamical KMS symmetry imposes  $\beta_1 =2\pi \gamma T$ where $T$ is the environment temperature \cite{Hongo:2018ant}, such that in the large dissipation regime ($\gamma \gg H$)
	\begin{align}
		\Delta^2_\zeta \propto \frac{T}{H}\frac{H^4}{f_\pi^4} \sqrt{\frac{\gamma}{H}}
	\end{align}
	which reproduces the warm inflation expectation \cite{Berera:1995ie, Berera:1995wh, Berera:1999ws, Berera:2008ar, Ballesteros:2023dno, Montefalcone:2023pvh}.
\end{tcolorbox}

The two other contributions are given by similar expressions -- with different numerical coefficients and different $\nu_\gamma$-dependence -- the details of which can be found in Appendix C of \cite{Salcedo:2024smn}.
The resulting contributions may again satisfy the observational constraint $\Delta^2_\zeta = 10^{-9}$ by imposing some hierarchy between the various scales.

\subsubsection{Primordial non-Gaussianities}

The computation of primordial non-Gaussianities closely follows the flat space example given in \Sec{subsec:flatinter}. Remembering that the curvature perturbations are defined through $\zeta = - H \pi/f_\pi^2$ in the spatially flat gauge, the primordial bispectrum reads 
\begin{align}
	\langle \zeta_{\bmk_1}  \zeta_{\bmk_2}  \zeta_{\bmk_3} \rangle  = - \frac{H^3}{f_\pi^6}	\langle \pi_{\bmk_1}  \pi_{\bmk_2}  \pi_{\bmk_3} \rangle \equiv (2\pi)^3 \delta(\bmk_1 + \bmk_2 + \bmk_3) B(k_1,k_2,k_3).
\end{align} 
Quantities of interest to characterise the non-Gaussian signatures are the amplitude of the signal
\begin{align}
	f_{\mathrm{NL}}(k_1,k_2,k_3) \equiv \frac{5}{6} \frac{B(k_1,k_2,k_3)}{P(k_1) P(k_2) +  P(k_1) P(k_3)  + P(k_2) P(k_3) }\,,
\end{align}
(discussed for specific configurations below) and the shape function defined in \Eq{eq:shaperef}.

One can proceed just as in the flat space case presented in \Sec{subsec:flatB} to evaluate contact bispectra. The generic structure of the integrals is
\begin{align}\label{eq:Bexpref}
	&B(k_1,k_2,k_3) = (-1)^{4-n_d}(-i)^{n_K + n_R + 1}\frac{H^3}{f_\pi^6} \frac{g}{H^{4-n_d}} \int_{-\infty(1 \pm i \epsilon)}^{0^-} \frac{\dd \eta}{\eta^{4-n_d}} \nonumber \\
	&\qquad \widehat{\mathcal{D}}(\{\bmk_i\}, \partial_\eta)\left[ G^{K/R}(k_1, 0, \eta) G^{K/R}(k_2, 0, \eta) G^R(k_3, 0, \eta) + 5~\mathrm{perms}. \right]
\end{align}
where $n_K$ counts the number of Keldysh propagators, $n_R$ the number of retarded ones and $n_d$ the number of (spatial and temporal) derivatives. $\widehat{\mathcal{D}}(\{\bmk_i\}, \partial_\eta)$ is a differential operator schematically representing the $n_d^{\mathrm{th}}$ spatial and temporal derivatives acting on the propagators. Note that there is always at least one $G^R$ due to at least linearity in $\pia$ inherited from \Eq{eq:norm}. The bulk-to-boundary propagators $G^R(k, 0, \eta)$, $G^K(k, 0,\eta)$ and their derivatives can be expressed analytically in terms of Bessel and generalised hypergeometric functions.

It is then a matter of evaluating the time integral of \Eq{eq:Bexpref} injecting the expressions for the propagators.
This task is analytically hard in full generality, therefore below we mostly rely on numerical integration; an analytical treatment is useful only in some specific regimes.

\subsubsection{Numerical results}

\begin{figure}[t]
    \centering
    \begin{subfigure}{0.4\textwidth}
        \includegraphics[width=\textwidth]{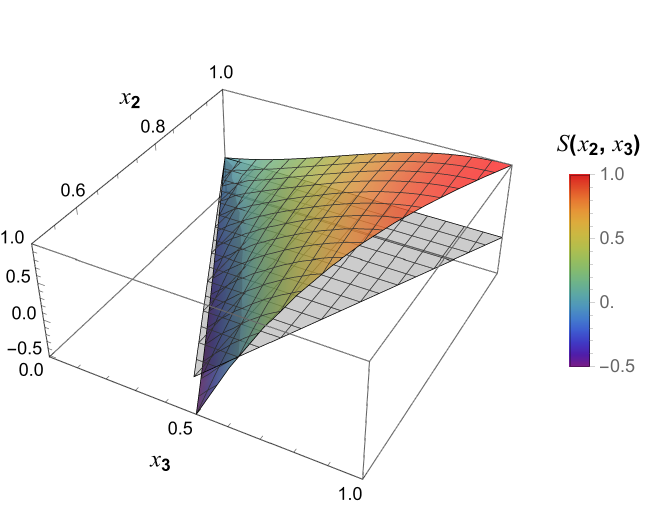}
        \caption{$\left(\partial_i \pir \right)^2 \pia$ operator.}
        \label{fig:c}
    \end{subfigure}
    \begin{subfigure}{0.4\textwidth}
        \includegraphics[width=\textwidth]{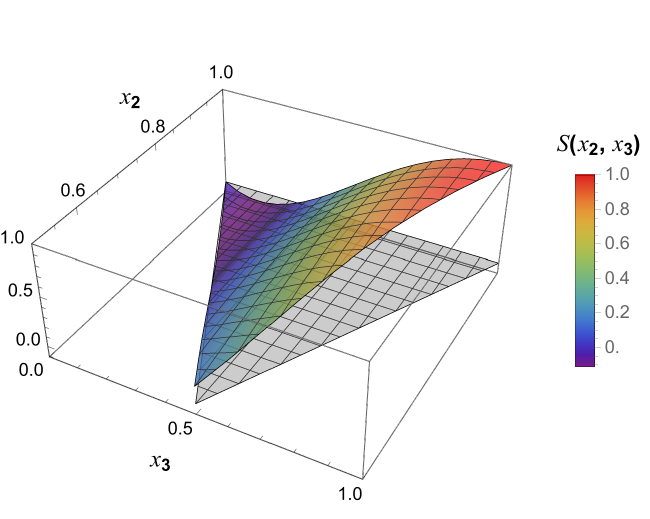}
        \caption{$\dot{\pi}^2_{r} \pia$ operator.}
        \label{fig:d}
    \end{subfigure}
    \caption{Shapes of the bispectrum at large dissipation $\gamma = 5H$. In this regime, the signal peaks in the equilateral configuration $x_2 = x_3 = 1$. Consistency relations hold in the squeezed limit $x_3 \ll x_2 = 1$.}
	\label{fig:shapes_high}
\end{figure}

\begin{figure}[t]
    \centering
    \begin{subfigure}{0.4\textwidth}
        \includegraphics[width=\textwidth]{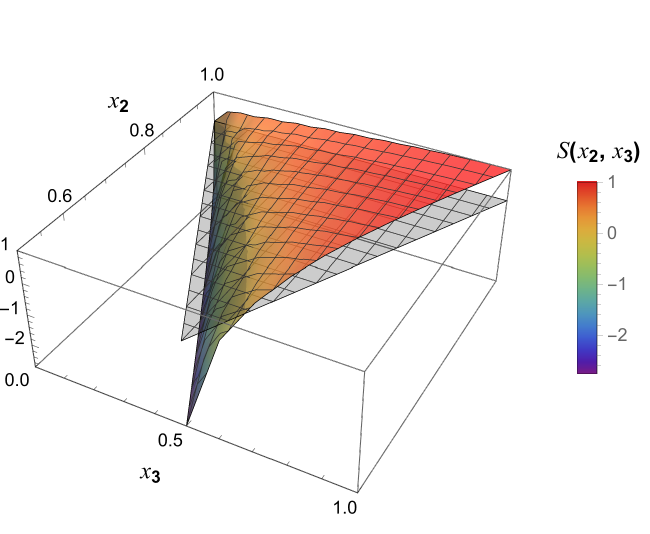}
        \caption{$\left(\partial_i \pir \right)^2 \pia$ operator.}
        \label{fig:cbis}
    \end{subfigure}
    \begin{subfigure}{0.4\textwidth}
        \includegraphics[width=\textwidth]{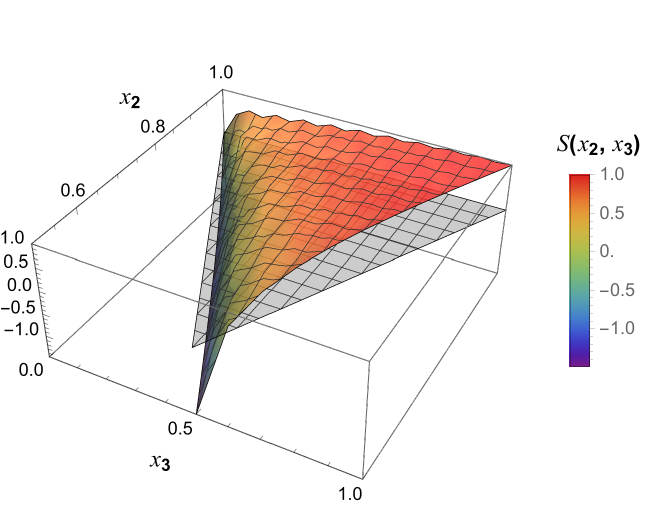}
        \caption{$\dot{\pi}^2_{r} \pia$ operator.}
        \label{fig:dbis}
    \end{subfigure}
    \caption{Shapes of the bispectrum at low dissipation $\gamma= 0.001H$. In this regime, the signal peaks near folded configurations $x_2 + x_3 = 1$. Consistency relations still hold in the squeezed limit $x_3 \ll x_2 = 1$. The tiny oscillations are artefacts of the numerical integration over a finite range.}
	\label{fig:shapes_low}
\end{figure}

In this section, we summarise the main phenomenological implications of the contact bispectra generated by the cubic operators of \Eqs{eq:L13}, \eqref{eq:L23} and \eqref{eq:L33}.
Many of the sixteen cubic operators appearing in \Eqs{eq:L13}, \eqref{eq:L23} and \eqref{eq:L33} lead to similar signatures for the contact bispectrum. Hence, we only display the results for two of these operators.
The shapes of the contact bispectrum generated by them are displayed in \Figs{fig:shapes_high} and \ref{fig:shapes_low}. Just as for the flat space case, different behaviours emerge in the large ($\gamma \gg H$) and small ($\gamma \ll H$) dissipation regime. While the former peaks in the equilateral configuration as already noted in \cite{LopezNacir:2011kk}, the latter reaches an extremum near the folded region. 

This smoking gun of open dynamics might seem degenerate with other classes of models that also lead to a signal in the folded triangles such as non-Bunch Davies initial states \cite{Holman:2007na, Chen:2006nt, Meerburg:2009ys, Agullo:2010ws, Ashoorioon:2010xg, Agarwal:2012mq, Ashoorioon:2013eia,  Albrecht:2014aga, Green:2020whw}. A crucial difference, which appears in our numerical treatment and is briefly discussed below, is that dissipation regularises the divergence by smoothing the peak and displacing it from the edge of the triangular configurations, leading to finite values of the bispectrum for any physical configuration.   
In particular, it implies no divergence in the squeezed limit of the bispectrum $k_1 \simeq k_2 \gg k_3$. Small values of $\gamma/H$ may eventually lead to an intermediate peak due to the regularised folded singularity, yet consistency relations hold \cite{Maldacena:2002vr, Creminelli:2012ed} and the squeezed limit goes to zero because of the symmetries of the theory. Notice that operators such as $\pia^3$ follow this trend despite what one may have naively thought in the absence of derivatives. This is because of the modified propagators compared to the free theory, which for instance ensure IR convergence.	

\begin{figure}[t]
	\begin{minipage}{6in}
		\centering
		\raisebox{-0.5\height}{\includegraphics[width=.48\textwidth]{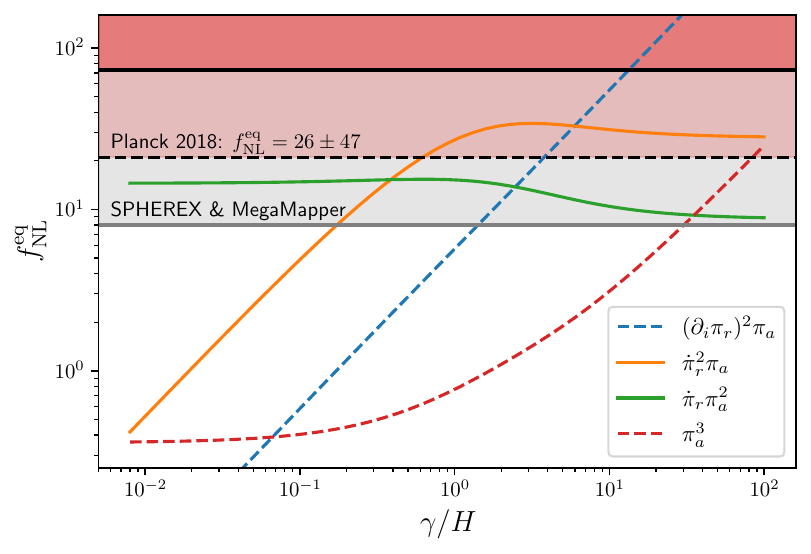}}
		\raisebox{-0.5\height}{\includegraphics[width=.48\textwidth]{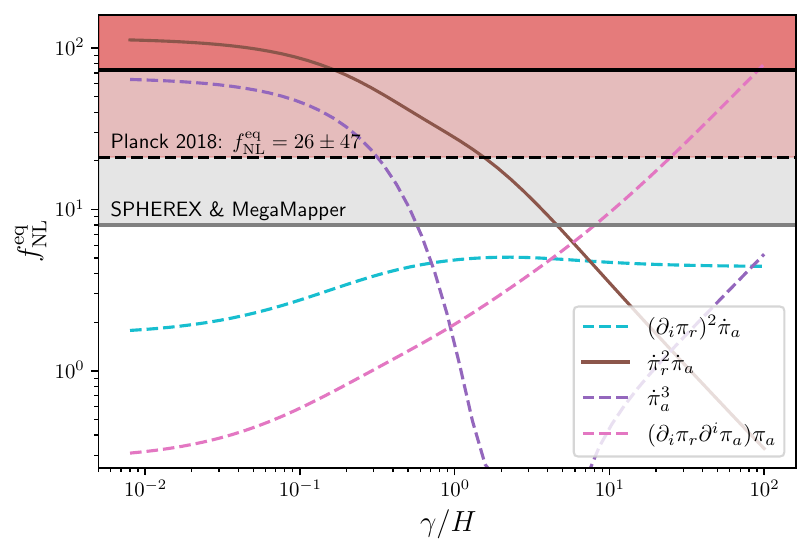}}
	\end{minipage}
	\caption{Amplitude in the equilateral configuration as a function of the dissipation parameter $\gamma/H$. Continuous (dashed) lines are positive (negative) values. \textit{Left:} The four cubic operators appearing in the matching with \cite{Creminelli:2023aly} performed in \Sec{sec:matching}. All the curves well reproduce the results from \cite{Creminelli:2023aly}. The otherwise arbitrary values of the EFT coefficients were chosen to qualitatively reproduce the results from \cite{Creminelli:2023aly}. \textit{Right:} New operators of the open EFToI that may lead to specific signatures on the non-Gaussian signal. Again, numerical values of the EFT coefficients are arbitrary, chosen to be comparable to the \textit{Left} panel. All scalings with $\gamma$ are consistent with the heuristic estimates of \Sec{subsec:energy}. }
	\label{fig:fnleq}
\end{figure}

The amplitude in the equilateral configuration is controlled by 
\begin{align}
	f_{\mathrm{NL}}^{\mathrm{eq}} \equiv \frac{10}{9} \frac{k^6}{(2\pi)^4}\frac{B(k,k,k)}{\Delta_\zeta^4}.
\end{align} 
In \Fig{fig:fnleq}, we display the dependence of $f_{\mathrm{NL}}^{\mathrm{eq}}$ on the dissipation parameter $\gamma/H$ from numerical integration of the bispectrum $B(k,k,k)$ for the operators considered above. One can use observational constraints $f_{\mathrm{NL}}^{\mathrm{eq}} = 26 \pm 47$ from \cite{Planck:2019kim} to place bounds on the EFT parameters in the large dissipation regime. For instance, a numerical fit of the $(\partial_i \pir)^2\pia$ contributions leads to $f_{\mathrm{NL}}^{\mathrm{eq}} \simeq - \gamma/(4H)$. It naively implies that $\gamma / H < 80$ at $68\%$ confidence. Of course, this is more of a proof of principle than a realistic estimate due to the cumulative effects of different cubic operators that cannot be disentangled one from another. Yet, it demonstrates how this class of model can be confronted to data, as done in \cite{Salcedo:2026sdn}. Such observational bounds could be tightened thanks to future LSS experiments such as SPHEREX \cite{SPHEREx:2014bgr} or MegaMapper \cite{Cabass:2022epm, Braganca:2023pcp}, further constraining this class of models.   

\paragraph{\it Folded singularities}
An interesting question is the fate of folded singularities in an expanding background. In general, a folded singularity appears when two modes resonate with a third for an infinite amount of time. For this reason, the origin of the divergence can be derived from the early time oscillating behaviour of the propagators.

What happens here is the analogue of the flat space case discussed in \Eq{eq:M4int}: the singularity is regulated by the presence of dissipation.
In the limit of vanishing dissipation, we recover the log-folded singularity familiar to non-Bunch Davies initial states \cite{Holman:2007na, Chen:2006nt, Meerburg:2009ys, Agullo:2010ws, Ashoorioon:2010xg, Agarwal:2012mq, Ashoorioon:2013eia,  Albrecht:2014aga, Green:2020whw}.
\subsubsection{Matching with a gauge-inflation model}
\label{sec:matching}

For an EFT construction to be of interest, it has to encompass a class of physically motivated models. For this reason, we consider in this section the matching of the open EFT of inflation to a concrete ``ultra-violet'' model constructed and studied in \cite{Creminelli:2023aly}. This is not the first model of dissipative dynamics, but it has the distinguishing feature of leading to a \textit{local} low-energy dynamics around and below the Hubble scale. This is related to the fact that the window of instability for particle production involves wavelengths that are parametrically sub-Hubble. 
In addition to the inflaton field $\phi$, the model features a massive scalar field $\chi$ with a softly-broken $U(1)$ symmetry. The action of the model is given by
\begin{align}\label{eq:actionmatch}
	S= \int \dd^4 x &\sqrt{- g} \bigg[ \frac{1}{2} M_{\mathrm{Pl}}^2  R - \frac{1}{2} \left(\partial \phi \right)^2 - V(\phi) - \left|\partial \chi \right|^2 + M^2 \left| \chi \right|^2 \nonumber \\
	- &\frac{\partial_\mu \phi}{f}\left( \chi \partial^\mu \chi^* - \chi^* \partial^\mu \chi\right) - \frac{1}{2}m^2\left(\chi^2 + \chi^{*2}\right)\bigg].
\end{align}
The last term in \Eq{eq:actionmatch} breaks the $U(1)$ symmetry $\chi \rightarrow \ee^{i \alpha} \chi$ for $\alpha \in \mathbb{R}$. An important parameter for the dynamics of the system is $\rho \equiv \dot{\phi}_0(t)/f$, which controls the mixing of $\phi$ with $\chi$. Here $\dot{\phi}_0(t)$ is the time derivative of the inflaton background $\phi_0(t)$. The hierarchy of scales for which the treatment of \cite{Creminelli:2023aly} is valid is
\begin{align}
	f \gg \rho \gtrsim M \gg m \gg H.
\end{align}
The model exhibits a narrow instability band in the sub-Hubble regime, during which particle production occurs. The amount of particle production is controlled by the dimensionless parameter
\begin{align}
	\xi \simeq \frac{m^4}{8 H \rho M^2} \gtrsim \mathcal{O}(1).
\end{align}
The inflaton fluctuations experience a dissipative dynamics due to the presence of an environment of $\chi$ particles generated by the instability. The inflaton dynamics is effectively described in terms of a non-linear Langevin equation \cite{Creminelli:2023aly}
\begin{align}\label{eq:Paolo}
	\pi'' + \left( 2H + \gamma\right)a\pi' - \partial_i^2 \pi &\simeq \frac{\gamma}{2\rho f} \left[ \left( \partial_i \pi\right)^2 - 2 \pi \xi \pi^{\prime2}\right] - \frac{a^2m^2}{f} \left(1+2\pi\xi \frac{\pi'}{a \rho f} \right) \delta \mathcal{O}_S\,,
\end{align}
where we neglected the inflaton potential contribution $a^2 V'' \pi$ which is slow-roll suppressed. The dissipation parameter $\gamma$ (with units of mass) is determined by the following combination of microphysical parameters
\begin{align}\label{eq:gammaCrem}
	\gamma \simeq \frac{\xi m^4}{\pi M f^2} \ee^{2 \pi \xi}.
\end{align}
The effect of noise is captured in terms of its two- and three-point statistics
\begin{align}
	\langle  \delta \mathcal{O}_S(\bmk, \eta) \delta \mathcal{O}_S(\bmk',\eta')\rangle &\simeq (2\pi)^3 \delta(\bmk + \bmk') \delta(\eta - \eta')H^4 \eta^4 \nu_{\mathcal{O}}\,, \\
	\langle  \delta \mathcal{O}_S(\bmk, \eta) \delta \mathcal{O}_S(\bmk',\eta')  \delta \mathcal{O}_S(\bmk'',\eta'')\rangle &\simeq (2\pi)^3 \delta(\bmk + \bmk' +\bmk'') \delta(\eta - \eta') \delta(\eta - \eta'')H^8 \eta^8 \nu_{\mathcal{O}^3} \,,\label{eq:NGnoiseCrem}
\end{align}
with the amplitudes of the noises being controlled by 
\begin{align}
	\nu_{\mathcal{O}} \simeq \frac{M}{m} \frac{\ee^{4\pi \xi}}{4\pi^2} \qquad \mathrm{and} \qquad \nu_{\mathcal{O}^3} \simeq \frac{\ee^{6\pi \xi}}{\pi^2 m^2} .
\end{align}
The parameters at play are summarised in the first line of Table \ref{tab:matching}.

As demonstrated in \cite{Colas:2025lectures}, the low-energy dynamics of this model is equivalently described in terms of
\begin{align}\label{eq:Seffmatch}
	\quad S_{\mathrm{eff}} &= \int \dd^4 x \Big[ a^2 \pi'_{r} \pi'_{a} - c_{s}^{2} a^2 \partial_i \pir  \partial^i \pia   -  a^3 \gamma \pi'_{r} \pia + i \beta_1 a^4 \pia^2 \nonumber \\
	+ &\frac{ \left(8\gamma_2 - \gamma\right)}{2f^{2}_{\pi}}  a^2\pir^{\prime2} \pia	+ \frac{\gamma}{2f_{\pi}^{2}} a^2	\left(\partial_i \pir \right)^2 \pia - 2i\frac{\beta_{5}}{f_\pi^2} a^3 \pir^{\prime} \pia^2 +\frac{\delta_1}{f_\pi^2} a^4 \pia^3  \Big],
\end{align}
upon matching the EFT coefficients to the microphysical parameters of the model through the relations of Table \ref{tab:matching}. 

\begin{table}
	\begin{center}
		\begin{tabular}{ |P{4cm} P{4cm} |  }
			\hline 
			$\Bigg.$ \textit{Matching:} & $f^2_\pi = \rho f $ \\
			$\Bigg.$ $c_s = 1$ 	&	$\gamma = \frac{\xi m^4}{\pi M f^2} \ee^{2 \pi \xi}$ \\
			$\Bigg.$ $8\gamma_2 =\left(1-2 \pi \xi\right) \gamma$ & $\beta_1 = \frac{\nu_{\mathcal{O}}}{2 \rho f} \frac{m^4}{f^2} $ \\
			$\Bigg.$ $\beta_{5}=  2\pi\rho f\xi \beta_1$ & 		 
			$6\delta_1 = \rho f \frac{m^6}{f^3} \nu_{\mathcal{O}^3} $  \\
			\hline
		\end{tabular}
	\end{center}
	\caption{Matching of the EFT coefficients appearing in \Eq{eq:Seffmatch} to the microphysical parameters of \cite{Creminelli:2023aly}. The matching is obtained either at the level of the power spectrum and contact bispectra or by deriving the Langevin equation for the fluctuations.}
	\label{tab:matching}
\end{table}


\subsection{\textit{Problem set}}\label{subsec:prob3}

\begin{exbox}
\paragraph{\it The decoupling limit of the EFT of inflation}

We want to understand under which conditions 
\begin{align}\label{eq:Starget}
	S = \int \dd^4 x \sqrt{-g} \left[ \frac{M^2_{\mathrm{Pl}}}{2} R - \Lambda(t) - c(t) g^{00} + \sum_{n=2}^\infty M^4_n(t) (1 + g^{00})^n\right],
\end{align}
simplifies to 
\begin{align}\label{eq:Stargetpi}
	S_\pi = \int \dd^4 x \sqrt{-g} \left\{ \epsilon \Mpl^2 H^2 \left[ \dot{\pi}^2 - \frac{(\partial_i \pi)^2}{a^2} \right] + 2 M_2^4 \left[ \dot{\pi}^2 + \dot{\pi}^3 - \dot{\pi}\frac{(\partial_i \pi)^2}{a^2} \right] - \frac{4}{3} M_3^4 \dot{\pi}^3 + \cdots \right\}, 
\end{align}
It will be useful to consider the Friedmann equations
\begin{align}
	3\Mpl^2 H^2 = \Lambda(t) + c(t), \qquad 2\Mpl^2 \dot{H} = -2c(t),
\end{align}
and to define the first and second slow-roll parameters $\epsilon \equiv - \dot{H}/H^2 $ and $\eta \equiv \dot{\epsilon}/(\epsilon H)$, together with the speed of sound $c_s^{-2} \equiv [c(t) + 2 M_2^4(t)]/c(t)$.
\begin{enumerate}
	\item Perform a \stuck trick to \Eq{eq:Starget} to reintroduce $\pi$. 
	\item We expand the metric $g_{\mu\nu} = \bar{g}_{\mu\nu} + \delta g_{\mu\nu}$, where $\bar{g}^{\mu\nu} = \mathrm{diag}(-1, a^{2}, a^{2}, a^{2})$. Identify the four operators that generate mixing between $\pi$ and $\delta g_{\mu\nu}$ at quadratic order.
\end{enumerate}   
From now on, we work in the flat gauge where
	\begin{equation}
		\delta g_{ij}^{\rm scalar}=0\quad,\quad\delta g_{00}=-2\phi=-\delta g^{00}\quad,\quad \delta g_{0i}=a(t)\partial_{i}F= a^2(t)\delta g^{0i}.
	\end{equation}
	The $00$ and $0i$ linear Einstein equations are given by
	\begin{align}
		\frac{\Mpl^{2}}{2}\delta G_{00}&+\left[\frac{\Lambda(t)}{2} - M_2^4(t)\right]  \delta g_{00} - \left[c(t) + 2M_2^4(t) \right]\dpir+3Hc(t)\pi=0, \label{eq:E1}\\
		\frac{\Mpl^{2}}{2}\delta G_{0i}&+\left[\frac{\Lambda(t)}{2}-\frac{c(t)}{2}\right]\delta g_{0i}-2c(t)\partial_{i}\pi=0, \label{eq:E2}
	\end{align}
	where $\delta G_{\mu\nu}$ is the perturbed Einstein tensor, reading
	 \begin{align}
		\delta G_{00} &= - \frac{2}{a(t)} H \nabla^2 F, \Bigg. \label{eq:G00}\\
		\delta G_{0i} &= \partial_i \left[ 2H\phi  - \left(2 \dot{H} + 3H^2 \right) a(t) F\right], \Bigg. \label{eq:G0i} 
	\end{align}
\begin{enumerate}\setcounter{enumi}{2}
	\item Solve for the constraints $\phi$ and $\nabla^2 F$ as a function of $\pi$ and $\dot{\pi}$. 
	\item Injecting the constraints in the linear mixing operators identified above, under which conditions are these terms sub-dominant compared to the kinetic term of $\pi$? You may need the heuristic estimate $\dot{\pi} \sim H \pi$ during inflation.  
	\item What about the quadratic operators generated by the time derivatives of $\Lambda(t + \pi)$ and $c(t + \pi)$? Under which conditions are they sub-dominant compared to the kinetic term of $\pi$? 
	\item What are the operators that generate mixing between $\pi$ and $\delta g_{\mu\nu}$ at cubic order? Compare them with the cubic operators appearing in $\pi$. You may need the heuristic estimate $\cs k = a H$ during inflation.
	\item Under which conditions does the decoupling limit hold during inflation? Give the expression of $\delta g^{00} = 1 + g^{00}$ in the decoupling limit. This constitutes the building block of the theory of a shift symmetric scalar in rigid de Sitter.
\end{enumerate}
\end{exbox}

\paragraph{\textit{Solution}}

\begin{enumerate}
    \item Under a time diffeomorphism $t \mapsto \tilde{t} = t + \xi^0(x)$ the $00$ component of the inverse metric transforms as
        \begin{equation}
            g^{00} \mapsto \tilde{g}^{00}(\tilde{x}) = \frac{\partial \tilde{x}^0}{\partial x^\mu}(x) \frac{\partial \tilde{x}^0}{\partial x^\nu}(x) g^{\mu\nu}(x)
        \end{equation}
    We simply have to plug these expressions for $\tilde{t}$ and $\tilde{g}^{00}(\tilde{x})$ into the action \Eq{eq:Starget}:
        \begin{align}
            \begin{split}
                S = \int &\dd^4 \tilde{x} \sqrt{-\tilde{g}(\tilde{x})} \left[ \frac{M^2_{\mathrm{Pl}}}{2} R - \Lambda\left(\tilde{t}-\xi^0(x)\right) - c\left(\tilde{t}-\xi^0(x)\right) \frac{\partial \left(\tilde{t}-\xi^0(x)\right)}{\partial x^\mu} \frac{\partial \left(\tilde{t}-\xi^0(x)\right)}{\partial x^\nu} g^{\mu\nu}(\tilde{x}) +\right.\\
                &\left.+\sum_{n=2}^\infty M^4_n\left(\tilde{t}-\xi^0(x)\right) \left(1 + \frac{\partial \left(\tilde{t}-\xi^0(x)\right)}{\partial x^\mu} \frac{\partial \left(\tilde{t}-\xi^0(x)\right)}{\partial x^\nu} g^{\mu\nu}(\tilde{x})\right)^n \right] \,.
            \end{split}
        \end{align}
    We then reintroduce $\pi$ via the substitution $\xi^0(x) \mapsto \tilde{\pi} (\tilde{x})$, obtaining the action (we drop all tildes for simplicity)
        \begin{align}
            \begin{split}
                S = \int &\dd^4 \tilde{x} \sqrt{-g} \left[ \frac{M^2_{\mathrm{Pl}}}{2} R - \Lambda\left(t+\pi(x)\right) - c\left(t+\pi(x)\right) \frac{\partial \left(t+\pi(x)\right)}{\partial x^\mu} \frac{\partial \left(t+\pi(x)\right)}{\partial x^\nu} g^{\mu\nu}(x) +\right.\\
                &\left.+\sum_{n=2}^\infty M^4_n\left(t+\pi(x)\right) \left(1 + \frac{\partial \left(t+\pi(x)\right)}{\partial x^\mu} \frac{\partial \left(t+\pi(x)\right)}{\partial x^\nu} g^{\mu\nu}(x)\right)^n \right]
                \end{split}
        \end{align}
    
    thus restoring the full diffeomorphism invariance of the action provided that the Goldstone transforms as $\pi(x) \mapsto \tilde{\pi}(\tilde{x}) = \pi(x) - \xi^0(x)$.
    \item At quadratic order in $\pi$ and $\delta g_{\mu\nu}$ the mixing terms are:
    \begin{align}
        -2c(t)\dot{\pi}\delta g^{00}, && -2c(t)\partial_i\pi \delta g^{0i}, && -\dot{c}(t)\pi\delta g^{00} &&-c(t)\dot{\pi}\bar{g}^{00}\bar{g}^{ii}\delta g_{ii}
    \end{align}
    \item The $00$ Einstein equation and the spatial derivative of the $0i$ Einstein equation form a linear system with unknowns $\nabla^2 F$ and $\phi$, whose solution is:
    \begin{align}
        \phi = 2\epsilon H\pi, && \nabla^2F = -\frac{a\epsilon H}{c_s^2} \left[\dot{\pi} + \left(3c_s^2 - 2\epsilon\right)H\pi\right]
    \end{align}
    where $c_s^{-2} = 1 + 2M_2^4(t)/c(t)$.
    \item We can estimate the mixing terms as
        \begin{equation}
            c(t)\dot{\pi}\delta g^{00} \sim \epsilon M^2_{\mathrm{Pl}} H^2 H\pi \phi \sim \epsilon^2 M^2_{\mathrm{Pl}} H^4 \pi^2
        \end{equation}
        \begin{equation}
            c(t)\partial_i\pi \delta g^{0i} \sim \epsilon M^2_{\mathrm{Pl}} H^2 \pi a^{-1}\nabla^2 F \sim \epsilon^2 M^2_{\mathrm{Pl}} H^4 \pi^2 \frac{1 + 3c_s^2 -2\epsilon}{c_s^2}
        \end{equation}
        \begin{equation}
            \dot{c}(t)\pi\delta g^{00} \sim (\#\epsilon+\#\eta) \times \epsilon M^2_{\mathrm{Pl}} H^2 \pi \phi \sim (\#\epsilon + \#\eta) \times \epsilon^2 M^2_{\mathrm{Pl}} H^4 \pi^2
        \end{equation}
        (where $(\#\epsilon+\#\eta)$ stands for some linear combination of $\epsilon$ and $\eta$), while the kinetic term is
        \begin{equation}
            c(t)\dot{\pi}^2 \sim \epsilon M^2_{\mathrm{Pl}} H^4 \pi^2
        \end{equation}
        We see that mixing is suppressed when the first slow-roll parameter is small $\epsilon \ll 1$.
    \item We can estimate
        \begin{equation}
            \ddot{\Lambda}(t)\pi^2 \sim (\#\epsilon + \#\eta) \times \epsilon M^2_{\mathrm{Pl}} H^4 \pi^2
        \end{equation}
        therefore this is suppressed when $\epsilon, \eta \ll 1$. Since $c(t) \sim \epsilon \Lambda(t)$, the operator generated by $c(t)$ is even more suppressed.
    \item At cubic order we have mixing operators such as
        \begin{align}
            c(t)\dot{\pi}^2 \delta g^{00}, && 
            \ddot{\Lambda}(t)\pi^2 \delta g^{00}, && \ddot{c}(t)\pi^2 \delta g^{00}, && c(t)\dot{\pi}\partial_i\pi \delta g^{0i}
        \end{align}
        and we should compare them to the cubic operators for the Goldstone
        \begin{align}
            \dot{c}(t) \dot{\pi}^2\pi, && M_2^4 \dot{\pi}^3, && M_3^4 \dot{\pi}^3, && M_2^4 a^{-2}\dot{\pi} (\partial_i\pi)^2
        \end{align}
        Using the previously found results, we can easily estimate\footnote{For the last operator, we used
        \begin{equation}
            \partial_i \delta g^{0i} \sim a^{-1} \nabla^2F \sim \frac{\epsilon H}{c_s^2} \left[\dot{\pi} + (3c_s^2 - 2\epsilon) H\pi \right] \sim \epsilon H^2\pi \left(3 +\frac{1-2\epsilon}{c_s^2} \right)\,.
        \end{equation}}
        \begin{align}
           c(t)\dot{\pi}^2 \delta g^{00} &\sim \epsilon M^2_{\mathrm{Pl}} H^4 \pi^2 \times \epsilon H\pi \sim \epsilon^2 M^2_{\mathrm{Pl}} H^5 \pi^3 \\
           \ddot{\Lambda}(t)\pi^2 \delta g^{00} &\sim (\#\epsilon + \#\eta) \times \epsilon M^2_{\mathrm{Pl}} H^4 \pi^2 \times \epsilon H\pi \sim (\text{slow roll})^3 M^2_{\mathrm{Pl}} H^5\pi^3 \\
            \ddot{c}(t)\pi^2 \delta g^{00} &\sim (\text{slow roll})^3 M^2_{\mathrm{Pl}} H^2 \times H^2\pi^2 \times \epsilon H\pi \sim (\text{slow roll})^4 M^2_{\mathrm{Pl}} H^5\pi^3 \\
            c(t)\dot{\pi}\partial_i\pi \delta g^{0i} &\sim \epsilon^2 M^2_{\mathrm{Pl}} H^5\pi^3 \left( 3-c_s^{-2}\right)
        \end{align}

        Let us turn to the cubic operators for the $\pi$. The first one, 
        \begin{equation}
            \dot{c}(t) \dot{\pi}^2\pi \sim (\text{slow roll})^2 M^2_{\mathrm{Pl}}H^2 \times H^2\pi^3 \sim (\text{slow roll})^2 M^2_{\mathrm{Pl}}H^5\pi^3 
        \end{equation}
        is indeed slow-roll suppressed. It illustrates how EFT coefficients can be considered as constant in the decoupling limit of the EFT of inflation, up to slow-roll suppressed corrections. On the contrary, $M_2^4$ and $M_3^4$ have no reason to be slow-roll suppressed. Hence, in the limit where $\epsilon \ll 1$, they largely dominate the non-linear dynamics of $\pi$.  

    \item The decoupling holds as long as the slow roll parameters are small, that is when the slow-roll hierarchy applies. In the decoupling limit,
        \begin{align}
            \delta g^{00} &= 1 + g^{00} = 1 + (1+\dot{\pi})^2 \bar{g}^{00} + \partial_i\pi \partial_j\pi \bar{g}^{ij}=-2\dot{\pi} - \dot{\pi}^2 + a^{-2} (\partial_i\pi)^2\,.
        \end{align}
\end{enumerate}



\section{Outlook: Open gauge theories and gravity}\label{sec:lec4}

Having constructed the open effective field theory for scalar perturbations during inflation, we now turn our attention toward the current frontier of open quantum systems: gauge theories and gravity. Extending the Schwinger-Keldysh formalism to describe open photons and stochastic gravitons introduces profound conceptual challenges. In this concluding section, we discuss the relation between gauge invariance and dissipation and noise, and summarise the current theoretical developments in the field. For a more detailed exploration of these topics, we refer the reader to \cite{Liu:2018kfw, Landry:2022nog, Akyuz:2023lsm, Akyuz:2025bco, Salcedo:2024nex, Salcedo:2025ezu, Firat:2025upx, Kaplanek:2025moq, Kaplanek:2026kpp}.

\subsection{The subtlety of gauge symmetries in open systems}

In the study of open scalar theories, we observed a clear symmetry-breaking pattern. The environment explicitly breaks the independent forward and backward global symmetries of the closed time path, reducing them to the diagonal (retarded) subgroup: $G_+ \times G_- \xrightarrow{\text{open}} G_r$. This is known as a \textit{strong-to-weak symmetry breaking} \cite{Hauser:2026sgr}.

However, gauge redundancies and diffeomorphism invariance behave differently from global symmetries. They are redundancies of our description, that cannot be broken \cite{Firat:2025upx, Kaplanek:2025moq}. Hence, the two copies of the gauge symmetry in the Schwinger-Keldysh contour must always remain intact. 

The interplay between the breaking of the global symmetries and the conservation of the local gauge symmetries is subtle, especially in situations where fields non-linearly realise a symmetry. Depending on the phase of matter described by the EFT, for instance insulator \cite{Salcedo:2024nex, Kaplanek:2025moq} or conductor \cite{Landry:2022nog}, different descriptions can emerge, which has led to a rich development of the literature on the subject in recent years.

\subsection{Open electromagnetism}

It is instructive to explicitly construct the influence functional for an open electromagnetic field in a simple medium, such as an insulator. Following the top-down framework of \cite{Kaplanek:2025moq}, the quadratic influence functional takes the general non-local form
\begin{equation}
\begin{aligned}
S_{\text{IF}} &= \int d^4x \left[ -\frac{1}{2} F_{r\mu\nu}(x) F_a^{\mu\nu}(x) \right] \\
&\quad + \int d^4x \int d^4y \left[ F_a^{\mu\nu}(x) P_{\mu\nu\alpha\beta}(x,y) F_r^{\alpha\beta}(y) + \frac{i}{2} F_a^{\mu\nu}(x) K_{\mu\nu\alpha\beta}(x,y) F_a^{\alpha\beta}(y) \right]\,,
\end{aligned}
\label{eq:nonlocal_influence_functional}
\end{equation}
where the retarded and advanced field strength $F_{r,a}^{\mu\nu}$ are constructed from the sum and the difference of $F_{\pm}^{\mu\nu} = \partial^\mu A^\nu_\pm - \partial^\nu A^\mu_\pm$. The first line corresponds to the standard Maxwell term $F^{\mu\nu}F_{\mu\nu}$ written in the Keldysh basis, while the second line captures dissipation through $P_{\mu\nu\alpha\beta}(x,y)$ and noise through $K_{\mu\nu\alpha\beta}(x,y)$ respectively. 
Because this functional is written entirely using fully gauge-invariant field strengths ($F_a^{\mu\nu};\, F^{\rho\sigma}_r$), the exact advanced and retarded gauge symmetries remain manifest. In some regime, when a suitable scale hierarchy exists, a local limit of \Eq{eq:nonlocal_influence_functional} can emerge. 

However, the situation drastically changes for a conducting plasma in which charges move freely, and become an integral part of the EFT. A theory of charge diffusion was proposed in \cite{Crossley:2015evo} (see \cite{Liu:2018kfw} for a pedagogical introduction). The spontaneous breaking of the global $U(1)_+ \times U(1)_-$ to either $U(1)_r$ or no symmetry at all leads to the emergence of St\"uckelberg fields $\varphi_{r,a}$ that non-linearly realise the gauge symmetries. The field $\rho_r \equiv \partial_t \varphi_r$ physically represents the charge density \cite{ Akyuz:2025bco, Firat:2025upx}. The building blocks of this construction now become $B_{r,a}^\mu = A^\mu_{r,a} + \partial^\mu \varphi_{r,a}$ such that the two copies of the gauge symmetries are preserved, though in a non-trivial way. For a systematic discussion of how generalised 1-form global symmetries naturally organise these fluid phases without the need for gauge fixing, see \cite{2016RPPh...79i6001S, Akyuz:2023lsm}.

\begin{figure}[t]
	\centering
	\includegraphics{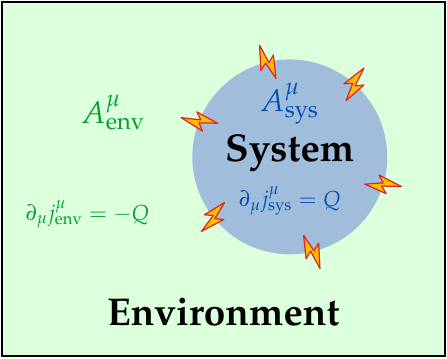}
	\caption{Illustration of the lack of conservation of the system's current. This effect is a mere consequence of the total charge conservation $\partial_\mu J^\mu = 0$. It simply illustrates the fact that charges initially located in the system sink in the environment. Same for gravity, with the total conservation of the stress-energy tensor $\nabla_\mu T^{\mu\nu} = 0$.}
	\label{fig:sink}
\end{figure}

Physically, the breaking of the global symmetries to their diagonal subgroup leads to an apparent, anomalous ``leaky'' current where charges in the system seem to not be conserved. However, the standard total current of both the system and the environment is exactly conserved ($\partial^\mu j_{\mathrm{sys}}^\mu = - \partial^\mu j_{\mathrm{env}}^\mu$). This entails the open nature of the system's dynamics. The apparent ``leaky'' current for the system illustrated in Figure \ref{fig:sink} is then perfectly resolved: the charge was never lost; it simply sank into the macroscopic environment. 

\subsection{Open gravity and stochastic gravitational waves}

The lessons learned from open electromagnetism bring some insight to tackle the open problem of gravity in a medium. During inflation, primordial gravitational waves (tensor perturbations of the metric) do not necessarily propagate in a pure vacuum. They can travel through a dense, interacting environment made of scalar fields, gauge sectors, or other forms of matter present in the early universe on an accelerating time-dependent Friedmann-Lemaître-Robertson-Walker (FLRW) background. 

Consequently, the effective coefficients of the open theory must in general be explicitly time-dependent, much like the setup discussed in Section \ref{subsec:open}. This introduces a major theoretical hurdle: unlike static thermal states in flat spacetime which provide a clear, constant energy scale (such as temperature $T$), dynamical cosmological backgrounds lack a single, definitive energy scale. Figuring out the correct organising principles of gravitational effective field theories in such explicitly time-dependent backgrounds, with a systematic derivative expansion and power counting of operators, remains one of the main open challenges in the field. 

\paragraph{\it Dissipative gravitational dynamics.}

The presence of a medium modifies the linear dynamics of gravitational waves through three dissipative and stochastic effects: friction, birefringence, and noise \cite{Salcedo:2025ezu}.
Just as a plasma exerts friction on a photon, the inflationary environment can exert a \textit{shear viscosity} on the propagating gravitational wave \cite{Weinberg:2003ur, Salcedo:2026cqb}. An open effective field theory of gravity must therefore introduce dissipative friction effects that operate alongside the standard Hubble friction. Furthermore, the open system allows for parity-violating operators at leading derivative order. This can introduce chiral birefringence, forcing distinct helicities of the stochastic graviton to evolve differently and experience different rates of dissipation \cite{Creminelli:2014wna}. At last, the tidal deformation of the environment can generate a stochastic source of gravitational waves that can have important phenomenological consequences such as the generation of stochastic gravitational wave backgrounds.

\paragraph{\it Fluid clocks and rulers.}

The challenges encountered in gravity are conceptually similar to those arising in the electromagnetic case, but are technically more involved due to the tensorial nature of the theory.
In Sec.~\ref{subsec:EFToI}, we presented the EFT of Inflation as the spontaneous symmetry breaking of time translation symmetry \cite{Cheung:2007st}, encompassed through $\text{Poincar\'e} \rightarrow \mathrm{SO}(3) \times \mathbb{R}^3$. For a conservative theory in the Schwinger-Keldysh contour, this is doubled to 
\begin{align}
    \text{Poincar\'e}_+ \times  \text{Poincar\'e}_-\rightarrow [\mathrm{SO}(3) \times \mathbb{R}^3]_+ \times [\mathrm{SO}(3) \times \mathbb{R}^3]_-
\end{align}
Just as above, the presence of dissipative and stochastic effects breaks the two copies of global symmetries to their diagonal subgroup, which in the end reduces to 
\begin{align}
    \text{Poincar\'e}_+ \times  \text{Poincar\'e}_-\rightarrow [\mathrm{SO}(3) \times \mathbb{R}^3]_r\,.
\end{align}
This structure may obscure how advanced diffeomorphisms are realised. However, to restore manifest diffeomorphism invariance across the entire Schwinger-Keldysh contour, one may introduce four St\"uckelberg fields $X^\mu_{r,a}(x)$ that non-linearly transform under spacetime translations and boosts  \cite{Liu:2018kfw}. These fields represent the ``clocks and rulers'' defining the rest frame of a dissipative fluid \cite{Crossley:2015evo}. More details about this construction can be found in \cite{Liu:2018kfw}. Its concrete application to cosmology is still under investigation. 

Just as the electromagnetic current, the gravitational stress-energy tensor obeys modified transport equations reflecting the exchange of energy and entropy with the environment ($\nabla_\mu T^{\mu\nu}_{\text{sys}}  = - \nabla_\mu T^{\mu\nu}_{\text{env}}$ as described in Fig.~\ref{fig:sink}). The present construction guarantees that the total stress-energy tensor of both the system and the environment is still conserved. 

\paragraph{\it Concluding remarks.}

The physics of the early universe often exhibits dynamics similar to those of open quantum systems. Whether we are tracking the curvature perturbation through a hidden matter sector, or watching stochastic gravitons damped and diffused during their propagation toward our detectors, the environment can fundamentally alter the dynamics of the cosmological probes and their observational signatures. 

While bottom-up EFT approaches provide us with the practical Langevin dynamics needed to confront observational data, the top-down requirement of gauge and diffeomorphism invariance forces us to acknowledge the rich phase structure of non-equilibrium systems in the Schwinger-Keldysh contour.
Bridging these two perspectives remains a vibrant area  at the crossroad of modern cosmology and condensed matter.


\section{Conclusion}

Cosmology presents phenomena that go beyond the familiar framework of flat-space quantum field theory in the vacuum. It compels us to broaden our theoretical toolkit to address non-equilibrium and open-system dynamics. The material discussed in this chapter aims to provide a starting point for those interested in applying the Schwinger–Keldysh formalism to cosmological settings. 

Beyond cosmology, SK-EFTs hold exciting potential in areas such as quark–gluon plasma, black hole and neutron-star physics, holography, and dissipative hydrodynamics. We hope that this framework will contribute to building bridges among these fields. The study of SK-EFTs is still in its infancy, and cosmology stands to benefit greatly from cross-disciplinary exchanges on topics as diverse as non-equilibrium renormalisation group or non-Abelian strong-to-weak symmetry breaking.

The \textit{Disordered Universe Summer School 2025}, where the content of this chapter was presented, has been a fertile ground for stimulating discussions among the participants. Far from being exhaustive, the following is a selection of topics highlighted throughout the lectures that may inspire future investigations:
\begin{itemize}
	\item \textbf{Beyond correlators:} Can the SK-EFT formalism be extended to compute information-spreading measures such as out-of-time-ordered correlators (OTOCs), entropy measures, or off-diagonal density-matrix elements? 
	\item \textbf{Open scalar theory in Minkowski:} What is the single-exchange, tree-level trispectrum in the dissipative theory? How does the framework simplify when local thermodynamic equilibrium is imposed through the KMS symmetry? 
	\item \textbf{Symmetries in the SK contour:} How can we extend the discussion to non-linear realisations of diffeomorphisms? Can the in-in coset construction be adapted to cosmology? Is there a differential-geometric formulation of gauge symmetries within the SK framework? 
	\item \textbf{EFT construction:} Can we develop non-equilibrium renormalisation-group techniques for open EFTs to clarify their power counting and cutoff structure? Is it possible to constrain EFT coefficients using entropy bounds, stability requirements, or microcausality? 
	\item \textbf{Phenomenology:} How can the Open EFT of Inflation be matched to gauge-inflation or warm-inflation models? Could the theory of gravity in a medium help describe dissipative dark sectors? Can similar methods model the dissipative propagation of gravitational waves in the interstellar medium? What observational limits can be placed on EFT coefficients using current data? 

\end{itemize}

The material we covered serves as an initial step towards acquiring the tools needed to address these exciting questions and to develop a new generation of EFTs, capable of extracting novel physics from the wealth of forthcoming data. Above all, we hope to have conveyed that this is a vibrant and accessible area of research — rich with open questions at nearly every level — well within reach for researchers trained in general relativity and quantum field theory.

\begin{acknowledgements}
	-- The \href{https://indico.sns.it/event/91/}{{\it La Ricotta} Summer School  {\it The Disordered Universe 2025}} received financial support from the ERC grant NOTIMEFORCOSMO (agreement n${}^{\circ}$ 1126304)), by Scuola Normale and by the Instituto Galego de F{\'i}sica das Altas Enerx{\'i}as (IGFAE-USC/Xunta de Galicia) through the CIGUS network and EU cofinancing program.
\end{acknowledgements}

\bibliographystyle{utphys}
\bibliography{2025-DU-OEFT-biblio}

\providecommand{\href}[2]{#2}\begingroup\raggedright\begin{thebibliography}{10}

\bibitem{Benincasa:2027tdu}
P.~Benincasa and G.~L. Pimentel, eds., {\em {The Disordered Universe. Open
  Systems, Scattering Amplitudes, de Sitter Space, and Disordered Systems in
  Cosmology.}}
\newblock Springer Lecture Notes in Physics, 2027 (in press).

\bibitem{Cheung:2007st}
C.~Cheung, P.~Creminelli, A.~L. Fitzpatrick, J.~Kaplan, and L.~Senatore, ``The
  {Effective} {Field} {Theory} of {Inflation},''
  \href{http://dx.doi.org/10.1088/1126-6708/2008/03/014}{{\em JHEP} {\bf 03}
  (2008)  014}. 0709.0293.

\bibitem{Gubitosi:2012hu}
G.~Gubitosi, F.~Piazza, and F.~Vernizzi, ``{The Effective Field Theory of Dark
  Energy},'' \href{http://dx.doi.org/10.1088/1475-7516/2013/02/032}{{\em JCAP}
  {\bf 02} (2013)  032}, \href{http://arxiv.org/abs/1210.0201}{{\tt
  arXiv:1210.0201 [hep-th]}}.

\bibitem{breuerTheoryOpenQuantum2002}
H.~P. Breuer and F.~Petruccione,
  \href{http://dx.doi.org/10.1093/acprof:oso/9780199213900.001.0001}{{\em The
  theory of Open Quantum Systems}}.
\newblock Oxford University Press, 2002.

\bibitem{Baumann:2009ds}
D.~Baumann, \href{http://dx.doi.org/10.1142/9789814327183_0010}{``Inflation,''}
  in {\em Theoretical {Advanced} {Study} {Institute} in {Elementary} {Particle}
  {Physics}: {Physics} of the {Large} and the {Small}}, pp.~523--686.
\newblock 2011.
\newblock 0907.5424.

\bibitem{Dodelson:2003ft}
S.~Dodelson, {\em {Modern Cosmology}}.
\newblock Academic Press, Amsterdam, 2003.

\bibitem{Chen:2017ryl}
X.~Chen, Y.~Wang, and Z.-Z. Xianyu, ``Schwinger-{Keldysh} {Diagrammatics} for
  {Primordial} {Perturbations},''
  \href{http://dx.doi.org/10.1088/1475-7516/2017/12/006}{{\em JCAP} {\bf 12}
  (2017)  006}. 1703.10166.

\bibitem{Maldacena:2002vr}
J.~M. Maldacena, ``Non-{Gaussian} features of primordial fluctuations in single
  field inflationary models,''
  \href{http://dx.doi.org/10.1088/1126-6708/2003/05/013}{{\em JHEP} {\bf 05}
  (2003)  013}. astro-ph/0210603.

\bibitem{Colas:2024ysu}
T.~Colas, J.~Grain, G.~Kaplanek, and V.~Vennin, ``{In-in formalism for the
  entropy of quantum fields in curved spacetimes},''
  \href{http://dx.doi.org/10.1088/1475-7516/2024/08/047}{{\em JCAP} {\bf 08}
  (2024)  047}, \href{http://arxiv.org/abs/2406.17856}{{\tt arXiv:2406.17856
  [hep-th]}}.

\bibitem{Weinberg:2005vy}
S.~Weinberg, ``{Quantum contributions to cosmological correlations},''
  \href{http://dx.doi.org/10.1103/PhysRevD.72.043514}{{\em Phys. Rev. D} {\bf
  72} (2005)  043514}, \href{http://arxiv.org/abs/hep-th/0506236}{{\tt
  arXiv:hep-th/0506236}}.

\bibitem{Weinberg:2006ac}
S.~Weinberg, ``Quantum contributions to cosmological correlations. {II}. {Can}
  these corrections become large?,''
  \href{http://dx.doi.org/10.1103/PhysRevD.74.023508}{{\em Phys. Rev. D} {\bf
  74} (2006)  023508}. hep-th/0605244.

\bibitem{Boyanovsky:2015xoa}
D.~Boyanovsky, ``{Effective Field Theory out of Equilibrium: Brownian quantum
  fields},'' \href{http://dx.doi.org/10.1088/1367-2630/17/6/063017}{{\em New J.
  Phys.} {\bf 17} (2015) no.~6, 063017},
  \href{http://arxiv.org/abs/1503.00156}{{\tt arXiv:1503.00156 [hep-ph]}}.

\bibitem{Colas:2025lectures}
T.~Colas, ``{Lectures on Open Effective Field Theories},''
\newblock 9, 2025.
\newblock \href{http://arxiv.org/abs/2510.00140}{{\tt arXiv:2510.00140
  [hep-th]}}.

\bibitem{Donath:2024utn}
Y.~Donath and E.~Pajer, ``{The In-Out Formalism for In-In Correlators},''
  \href{http://arxiv.org/abs/2402.05999}{{\tt arXiv:2402.05999 [hep-th]}}.

\bibitem{Kaya:2018jdo}
A.~Kaya, ``{On $i\epsilon$ Prescription in Cosmology},''
  \href{http://dx.doi.org/10.1088/1475-7516/2019/04/002}{{\em JCAP} {\bf 04}
  (2019)  002}, \href{http://arxiv.org/abs/1810.12324}{{\tt arXiv:1810.12324
  [gr-qc]}}.

\bibitem{Pajer:FTC}
E.~Pajer, ``{Lecture Notes on Field Theory in Cosmology},''.
\newblock \url{https://www.damtp.cam.ac.uk/user/ep551/FTC.html}.

\bibitem{Colas:2025ind}
T.~Colas, Z.~Qin, and X.~Tong, ``{Open Effective Field Theory and the Physics
  of Cosmological Collider Signals},''
  \href{http://arxiv.org/abs/2512.07941}{{\tt arXiv:2512.07941 [hep-th]}}.

\bibitem{manzanoShortIntroductionLindblad2020}
D.~Manzano, ``{A short introduction to the Lindblad master equation},''
  \href{http://dx.doi.org/10.1063/1.5115323}{{\em AIP Adv.} {\bf 10} (2020)
  no.~2, 025106}, \href{http://arxiv.org/abs/1906.04478}{{\tt arXiv:1906.04478
  [quant-ph]}}.

\bibitem{PhysRevA.8.423}
P.~C. Martin, E.~D. Siggia, and H.~A. Rose, ``{Statistical Dynamics of
  Classical Systems},'' \href{http://dx.doi.org/10.1103/PhysRevA.8.423}{{\em
  Phys. Rev. A} {\bf 8} (1973)  423--437}.

\bibitem{Caldeira:1981rx}
A.~O. Caldeira and A.~J. Leggett, ``Influence of dissipation on quantum
  tunneling in macroscopic systems,''
  \href{http://dx.doi.org/10.1103/PhysRevLett.46.211}{{\em Phys. Rev. Lett.}
  {\bf 46} (1981)  211}.

\bibitem{Caldeira:1982iu}
A.~O. Caldeira and A.~J. Leggett, ``Path integral approach to quantum
  {Brownian} motion,''
  \href{http://dx.doi.org/https://doi.org/10.1016/0378-4371(83)90013-4}{{\em
  Physica A: Statistical Mechanics and its Applications} {\bf 121} (1983)
  no.~3, 587--616}.
  \url{https://www.sciencedirect.com/science/article/pii/0378437183900134}.

\bibitem{Caldeira:1982uj}
A.~O. Caldeira and A.~J. Leggett, ``Quantum tunneling in a dissipative
  system,'' \href{http://dx.doi.org/10.1016/0003-4916(83)90202-6}{{\em Annals
  Phys.} {\bf 149} (1983)  374--456}.

\bibitem{Lau:2024mqm}
P.~H.~C. Lau, K.~Nishii, and T.~Noumi, ``{Gravitational EFT for dissipative
  open systems},'' \href{http://arxiv.org/abs/2412.21136}{{\tt arXiv:2412.21136
  [hep-th]}}.

\bibitem{Lindblad:1975ef}
G.~Lindblad, ``{On the Generators of Quantum Dynamical Semigroups},''
  \href{http://dx.doi.org/10.1007/BF01608499}{{\em Commun. Math. Phys.} {\bf
  48} (1976)  119}.

\bibitem{Schwinger:1960qe}
J.~S. Schwinger, ``{Brownian motion of a quantum oscillator},''
  \href{http://dx.doi.org/10.1063/1.1703727}{{\em J. Math. Phys.} {\bf 2}
  (1961)  407--432}.

\bibitem{Keldysh:1964ud}
L.~V. Keldysh, ``{Diagram technique for nonequilibrium processes},'' {\em Zh.
  Eksp. Teor. Fiz.} {\bf 47} (1964)  1515--1527.

\bibitem{10.1063/1.522979}
V.~Gorini, A.~Kossakowski, and E.~C.~G. Sudarshan, ``{Completely Positive
  Dynamical Semigroups of N Level Systems},''
  \href{http://dx.doi.org/10.1063/1.522979}{{\em J. Math. Phys.} {\bf 17}
  (1976)  821}.

\bibitem{FEYNMAN1963118}
R.~Feynman and F.~Vernon, ``The theory of a general quantum system interacting
  with a linear dissipative system,''
  \href{http://dx.doi.org/https://doi.org/10.1016/0003-4916(63)90068-X}{{\em
  Annals of Physics} {\bf 24} (1963)  118--173}.
  \url{https://www.sciencedirect.com/science/article/pii/000349166390068X}.

\bibitem{Calzetta:2008iqa}
E.~A. Calzetta and B.-L.~B. Hu,
  \href{http://dx.doi.org/10.1017/CBO9780511535123}{{\em Nonequilibrium
  {Quantum} {Field} {Theory}}}.
\newblock Cambridge {Monographs} on {Mathematical} {Physics}. Cambridge
  University Press, Sept., 2008.

\bibitem{Liu:2018kfw}
H.~Liu and P.~Glorioso, ``{Lectures on non-equilibrium effective field theories
  and fluctuating hydrodynamics},''
  \href{http://dx.doi.org/10.22323/1.305.0008}{{\em PoS} {\bf TASI2017} (2018)
  008}, \href{http://arxiv.org/abs/1805.09331}{{\tt arXiv:1805.09331
  [hep-th]}}.

\bibitem{Kamenev:2009jj}
A.~Kamenev and A.~Levchenko, ``{Keldysh technique and nonlinear sigma-model:
  Basic principles and applications},''
  \href{http://dx.doi.org/10.1080/00018730902850504}{{\em Adv. Phys.} {\bf 58}
  (2009)  197}, \href{http://arxiv.org/abs/0901.3586}{{\tt arXiv:0901.3586
  [cond-mat.other]}}.

\bibitem{2016RPPh...79i6001S}
L.~M. {Sieberer}, M.~{Buchhold}, and S.~{Diehl},
  \href{http://dx.doi.org/10.1088/0034-4885/79/9/096001}{``{Keldysh field
  theory for driven open quantum systems},''{\em Reports on Progress in
  Physics} {\bf 79} (Sept., 2016)  096001},
  \href{http://arxiv.org/abs/1512.00637}{{\tt arXiv:1512.00637
  [cond-mat.quant-gas]}}.

\bibitem{Berges:2012ty}
J.~Berges and D.~Mesterhazy, ``{Introduction to the nonequilibrium functional
  renormalization group},''
  \href{http://dx.doi.org/10.1016/j.nuclphysbps.2012.06.003}{{\em Nucl. Phys. B
  Proc. Suppl.} {\bf 228} (2012)  37--60},
  \href{http://arxiv.org/abs/1204.1489}{{\tt arXiv:1204.1489 [hep-ph]}}.

\bibitem{Rosenhaus:2025bgy}
V.~Rosenhaus and M.~Smolkin, ``{Renormalization Group in far-from-equilibrium
  states},'' \href{http://arxiv.org/abs/2504.06243}{{\tt arXiv:2504.06243
  [hep-th]}}.

\bibitem{Radovskaya:2020lns}
A.~A. Radovskaya and A.~G. Semenov, ``{Semiclassical approximation meets
  Keldysh\textendash{}Schwinger diagrammatic technique: scalar $\varphi ^4$},''
  \href{http://dx.doi.org/10.1140/epjc/s10052-021-09382-4}{{\em Eur. Phys. J.
  C} {\bf 81} (2021) no.~8, 704}, \href{http://arxiv.org/abs/2003.06395}{{\tt
  arXiv:2003.06395 [hep-ph]}}.

\bibitem{Stratonovich1957}
R.~L. {Stratonovich}, ``{On a Method of Calculating Quantum Distribution
  Functions},''{\em Soviet Physics Doklady} {\bf 2} (July, 1957)  416.

\bibitem{Hubbard1959}
J.~{Hubbard}, \href{http://dx.doi.org/10.1103/PhysRevLett.3.77}{``{Calculation
  of Partition Functions},''{\em Phys. Rev. Letters} {\bf 3} (July, 1959)
  77--78}.

\bibitem{Kubo:1966fyg}
R.~Kubo, ``{The fluctuation-dissipation theorem},''
  \href{http://dx.doi.org/10.1088/0034-4885/29/1/306}{{\em Rept. Prog. Phys.}
  {\bf 29} (1966) no.~1, 255}.

\bibitem{PhysRev.115.1342}
P.~C. Martin and J.~Schwinger,
  \href{http://dx.doi.org/10.1103/PhysRev.115.1342}{``Theory of many-particle
  systems. i,''{\em Phys. Rev.} {\bf 115} (Sep, 1959)  1342--1373}.
  \url{https://link.aps.org/doi/10.1103/PhysRev.115.1342}.

\bibitem{Hongo:2018ant}
M.~Hongo, S.~Kim, T.~Noumi, and A.~Ota, ``{Effective field theory of
  time-translational symmetry breaking in nonequilibrium open system},''
  \href{http://dx.doi.org/10.1007/JHEP02(2019)131}{{\em JHEP} {\bf 02} (2019)
  131}, \href{http://arxiv.org/abs/1805.06240}{{\tt arXiv:1805.06240
  [hep-th]}}.

\bibitem{Salcedo:2024smn}
S.~A. Salcedo, T.~Colas, and E.~Pajer, ``{The Open Effective Field Theory of
  Inflation},'' \href{http://dx.doi.org/10.1007/JHEP10(2024)248}{{\em JHEP}
  {\bf 10} (2024)  248}, \href{http://arxiv.org/abs/2404.15416}{{\tt
  arXiv:2404.15416 [hep-th]}}.

\bibitem{Chakrabarty:2019aeu}
B.~Chakrabarty, J.~Chakravarty, S.~Chaudhuri, C.~Jana, R.~Loganayagam, and
  A.~Sivakumar, ``{Nonlinear Langevin dynamics via holography},''
  \href{http://dx.doi.org/10.1007/JHEP01(2020)165}{{\em JHEP} {\bf 01} (2020)
  165}, \href{http://arxiv.org/abs/1906.07762}{{\tt arXiv:1906.07762
  [hep-th]}}.

\bibitem{kamenev_2011}
A.~Kamenev, \href{http://dx.doi.org/10.1017/CBO9781139003667}{{\em Field Theory
  of Non-Equilibrium Systems}}.
\newblock Cambridge University Press, 2011.

\bibitem{Ema:2024hkj}
Y.~Ema and K.~Mukaida, ``{Cutting rule for in-in correlators and cosmological
  collider},'' \href{http://arxiv.org/abs/2409.07521}{{\tt arXiv:2409.07521
  [hep-th]}}.

\bibitem{Babich:2004gb}
D.~Babich, P.~Creminelli, and M.~Zaldarriaga, ``{The Shape of
  non-Gaussianities},''
  \href{http://dx.doi.org/10.1088/1475-7516/2004/08/009}{{\em JCAP} {\bf 08}
  (2004)  009}, \href{http://arxiv.org/abs/astro-ph/0405356}{{\tt
  arXiv:astro-ph/0405356}}.

\bibitem{Chen:2006nt}
X.~Chen, M.-x. Huang, S.~Kachru, and G.~Shiu, ``{Observational signatures and
  non-Gaussianities of general single field inflation},''
  \href{http://dx.doi.org/10.1088/1475-7516/2007/01/002}{{\em JCAP} {\bf 01}
  (2007)  002}, \href{http://arxiv.org/abs/hep-th/0605045}{{\tt
  arXiv:hep-th/0605045}}.

\bibitem{Baumann:2022jpr}
D.~Baumann, D.~Green, A.~Joyce, E.~Pajer, G.~L. Pimentel, C.~Sleight, and
  M.~Taronna, ``{Snowmass White Paper: The Cosmological Bootstrap},'' in {\em
  {Snowmass 2021}}.
\newblock 3, 2022.
\newblock \href{http://arxiv.org/abs/2203.08121}{{\tt arXiv:2203.08121
  [hep-th]}}.

\bibitem{Piazza:2013coa}
F.~Piazza and F.~Vernizzi, ``{Effective Field Theory of Cosmological
  Perturbations},''
  \href{http://dx.doi.org/10.1088/0264-9381/30/21/214007}{{\em Class. Quant.
  Grav.} {\bf 30} (2013)  214007}, \href{http://arxiv.org/abs/1307.4350}{{\tt
  arXiv:1307.4350 [hep-th]}}.

\bibitem{Creminelli:2017sry}
P.~Creminelli and F.~Vernizzi, ``{Dark Energy after GW170817 and GRB170817A},''
  \href{http://dx.doi.org/10.1103/PhysRevLett.119.251302}{{\em Phys. Rev.
  Lett.} {\bf 119} (2017) no.~25, 251302},
  \href{http://arxiv.org/abs/1710.05877}{{\tt arXiv:1710.05877 [astro-ph.CO]}}.

\bibitem{Salcedo:2025ezu}
S.~A. Salcedo, T.~Colas, L.~Dufner, and E.~Pajer, ``{An open system approach to
  gravity},'' \href{http://dx.doi.org/10.1007/JHEP02(2026)241}{{\em JHEP} {\bf
  02} (2026)  241}, \href{http://arxiv.org/abs/2507.03103}{{\tt
  arXiv:2507.03103 [hep-th]}}.

\bibitem{Finelli:2018upr}
B.~Finelli, G.~Goon, E.~Pajer, and L.~Santoni, ``{The Effective Theory of
  Shift-Symmetric Cosmologies},''
  \href{http://dx.doi.org/10.1088/1475-7516/2018/05/060}{{\em JCAP} {\bf 05}
  (2018)  060}, \href{http://arxiv.org/abs/1802.01580}{{\tt arXiv:1802.01580
  [hep-th]}}.

\bibitem{Anber:2009ua}
M.~M. Anber and L.~Sorbo, ``{Naturally inflating on steep potentials through
  electromagnetic dissipation},''
  \href{http://dx.doi.org/10.1103/PhysRevD.81.043534}{{\em Phys. Rev. D} {\bf
  81} (2010)  043534}, \href{http://arxiv.org/abs/0908.4089}{{\tt
  arXiv:0908.4089 [hep-th]}}.

\bibitem{Creminelli:2023aly}
P.~Creminelli, S.~Kumar, B.~Salehian, and L.~Santoni, ``{Dissipative Inflation
  via Scalar Production},'' \href{http://arxiv.org/abs/2305.07695}{{\tt
  arXiv:2305.07695 [hep-th]}}.

\bibitem{Akyuz:2023lsm}
C.~O. Akyuz, G.~Goon, and R.~Penco, ``{The Schwinger-Keldysh Coset
  Construction},'' \href{http://arxiv.org/abs/2306.17232}{{\tt arXiv:2306.17232
  [hep-th]}}.

\bibitem{Collins:2012nq}
H.~Collins, R.~Holman, and A.~Ross, ``{Effective field theory in time-dependent
  settings},'' \href{http://dx.doi.org/10.1007/JHEP02(2013)108}{{\em JHEP} {\bf
  02} (2013)  108}, \href{http://arxiv.org/abs/1208.3255}{{\tt arXiv:1208.3255
  [hep-th]}}.

\bibitem{LopezNacir:2011kk}
D.~Lopez~Nacir, R.~A. Porto, L.~Senatore, and M.~Zaldarriaga, ``{Dissipative
  effects in the Effective Field Theory of Inflation},''
  \href{http://dx.doi.org/10.1007/JHEP01(2012)075}{{\em JHEP} {\bf 01} (2012)
  075}, \href{http://arxiv.org/abs/1109.4192}{{\tt arXiv:1109.4192 [hep-th]}}.

\bibitem{Berera:1995ie}
A.~Berera, ``Warm inflation,''
  \href{http://dx.doi.org/10.1103/PhysRevLett.75.3218}{{\em Phys. Rev. Lett.}
  {\bf 75} (1995)  3218--3221}. astro-ph/9509049.

\bibitem{Berera:1995wh}
A.~Berera and L.-Z. Fang, ``Thermally induced density perturbations in the
  inflation era,'' \href{http://dx.doi.org/10.1103/PhysRevLett.74.1912}{{\em
  Phys. Rev. Lett.} {\bf 74} (1995)  1912--1915}. astro-ph/9501024.

\bibitem{Berera:1999ws}
A.~Berera, ``{Warm inflation at arbitrary adiabaticity: A Model, an existence
  proof for inflationary dynamics in quantum field theory},''
  \href{http://dx.doi.org/10.1016/S0550-3213(00)00411-9}{{\em Nucl. Phys. B}
  {\bf 585} (2000)  666--714}, \href{http://arxiv.org/abs/hep-ph/9904409}{{\tt
  arXiv:hep-ph/9904409}}.

\bibitem{Berera:2008ar}
A.~Berera, I.~G. Moss, and R.~O. Ramos, ``Warm {Inflation} and its
  {Microphysical} {Basis},''
  \href{http://dx.doi.org/10.1088/0034-4885/72/2/026901}{{\em Rept. Prog.
  Phys.} {\bf 72} (2009)  026901}. 0808.1855.

\bibitem{Ballesteros:2023dno}
G.~Ballesteros, A.~Perez~Rodriguez, and M.~Pierre, ``{Monomial warm inflation
  revisited},'' \href{http://arxiv.org/abs/2304.05978}{{\tt arXiv:2304.05978
  [astro-ph.CO]}}.

\bibitem{Montefalcone:2023pvh}
G.~Montefalcone, V.~Aragam, L.~Visinelli, and K.~Freese, ``{WarmSPy: a
  numerical study of cosmological perturbations in warm inflation},''
  \href{http://dx.doi.org/10.1088/1475-7516/2024/01/032}{{\em JCAP} {\bf 01}
  (2024)  032}, \href{http://arxiv.org/abs/2306.16190}{{\tt arXiv:2306.16190
  [astro-ph.CO]}}.

\bibitem{Holman:2007na}
R.~Holman and A.~J. Tolley, ``{Enhanced Non-Gaussianity from Excited Initial
  States},'' \href{http://dx.doi.org/10.1088/1475-7516/2008/05/001}{{\em JCAP}
  {\bf 05} (2008)  001}, \href{http://arxiv.org/abs/0710.1302}{{\tt
  arXiv:0710.1302 [hep-th]}}.

\bibitem{Meerburg:2009ys}
P.~D. Meerburg, J.~P. van~der Schaar, and P.~S. Corasaniti, ``{Signatures of
  Initial State Modifications on Bispectrum Statistics},''
  \href{http://dx.doi.org/10.1088/1475-7516/2009/05/018}{{\em JCAP} {\bf 05}
  (2009)  018}, \href{http://arxiv.org/abs/0901.4044}{{\tt arXiv:0901.4044
  [hep-th]}}.

\bibitem{Agullo:2010ws}
I.~Agullo and L.~Parker, ``{Non-gaussianities and the Stimulated creation of
  quanta in the inflationary universe},''
  \href{http://dx.doi.org/10.1103/PhysRevD.83.063526}{{\em Phys. Rev. D} {\bf
  83} (2011)  063526}, \href{http://arxiv.org/abs/1010.5766}{{\tt
  arXiv:1010.5766 [astro-ph.CO]}}.

\bibitem{Ashoorioon:2010xg}
A.~Ashoorioon and G.~Shiu, ``{A Note on Calm Excited States of Inflation},''
  \href{http://dx.doi.org/10.1088/1475-7516/2011/03/025}{{\em JCAP} {\bf 03}
  (2011)  025}, \href{http://arxiv.org/abs/1012.3392}{{\tt arXiv:1012.3392
  [astro-ph.CO]}}.

\bibitem{Agarwal:2012mq}
N.~Agarwal, R.~Holman, A.~J. Tolley, and J.~Lin, ``{Effective field theory and
  non-Gaussianity from general inflationary states},''
  \href{http://dx.doi.org/10.1007/JHEP05(2013)085}{{\em JHEP} {\bf 05} (2013)
  085}, \href{http://arxiv.org/abs/1212.1172}{{\tt arXiv:1212.1172 [hep-th]}}.

\bibitem{Ashoorioon:2013eia}
A.~Ashoorioon, K.~Dimopoulos, M.~M. Sheikh-Jabbari, and G.~Shiu,
  ``{Reconciliation of High Energy Scale Models of Inflation with Planck},''
  \href{http://dx.doi.org/10.1088/1475-7516/2014/02/025}{{\em JCAP} {\bf 02}
  (2014)  025}, \href{http://arxiv.org/abs/1306.4914}{{\tt arXiv:1306.4914
  [hep-th]}}.

\bibitem{Albrecht:2014aga}
A.~Albrecht, N.~Bolis, and R.~Holman, ``{Cosmological Consequences of Initial
  State Entanglement},'' \href{http://dx.doi.org/10.1007/JHEP11(2014)093}{{\em
  JHEP} {\bf 11} (2014)  093}, \href{http://arxiv.org/abs/1408.6859}{{\tt
  arXiv:1408.6859 [hep-th]}}.

\bibitem{Green:2020whw}
D.~Green and R.~A. Porto, ``Signals of a {Quantum} {Universe},''
  \href{http://dx.doi.org/10.1103/PhysRevLett.124.251302}{{\em Phys. Rev.
  Lett.} {\bf 124} (2020) no.~25, 251302}. 2001.09149.

\bibitem{Creminelli:2012ed}
P.~Creminelli, J.~Nore\~na, and M.~Simonovi\'c, ``{Conformal consistency
  relations for single-field inflation},''
  \href{http://dx.doi.org/10.1088/1475-7516/2012/07/052}{{\em JCAP} {\bf 07}
  (2012)  052}, \href{http://arxiv.org/abs/1203.4595}{{\tt arXiv:1203.4595
  [hep-th]}}.

\bibitem{Planck:2019kim}
Y.~Akrami and {others}, ``Planck 2018 results. {IX}. {Constraints} on
  primordial non-{Gaussianity},''
  \href{http://dx.doi.org/10.1051/0004-6361/201935891}{{\em Astron. Astrophys.}
  {\bf 641} (2020)  A9}. 1905.05697.

\bibitem{Salcedo:2026sdn}
S.~A. Salcedo, T.~Colas, P.~Suman, B.~Zhang, J.~Fergusson, and E.~P.~S.
  Shellard, ``{Primordial non-Gaussianity constraints on dissipative
  inflation},'' \href{http://arxiv.org/abs/2603.13473}{{\tt arXiv:2603.13473
  [astro-ph.CO]}}.

\bibitem{SPHEREx:2014bgr}
{\bf SPHEREx} Collaboration, O.~Dor\'e {\em et al.}, ``{Cosmology with the
  SPHEREX All-Sky Spectral Survey},''
  \href{http://arxiv.org/abs/1412.4872}{{\tt arXiv:1412.4872 [astro-ph.CO]}}.

\bibitem{Cabass:2022epm}
G.~Cabass, M.~M. Ivanov, O.~H.~E. Philcox, M.~Simonovic, and M.~Zaldarriaga,
  ``{Constraining single-field inflation with MegaMapper},''
  \href{http://dx.doi.org/10.1016/j.physletb.2023.137912}{{\em Phys. Lett. B}
  {\bf 841} (2023)  137912}, \href{http://arxiv.org/abs/2211.14899}{{\tt
  arXiv:2211.14899 [astro-ph.CO]}}.

\bibitem{Braganca:2023pcp}
D.~Braganca, Y.~Donath, L.~Senatore, and H.~Zheng, ``{Peeking into the next
  decade in Large-Scale Structure Cosmology with its Effective Field Theory},''
  \href{http://arxiv.org/abs/2307.04992}{{\tt arXiv:2307.04992 [astro-ph.CO]}}.

\bibitem{Landry:2022nog}
M.~J. Landry and H.~Liu, ``{A systematic formulation of chiral anomalous
  magnetohydrodynamics},'' \href{http://arxiv.org/abs/2212.09757}{{\tt
  arXiv:2212.09757 [hep-ph]}}.

\bibitem{Akyuz:2025bco}
C.~O. Akyuz and R.~Penco, ``{Effective Description of Ajar Systems with a
  $U(1)$ Symmetry},'' \href{http://arxiv.org/abs/2503.22840}{{\tt
  arXiv:2503.22840 [hep-th]}}.

\bibitem{Salcedo:2024nex}
S.~A. Salcedo, T.~Colas, and E.~Pajer, ``{An Open Effective Field Theory for
  light in a medium},'' \href{http://dx.doi.org/10.1007/JHEP03(2025)138}{{\em
  JHEP} {\bf 03} (2025)  138}, \href{http://arxiv.org/abs/2412.12299}{{\tt
  arXiv:2412.12299 [hep-th]}}.

\bibitem{Firat:2025upx}
E.~Firat, A.~Gomes, F.~Nardi, R.~Penco, and R.~Rattazzi, ``{Schwinger-Keldysh
  effective theory of charge transport: redundancies and systematic $\omega/T$
  expansion},'' \href{http://arxiv.org/abs/2508.18346}{{\tt arXiv:2508.18346
  [hep-th]}}.

\bibitem{Kaplanek:2025moq}
G.~Kaplanek, M.~Mylova, and A.~J. Tolley, ``{Gauging Open EFTs from the top
  down},'' \href{http://arxiv.org/abs/2512.17089}{{\tt arXiv:2512.17089
  [hep-th]}}.

\bibitem{Kaplanek:2026kpp}
G.~Kaplanek, M.~Mylova, and A.~J. Tolley, ``{Schwinger-Keldysh Path Integral
  for Gauge theories},'' \href{http://arxiv.org/abs/2604.26941}{{\tt
  arXiv:2604.26941 [hep-th]}}.

\bibitem{Hauser:2026sgr}
J.~Hauser, K.~Su, H.~Ha, J.~Lloyd, T.~G. Kiely, R.~Vasseur, S.~Gopalakrishnan,
  C.~Xu, and M.~P.~A. Fisher, ``{Strong-to-Weak Symmetry Breaking in Open
  Quantum Systems: From Discrete Particles to Continuum Hydrodynamics},''
  \href{http://arxiv.org/abs/2602.16045}{{\tt arXiv:2602.16045 [quant-ph]}}.

\bibitem{Crossley:2015evo}
M.~Crossley, P.~Glorioso, and H.~Liu, ``{Effective field theory of dissipative
  fluids},'' \href{http://dx.doi.org/10.1007/JHEP09(2017)095}{{\em JHEP} {\bf
  09} (2017)  095}, \href{http://arxiv.org/abs/1511.03646}{{\tt
  arXiv:1511.03646 [hep-th]}}.

\bibitem{Weinberg:2003ur}
S.~Weinberg, ``{Damping of tensor modes in cosmology},''
  \href{http://dx.doi.org/10.1103/PhysRevD.69.023503}{{\em Phys. Rev. D} {\bf
  69} (2004)  023503}, \href{http://arxiv.org/abs/astro-ph/0306304}{{\tt
  arXiv:astro-ph/0306304}}.

\bibitem{Salcedo:2026cqb}
S.~A. Salcedo, T.~Colas, L.~Dufner, and E.~Pajer, ``{Phenomenology of an Open
  Effective Field Theory of Dark Energy},''
  \href{http://arxiv.org/abs/2603.12321}{{\tt arXiv:2603.12321 [astro-ph.CO]}}.

\bibitem{Creminelli:2014wna}
P.~Creminelli, J.~Gleyzes, J.~Nore{\~n}a, and F.~Vernizzi, ``{Resilience of the
  standard predictions for primordial tensor modes},''
  \href{http://dx.doi.org/10.1103/PhysRevLett.113.231301}{{\em Phys. Rev.
  Lett.} {\bf 113} (2014) no.~23, 231301},
  \href{http://arxiv.org/abs/1407.8439}{{\tt arXiv:1407.8439 [astro-ph.CO]}}.

\end{thebibliography}\endgroup

\end{document}